\documentclass{article}

\usepackage{arxiv}

\usepackage[utf8]{inputenc} % allow utf-8 input
\usepackage[T1]{fontenc}    % use 8-bit T1 fonts
\usepackage{hyperref}       % hyperlinks
\usepackage{url}            % simple URL typesetting
\usepackage{booktabs}       % professional-quality tables
\usepackage{amsfonts}       % blackboard math symbols
\usepackage{nicefrac}       % compact symbols for 1/2, etc.
\usepackage{microtype}      % microtypography
\usepackage{lipsum}
\usepackage{graphicx}
\usepackage{hyperref}
\usepackage{epstopdf, epsfig}
\usepackage{bm}
\usepackage{amsthm}
\usepackage{amsmath,amssymb}
\usepackage[utf8]{inputenc}
\usepackage[T1]{fontenc}
\usepackage{mathptmx}
\usepackage[numbers]{natbib}
\usepackage{color}
\usepackage{subfigure}
\usepackage[ruled,linesnumbered]{algorithm2e}
\SetKwComment{Comment}{$\triangleright$\ }{}

\title{State estimation with limited sensors -- A deep learning based approach}

\author{
  Yash Kumar \\
  Department of Mechanical Engineering\\
  Delhi Technological University\\
  Shahbad Daulatpur, Main Bawana Road, Delhi-110042, India \\
  \texttt{yashk8481@gmail.com} \\
   \And
 Pranav Bahl \\
  Department of Mechanical Engineering\\
  Delhi Technological University\\
  Shahbad Daulatpur, Main Bawana Road, Delhi-110042, India \\
  \texttt{bahlpranav24@gmail.com} \\
  \And
 Souvik Chakraborty \\
  Department of Applied Mechanics\\
  Indian Institute of Technology Delhi\\
  Hauz Khas - 110042, New Delhi, India \\
  \texttt{souvik@am.iitd.ac.in} \\
}

\begin{document}
\maketitle

\begin{abstract}
The importance of state estimation in fluid mechanics is well-established; it is required for accomplishing several tasks, including design/optimization, active control, and future state prediction.
A common tactic in this regard is to rely on reduced-order models.
Such approaches, in general,  use measurement data of a one-time instance. However, often data available from sensors is sequential, and ignoring it results in information loss. In this paper, we propose a novel deep learning-based state estimation framework that learns from sequential data. 
The proposed model structure consists of the recurrent cell to pass information from different time steps, enabling this information to recover the full state.
We illustrate that utilizing sequential data allows for state recovery from minimal and noisy sensor measurements.
For efficient recovery of the state, the proposed approach is coupled with an auto-encoder based reduced-order model. 
We illustrate the performance of the proposed approach using three examples, and it is found to outperform other alternatives existing in the literature.
\end{abstract}

% keywords can be removed

\section{\label{sec:level1}Introduction}
The integration of deep learning has benefited modern algorithms in modeling, data processing, prediction, and control of various engineering systems. In fluid mechanics, work on machine learning implementation started last decade and has grown since then. \cite{MILANO20021} used a neural network to reconstruct turbulence flow fields and the flow in the near-wall region of a channel flow using wall information. 
\cite{ling_kurzawski_templeton_2016} and \cite{geneva2019quantifying}, in two separate works, used deep learning algorithms to improve a Reynolds-averaged Navier Stokes turbulence model.
Recently, \cite{geneva2020multi} proposed a multi-fidelity deep learning framework for turbulent flow. \cite{erichson2019shallow} used a shallow network for estimating 2D state from measurements over cylinder surface.
In this paper, we are particularly interested in applying deep learning for state estimation in fluid mechanics, and hence, the discussion hereafter is focused on the same.

State estimation is the ability to recover flow based on a few noisy measurements. It is an inverse problem and arises in many engineering applications such as remote sensing, medical imaging, ocean dynamics, reservoir modeling, and blood flow modeling. Controlling the flow and optimizing machine design in these applications depend upon the ability to predict the state with given sensors.
The challenge associated with state estimation is two-fold. Firstly, for almost all practical cases, state-estimation is an ill-posed problem, and hence, a unique solution to the problem does not exist \citep{rosasco2004learning}.
Secondly, for practical problems, the number of sensors available is often limited. As a consequence, one has to deal with a sparse data set \citep{nayek2018mass}. 

Attempts for state estimation dates back to 1960 when the Kalman filter based approaches were used for state estimation \citep{sarkka2013bayesian}.
This method assumes the system's dynamics to produce full state and updates it based on new measurements to reduce estimation error forming a closed feedback loop. 
However, the classical Kalman filter based approaches are only applicable for linear dynamical systems \citep{bishop2006pattern}.
Improvements to Kalman filter algorithms, such as Extended Kalman filter \citep{reif1999stochastic} and Unscented Kalman filter \citep{wan2001unscented} algorithms can also be found in the literature. \citep{nayek2019gaussian} attempts to generalize Kalman filters by using the Gaussian process. Approaches based on observer dynamical system uses a reduced-order model to predict the future based on the past while simultaneously corrected by receiving measurements. \cite{Jonathan} applies dynamic mode decomposition as a reduced-order model to Kalman smoother estimate to identify coherent structures.
\cite{BUFFONI20082626} used a nonlinear observer-based on Galerkin projection of Navier-Stokes equation to estimate POD coefficients.
The use of Bayes filters, such as the Kalman and particle filters, in conjugation with POD based ROM on various flow problems, can also be found in the literature \citep{Kikuchi_2015,MONS2016255,art}.

Another major category of approaches includes library-based approaches and stochastic approaches. Library-based approaches use offline data, and the library consists of generic modes such as Fourier, wavelet, discrete cosine transform basis or data specific POD or DMD modes, or training data. Library based approaches using sparse representation assumes state can be expressed as the combination of library elements. Sparse coefficients are obtained by solving pursuit problem \cite{4385788, 258082}. \cite{Callaham_2019} used sparse representation and training data as the library with localized reconstruction for reconstructing complex fluid flows. Gappy POD \citep{gappy} estimates POD coefficients in a least-square sense and uses a library of POD modes. However, it is prone to ill-conditioning and is dealt with using the best sensor placements \citep{Cohen} to improve the condition number\citep{WILLCOX2006208}. 

The most explored approach for state estimation is perhaps the one based on stochastic estimation. 
The idea was first proposed by \cite{Adrian} for a turbulence study where the conditional mean was approximated using a power series. 
In a linear setting, coefficients are computed using a two-point, second-order correlation tensor.
Other variants like quadratic stochastic estimation \citep{doi:10.1063/1.1389284} and spectral linear stochastic estimation \citep{10.1007/978-94-011-4601-2_33} can also be found in the literature. \cite{doi:10.1063/1.857396} proposed to include time delayed measurements to further improve accuracy. \cite{Bonnet1994} extended stochastic approach to estimate POD coefficients. A linear mapping between sensors and coefficients was assumed. Recently, \cite{nair_goza_2020} used a neural network to learn a non-linear mapping between sensor measurements and POD coefficients. These approaches allow more flexibility in sensor placements and have been applied for flow control over airfoil \citep{Aero} and analyzing isotropic turbulence \citep{tur, doi:10.1063/1.863130} and turbulent boundary layers \citep{doi:10.1063/1.857396,doi:10.1063/1.1389284}.

One limitation associated with all the approaches mentioned above resides in the fact that spatial information of a single sample is used to recover the full state, but often, data is sequential. Ignoring the sequence of the data during state information invariably results in information loss.
To address this apparent shortcoming, we propose a novel deep learning-based non-intrusive framework for state estimation that learns from sequential data.
The proposed framework couples recurrent neural network (RNN) with auto-encoder (AE).
While AE is used to learn the nonlinear manifold, RNN is employed to take advantage of the time series data.
We illustrate that by utilizing sequential data, the proposed framework is able to estimate the state in a more accurate fashion. Perhaps, more importantly, the number of sensors required is significantly less.
For showcasing the performance of the proposed framework, two benchmark problems involving periodic vortex shedding and transient flow past a cylinder are considered.
Results obtained are compared with those obtained using other state-of-the-art techniques.

The remainder of the paper is organized as follows.
In Section \ref{sec:ps}, details on the problem statement is provided.
A brief review of RNN and AE are furnished in Section \ref{sec:br}.
Details on the proposed approach are provided in Section \ref{sec:pa}.
Section \ref{sec:ne} presents two numerical examples to illustrate the performance of the proposed approach.
Finally, Section \ref{sec:conc} provides the concluding remarks.

\section{Problem statement}
\label{sec:ps}
We consider a dynamical system obtained by partial discretization of the governing differential equations:
\begin{equation}\label{eq:state_equn}
    \dot {\bm w} = \bm f \left( \bm w, t; \bm \theta \right), \;\;\; \bm w (t_n, \bm \theta) = \bm w^n \left( \bm \theta \right),
\end{equation}
where $\bm w \in \mathbb R^{N_w}$ represents the high-dimensional state vector that depends on parameters $\bm \theta \in \mathbb R^{N_d}$ and time $t \in \left[ 0, t_{max} \right]$.
$\bm f$ in Eq. (\ref{eq:state_equn}) is a nonlinear function that governs the dynamical evolution of the state vector $\bm w$.
Note that for brevity of representation, we have not shown the dependence of the state $\bm w$ on $t$ and $\bm \theta$, and the dependence of $\bm f$ on $\bm w, t$ and $\bm \theta$.
%The domain of Eq. (\ref{eq:state_equn}) in shown in Fig. \ref{fig:problem_domain}.

\begin{figure}
    \centering
    \includegraphics[width = \textwidth]{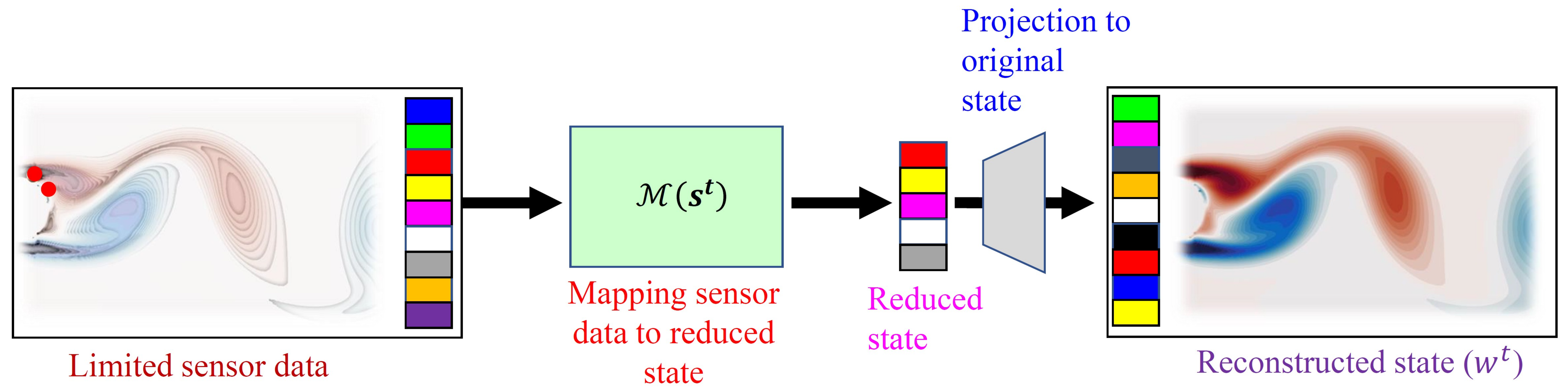}
    \caption{Schematic representation of state estimation framework. It consists of two components, (a) $\mathcal M (\bm s )$ - it maps the sensor data to a reduced state and (b) a projection operator - it maps the predicted reduced state to the original state. Performance of a state estimation framework is dependent on performance of this two components and the quality and quantity of data.}
    \label{fig:rom_se}
\end{figure}

We note that the state vector $\bm w$ is high-dimensional in nature and it is extremely difficult to directly work with $\bm w$.
A commonly used strategy in this regards is to 
approximate the high-dimensional state vector $\bm w$ on a low-dimensional manifold,
\begin{equation}\label{eq:rom}
    \bm w \left(t;\bm \theta \right) \approx \bm w_r \left( t; \bm \theta \right) = \bm \Phi \left( \bm a \left( t; \bm \theta \right) \right),
\end{equation}
where $\bm a \left( t; \bm \theta \right) \in \mathbb R^{N_a}$ represents the reduced space and $\bm \Phi \left( \cdot \right): \bm a \left( t; \bm \theta \right)  \mapsto \bm w \left( t; \bm \theta \right)  $ is the manifold. $N_a$ in Eq. (\ref{eq:rom}) represent the dimension of reduced space such that $N_a  \ll N_w$.
Substituting Eq. (\ref{eq:rom}) into Eq. (\ref{eq:state_equn}), we obtain
\begin{equation}\label{eq:state_equn_rom}
    \dot {\bm a} = \bm \Psi \left( \bm f \left( \bm \Phi\left( a \right), t;\bm \theta \right)\right), \;\bm a(t_n;\bm \theta) = \bm a^n(\bm \theta), 
\end{equation}
where 
$\bm \Psi (\cdot ) = \left( \bm \Lambda \bm \Gamma ^{-1} \right)  \circ \bm \Lambda \left( \cdot \right)$ and
$\bm \Lambda : \mathbb R^{N_w} \mapsto \mathbb R^{N_a}$.
In Eq. (\ref{eq:state_equn_rom}), we have assumed that $\bm \Phi (\bm a )$ is continuously differentiable such that 
$\bm \Gamma (\dot {\bm a}) = \dot{\bm \Phi}(\bm a)$; $\bm \Gamma: \mathbb R^{N_a} \mapsto \mathbb R^{N_w}$.
$a^n(\bm \theta)$ in Eq. (\ref{eq:state_equn_rom}) represents the initial condition.
With this representation, the objective in state-estimation reduces to estimating the reduced order state variable $\bm a^n \left( \bm \theta \right)$.
Generally, this is achieved by determining a  mapping $\mathcal M : \mathbb R^{N_s} \mapsto \mathbb R^{N_a}$ between the sensor measurements and reduced state,
\begin{equation}
    \bm a^n = \mathcal M (\bm s^n (\bm \theta)),
\end{equation}
where $\bm s^n \in \mathbb R^{N_s}$ represents the sensor measurements at time-step $n$ and
$N_s$ indicates the number of sensors present.
A schematic representation of the same is shown in Fig. \ref{fig:rom_se}.

The state estimation framework discussed above has two major limitations.
\begin{itemize}
    \item We note that the state estimation framework only relies on sensor responses at the current step for predicting the state variables.
    In other words, the sequential nature of the sensor measurements is ignored.
    This invariably results in loss of information, and hence, the accuracy of the state estimation is compromised.
    \item Secondly, as shown in Eq. (\ref{eq:rom}), the use of reduced-order model results in information loss. While completely avoiding this information loss is unavoidable, it is necessary to ensure that this information loss is minimized.
\end{itemize}
This paper aims to develop a deep learning-based framework for state estimation that addresses the two limitations discussed above.

\section{A brief review of AE and RNN}
\label{sec:br}
This section briefly reviews two poplar deep learning approaches, namely auto-encoders (AE) and recurrent neural networks (RNN).
It is to be noted that AE and RNN form the backbone of the proposed approach.
\subsection{Auto-encoders}
\label{subsec:ae}
\begin{figure}
    \centering
    \includegraphics[width = 0.7\textwidth]{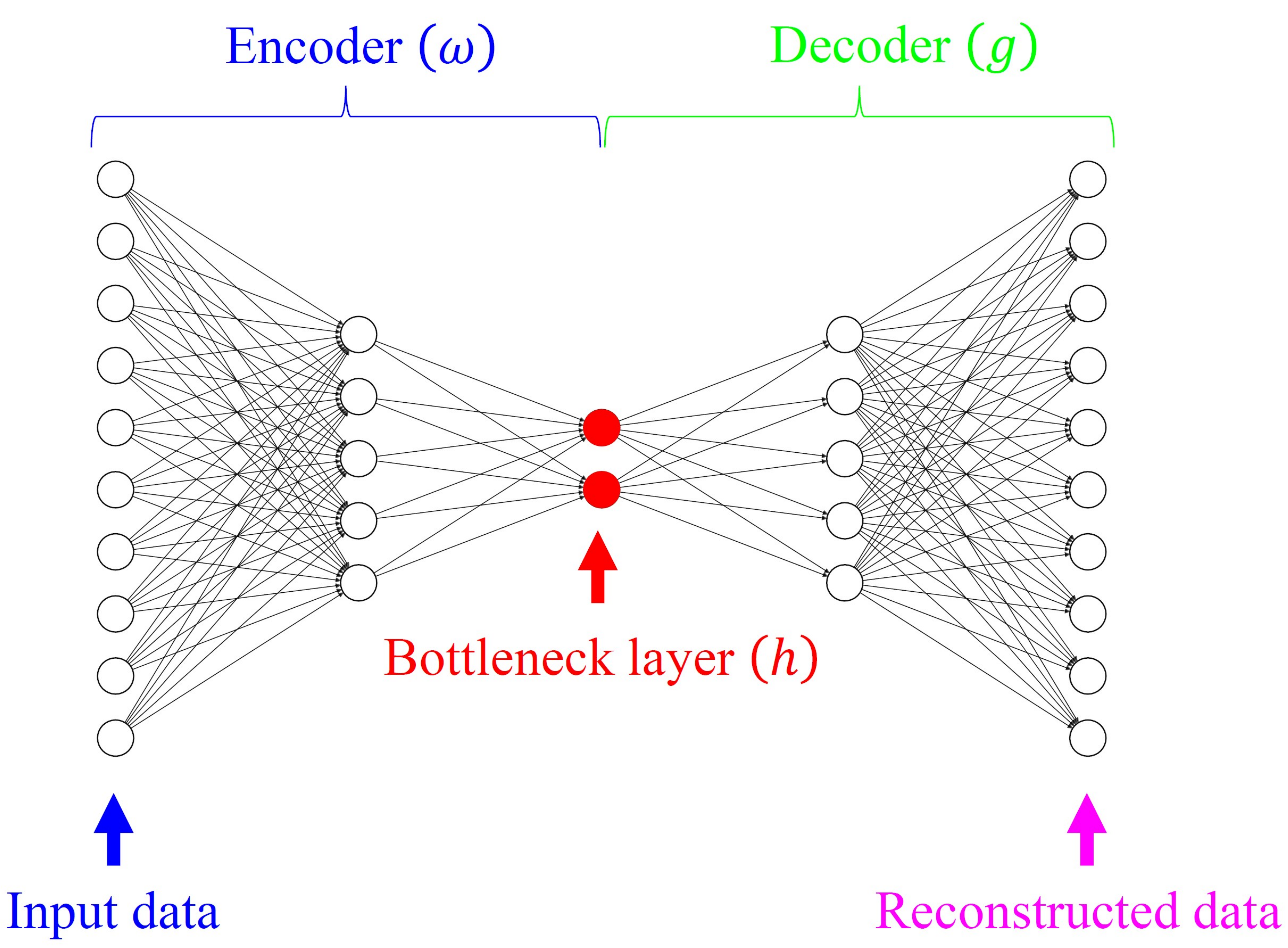}
    \caption{Schematic representation of an auto-encoder (AE). It consists of encoder ($\omega$) and ($g$). The red nodes represents the bottleneck layer.}
    \label{fig:ae}
\end{figure}

AE is a class of unsupervised deep learning techniques trained to copy the inputs to the output.
It consists of a latent space/hidden layer $\bm h \in \mathbb R^h$ that represents a compressed representation of the input; this is often referred to as the bottleneck layer.
The network architecture for an AE can be viewed as having two parts, an encoder that maps the input to the latent space $\bm h$ and a decoder that reconstructs the inputs from the latent space. Mathematically, this is represented as
\begin{equation} \label{eq:ae}
    \bm h = \bm \omega \left( \bm \xi \right),\; \hat {\bm \xi} = \bm g \left( \bm h \right),
\end{equation}
where $\bm \omega$ represents the encoder network and $\bm g$ represents the decoder network.
$\bm \xi \in \mathbb R^N$ in Eq. (\ref{eq:ae}) represents the input variables with $N$ being the number of variables.
A schematic representation of AE is shown in Fig. \ref{fig:ae}.

AE is integral to the neural network landscape and was initially developed for model reduction and feature extraction.
As far as training an AE is concerned, an adaptive learning rate optimization algorithm(ADAM) is a popular choice.
The learning in AE is generally expressed as
\begin{equation}\label{eq:ae_train}
    \bm \alpha = \arg \min_{\bm \alpha}\mathcal L(\bm \xi_t, \bm g\left( \bm \omega (\bm \xi) \right); \bm \alpha), 
\end{equation}
where $\mathcal L (\bm \xi_t, \bm g\left( \bm \omega (\bm \xi) \right); \bm \alpha)$
represents the loss-function and  $\bm \alpha = \left[ \bm \alpha_e, \bm \alpha_d \right]$
are the hyperparameters (weights and biases) of the neural network.
$\bm \alpha_e$ corresponds to the hyperparameters of the encoder while $\bm \alpha_d$ corresponds to the hyperparameters of the decoder.
Some important remarks on AE are furnished below

\textbf{Remark 1:} A situation where $\bm g\left( \bm \omega (\bm \xi) \right) = \bm \xi$ everywhere needs to be avoided \citep{murphy2012machine}. In other words, the training algorithm is designed in such a way to restrict direct copying of the input.

\textbf{Remark 2:} In AE, the dimensionality of the latent space $\bm h \in \mathbb R^h$ is generally much smaller than the dimensionality of the input variable $\bm \xi \in \mathbb R_N$, $h \ll N$. Therefore, $\bm h$ can be thought of as a reduced-order representation of $\bm \xi$.

\textbf{Remark 3:} When the decoder $\bm g$ is linear, and $\mathcal L$ is a mean-squared error, an AE learns to span the same subspace as principal component analysis.

\textbf{Remark 4:} AE with nonlinear encoder $\bm \omega$ and nonlinear decoder $\bm g$ learns a nonlinear reduced-order manifold; however, an AE with too much expressive capacity learns to copy the inputs (Remark 1).

It is to be noted that researchers are still working on developing AE for various types of tasks. Some of the popular AE available in the literature includes variational AE \citep{khemakhem2020variational}, sparse AE \citep{glorot2011deep}, stochastic AE \citep{bahuleyan2018stochastic} and capsule AE \citep{kosiorek2019stacked} among other.
For further details on different types of AE, interested readers may refer \citep{murphy2012machine}.

% Auto-encoders are unsupervised deep learning networks. They are generally used for data compression and decompression. Their structure forces data through bottleneck layers with a smaller number of nodes.
% The middle layer in auto-encoders is called the bottleneck layer. They are nonlinear extensions of POD and produce nonlinear modes of the system. Baldi and Hornik\cite{BALDI198953} showed that symmetric linear autoencoder is closely related to POD. Milano at al.\cite{MILANO20021} used them to reconstruct the near-wall velocity field in a turbulent channel flow using wall pressure and shear. Autoencoders can be used for dimensionality reduction as used by Eivazi et al. \cite{Eivazi_2020}, Wang et al.\cite{7572934}

% The network consists of two parts in which first maps the data input to a latent space with very few variables, and second, maps latent state variables back to the original input. 
% \begin{equation} \label{eu_eqn}g^{t} = \omega(u^{t}), \hat{u}^{t} = \psi(g^t)\end{equation}
% where $\omega$ represents encoder network and $\psi$ represents decoder network. $u^t$ in current case is vorticity field data of flow behind cylinder on 2D regular grid which is flattened to form a long vector.

\subsection{Recurrent neural networks}
\label{subsec:rnn}
\begin{figure}
    \centering
    \includegraphics[width=0.3\textwidth]{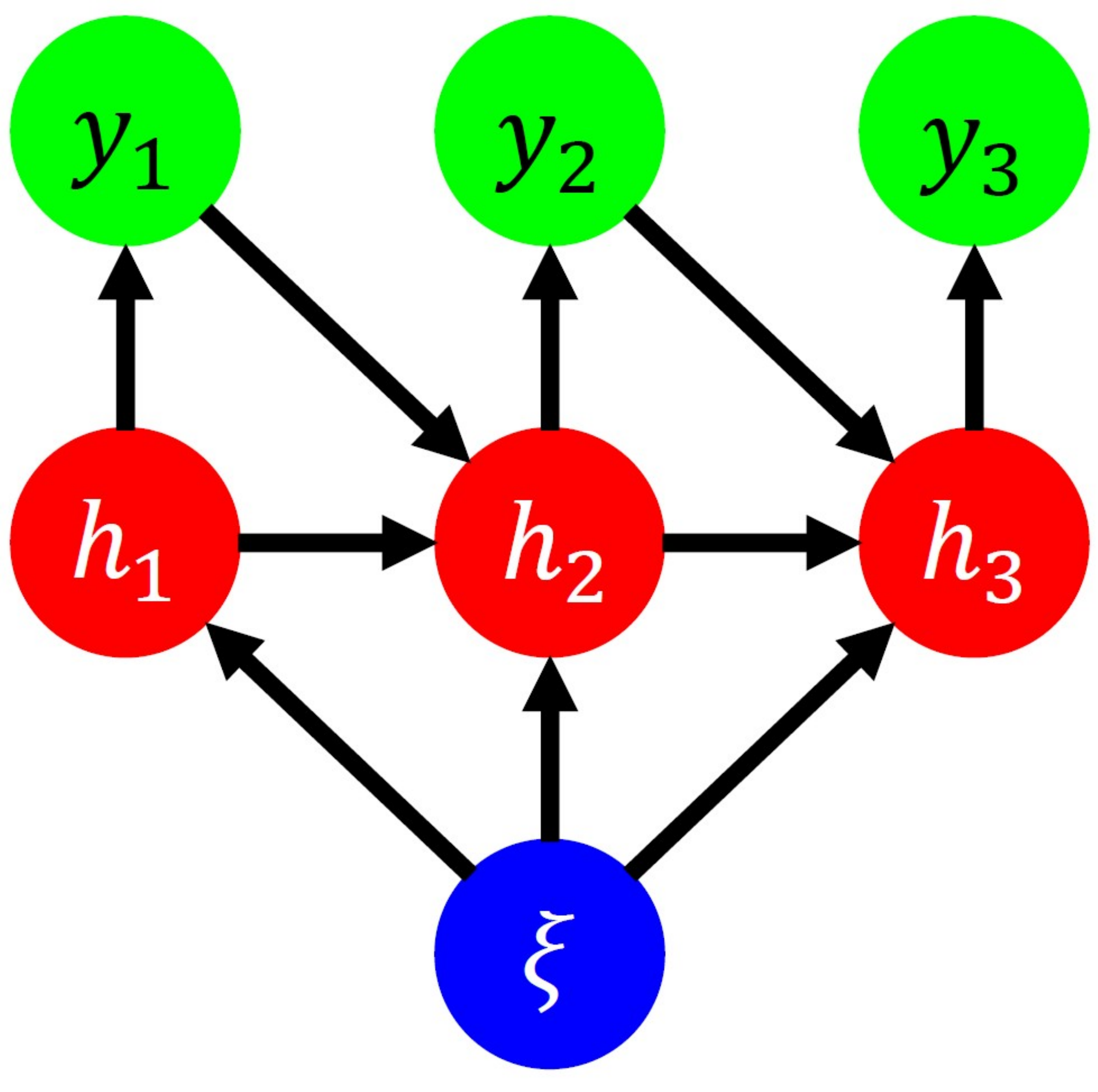}
    \caption{Schematic representation of a typical RNN. $\xi$ (blue node) represents the input, $h_i, \forall i$ (red node) represents the hidden layer and $y_i$ (green node) is the output sequence.}
    \label{fig:rnn}
\end{figure}

Many of the learning tasks involved in artificial intelligence necessitate handling data that are sequential in nature.
Examples of sequential data include image captioning, time-series forecasting, and speech synthesis, among others.
A recurrent neural network (RNN) is a type of neural network that is particularly suitable for sequential data.
RNN captures the time dynamics by using cycles in the graph. Consider $\bm \xi$ to be inputs and $\bm y_t$  to be the output at time $t$.
The output $\bm y_t$ in RNN is expressed as a function of $\bm \xi$ and $\bm h_t$; however, owing to the cyclic graph in RNN, the hidden state $\bm h_t$ is continuously updated as the sequence is processed.
A schematic representation of a simple RNN is shown in Fig. \ref{fig:rnn}.
Note that different variants to the classical RNN can be found in the literature.
In this work, we have used a Long Short-Term Memory \citep{LSTM}, and hence, the discussion hereafter is focused on the same.
Readers interested in other types of RNN may refer to \citep{lipton2015critical}.

\begin{figure}
    \centering
    \includegraphics[width=0.6\linewidth]{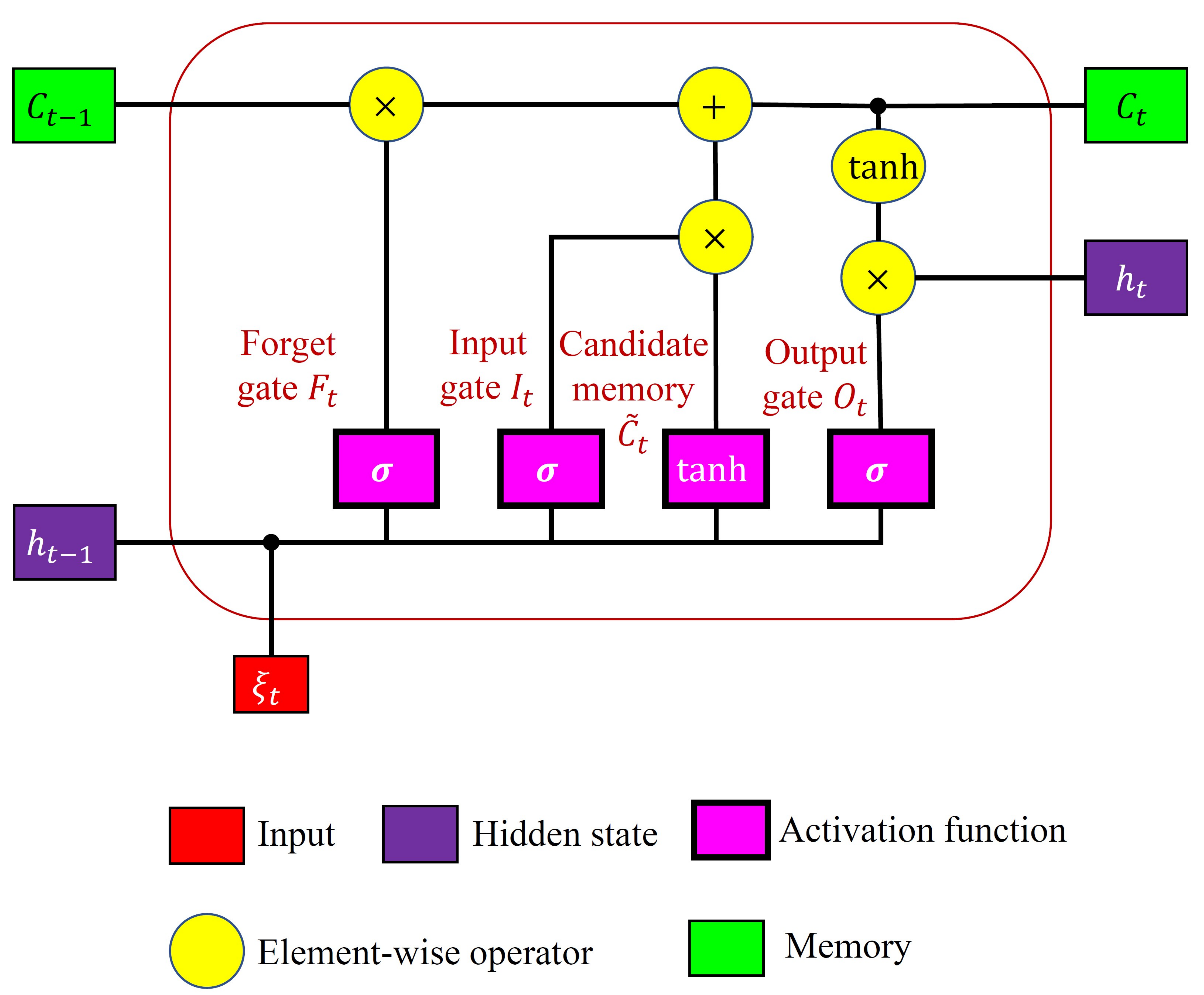}
    \caption{Schematic of the LSTM cell}
    \label{fig:lstmhh}
\end{figure}

LSTM, first proposed by \cite{LSTM}, is a type of gated RNN cell that overcomes the well-know issue of vanishing gradient with the help of the gates that control the flow of information, i.e., differentiates between the information to be updated and that to be deleted \citep{doi:10.1063/5.0012853, DBLP:journals/corr/abs-1912-13382, doi:10.1142/S0218488598000094}. 
%\cite{graves2007multidimensional} proposed multidimensional LSTM, which can handle multidimensional data efficiently. The inability to forget the information leads to propagation of increasing residual error in the Neural Net, which decreases its efficiency to handle long sequences \citep{DBLP:journals/corr/abs-1912-13382}. 
LSTM cell comprises a forget gate, input gate, output gate, and a cell state. Each of these has its significance. The cell state refers to the information that has to be transferred in the sequence, and the respective gates determine the information that has to updated or deleted from the cell state \cite{doi:10.1063/5.0020526}. 
The output of the current cell, also referred to as the hidden state, helps retain the short-term memory, and the cell state, on the other hand, is used to retain the Long-term Memory. 
Cell state in LSTM is multiplied with the forget gate in each cell along with the addition from the input gate, and this provides the opportunity for forgetting gate to eradicate the unimportant information and input gate to enhance state with useful information \cite{doi:10.1063/5.0020526}.
A schematic representation of LSTM cell is shown
in Fig. \ref{fig:lstmhh}.
Mathematical, the operations being carried out inside a LSTM cell is represented using the following equations.
\begin{subequations}
 \begin{equation}
     C_t = \mathbf F_t\odot \mathbf C_{t-1}+ \mathbf I_t\odot \mathbf{\tilde{C}}_t
 \end{equation}
 \begin{equation}
     \mathbf{\tilde{C}}_t = \tanh (\mathbf \xi_t \mathbf W_{x c} + \mathbf h_{t-1} \mathbf W_{h c} + \mathbf b_c
 \end{equation}
 \begin{equation}
     \mathbf I_t = \sigma( \mathbf \xi_t \mathbf \xi_{xi} + \mathbf h_{t-1} \mathbf W_{hi} + \mathbf b_i
 \end{equation}
 \begin{equation}
     \mathbf F_t = \sigma(\mathbf \xi_t \mathbf \xi_{x f} + \mathbf h_{t-1} \mathbf W_{hf} + \mathbf b),
 \end{equation}
\end{subequations}
where $\mathbf I_t$, $\mathbf F_t$ and $\tilde {\mathbf C}_t$ are respectively the input gate, forget gate, and a candidate cell state.
Note that the update is carried out is additive in nature; this allows long-term information to pass through and avoids the gradient from vanishing.
The short term state in LSTM is calculated as
\begin{subequations}
 \begin{equation}
     \mathbf O_t = \sigma \left( \mathbf \xi_t \mathbf \xi_{xo} + \mathbf h_{t-1} \mathbf W_{ho} + \mathbf b_o \right)
 \end{equation}
 \begin{equation}
     \mathbf h_t = \mathbf O_t \odot \tanh (\mathbf C_t),
 \end{equation}
\end{subequations}
where $\mathbf O_t$ represents the output gate.
It is to be noted that $\mathbf h_t$ is used as the output of the cell as well as the hidden state for the next time-step; this is responsible for the short term memory of LSTM.
$\mathbf C_t$ on the other hand is responsible for long term memory.

The use of RNN for complex dynamical systems has attracted significant interest from the research community; this is primarily because of its capability in capturing temporal dependencies  \cite{doi:10.1063/5.0012853}. 
Multiple architectures have been proposed for using RNN for accomplishing the task future state prediction of a dynamical system. 
Recently \cite{geneva2020transformers} and \cite{Eivazi_2020} used LSTM  and transformers as time integrator to predict flow evolution state. \cite{s18051316} used simple RNN for IMU modeling in deep Kalman filter. \cite{otto2019linearlyrecurrent} used autoencoder with linear recurrence to learns important dynamical features.
In this work also, we use LSTM for extracting useful information from sequential sensor data.
The next section provides more details on the same.

\section{Proposed approach}
\label{sec:pa}
In this section, we propose a novel deep learning based framework for state estimation.
The proposed approach integrates the two deep learning approaches discussed in 
Section \ref{sec:br}, namely AE and RNN.
Within the proposed framework, AE learns the reduced nonlinear subspace.
It helps in reducing the information loss due to the compressed representation.
RNN, on the other hand, extends the capability of the proposed approach and allows it to reuse sensor data collected at previous time-steps.
AE (by reducing the state variable) and RNN (by incorporating information from the past) also helps address the ill-posedness associated with solving a state estimation problem.

\begin{figure}
    \centering
    \includegraphics[width = \textwidth]{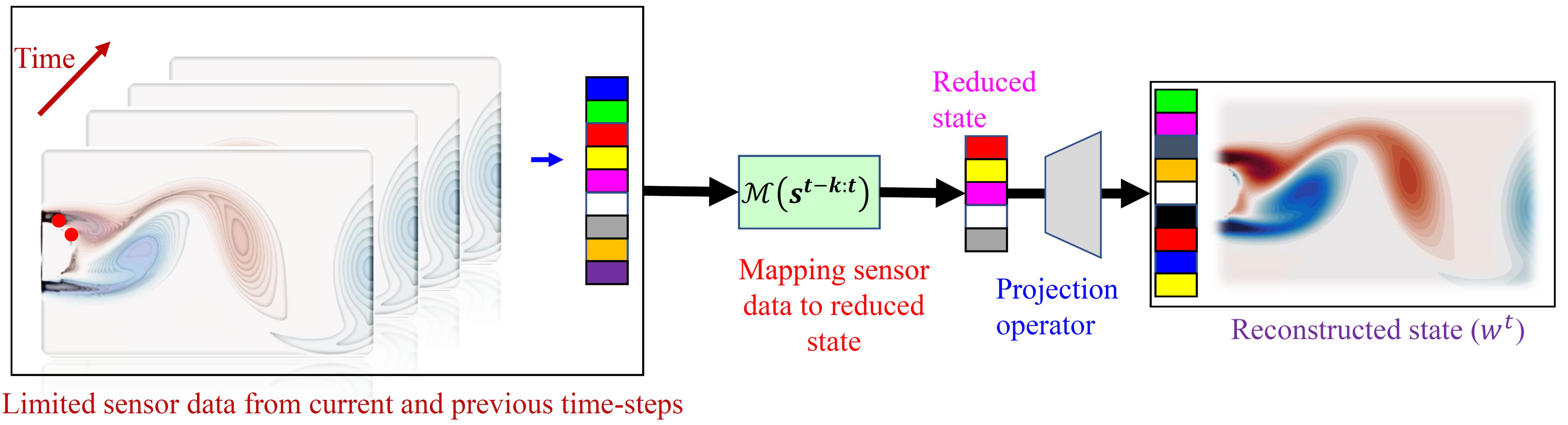}
    \caption{Schematic representation of the proposed approach. The difference with Fig. \ref{fig:rom_se} resides in the fact that information from previous time-steps are also utilized within the proposed framework. We propose to model $\mathcal M\left(\cdot;\bm \theta_M\right)$ using RNN and $g \left(\cdot; \bm \theta_g \right)$ by using AE.}
    \label{fig:pa1}
\end{figure}

Consider $\bm s = \left[ \bm s_1, \bm s_2, \ldots, \bm s_n \right] \in \mathbb R^{N_t \times N_s}$ represents the measurement data obtained from $N_s$ sensors over $N_t$ time-steps.
Also consider $\bm w \in \mathbb R_{N_w}$ to be state variables. We can express the state variable $\bm w^t$ at time $t$ as
\begin{equation}\label{eq:map}
    \bm w^t = g \left( \mathcal M \left(\bm s^{t-k:t}; \bm \theta_M \right); \bm \theta_g \right)
\end{equation}
where $\mathcal M \left(\cdot; \bm \theta_M\right)$ represents a mapping between
the sensor data and the reduced state variable and $g \left(\cdot;\bm \theta_g\right)$ projects the reduced state variable back to the original
space.
$\bm \theta_M$ and $\bm \theta_g$ represent parameters associated with $\mathcal M \left(\cdot; \bm \theta_M\right)$ and 
$g \left(\cdot;\bm \theta_g\right)$, respectively.
Unlike existing methods, sensor data corresponding to current and previous time-steps have been used for predicting $\bm w^t$ in Eq. (\ref{eq:map}).
A schematic representation of the same is shown in Fig. \ref{fig:pa1}.
We also note that the sensor data is sequential and hypothesize that modeling this sequential nature of the sensor data will improve the predictive capability of model $\mathcal M \left(\cdot; \bm \theta_M\right)$.
Therefore, we propose to model $\mathcal M \left(\cdot; \bm \theta_M\right)$ by using RNN.
As stated in Section \ref{subsec:rnn}, RNN is suited explicitly for modeling such sequential data.
Another aspect in Eq. (\ref{eq:map}) is associated with the projection
operator $g \left(\cdot;\bm \theta_g\right)$.
We reiterate that accuracy and efficiency of 
Eq. (\ref{eq:map}) is significantly dependent on
$g \left(\cdot;\bm \theta_g\right)$.
One popular choice among researchers is to use proper orthogonal decomposition for computing the projection operator $g$.
However, proper orthogonal decomposition being a
linear projection scheme has limited expressive capability.
In this work, we propose to use AE as $g \left(\cdot;\bm \theta_g\right)$.
Owing to the fact that AE is a nonlinear reduced order model, we expect the accuracy to enhance. However, one must note that training AE demands more computational effort as compared to proper orthogonal decomposition.
Hereafter, we refer to the proposed framework as Auto-encoder and Recurrent neural network based state Estimation (ARE) framework.
Next, details on network architecture and training algorithm for ARE are furnished.

\subsection{Network architecture and training}
\begin{figure}[t]
    \centering
    \subfigure[]{
    \includegraphics[width = 0.9\textwidth]{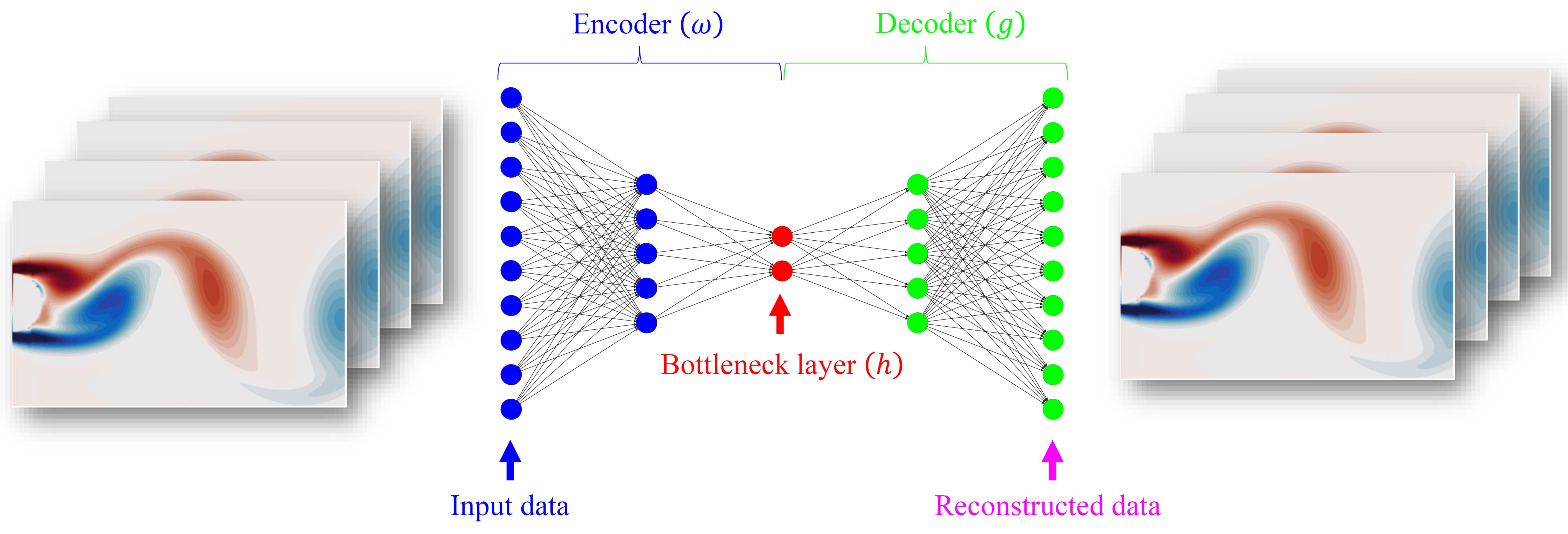}}
    \subfigure[]{
    \includegraphics[width = \textwidth]{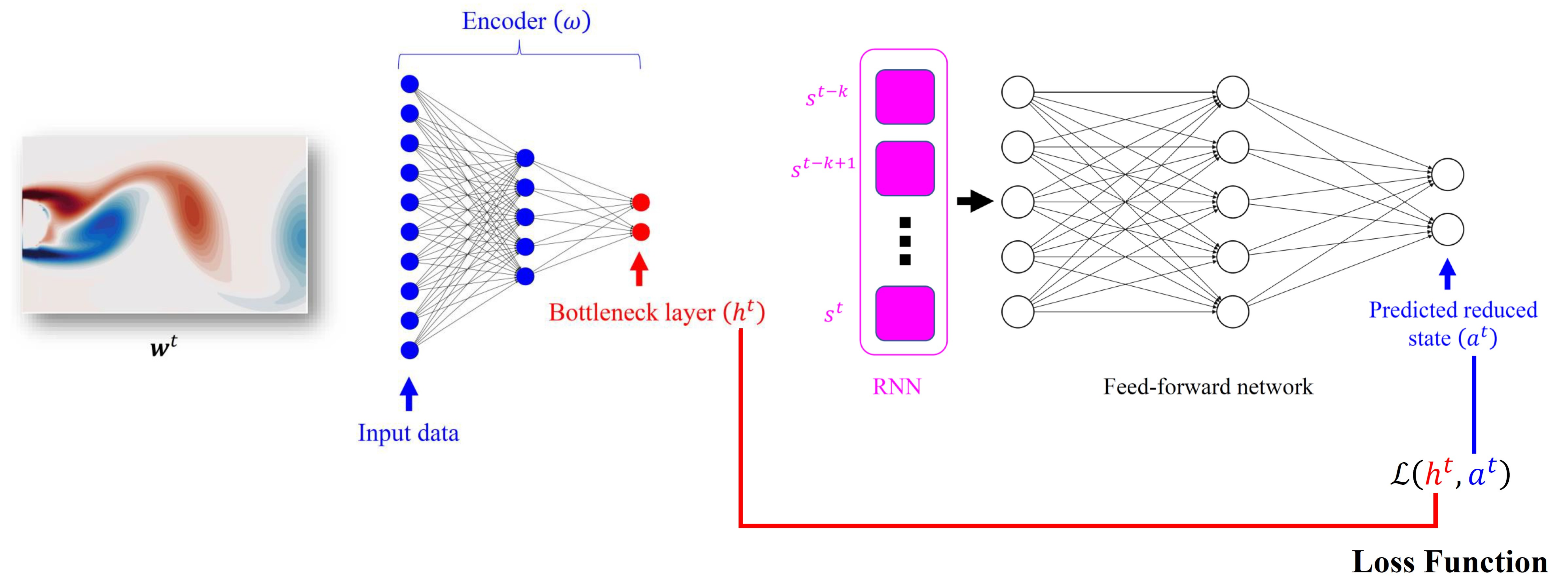}}
    \caption{Schematic representation of ARE (training phase) - (a) Training AE for reconstructing the flow-field. (b) Training the RNN for mapping the sensor data to the reduced state.}
    \label{fig:schematic}
\end{figure}

The ARE architecture proposed in this paper involves an AE and an RNN.
For state estimation, the trained AE is split into the encoder and the decoder parts.
The encoder part is used during training the RNN within the ARE framework.
The decoder part is used while estimating the state variable.
The AE and RNN networks are trained separately, with the latter following the former (see Fig. \ref{fig:schematic}). 
The AE architecture considered in this paper consists of 5 hidden layers, with the 3rd layer being the bottleneck layer.
The flow-field is vectorized before providing it as an input to the AE. 
Rectified linear unit (ReLU) activation function has been used for all but the last layer.
For the last layer, a linear activation function is used.
For training the network, ADAM \citep{Kingma} optimizer with a learning rate of $0.0007$ is used.
We denote the trained parameters of the AE as $\bm \theta_{AE} = \left[\bm \theta_w, \bm \theta_g \right]$, where $\bm \theta_w$ and  $\bm \theta_g$ corresponds to the network parameters for encoder and decoder part respectively.
For making the model robust to noisy data, two batch-norm layers and one dropout layer is added to AE. 
Batch normalization \citep{ioffe2015batch} is used to  normalizes the activation distribution. It reduces the model's sensitivity to learning rate \citep{arora2018theoretical}, reduces training time, and increases stability.
Additionally, this also acts as a regularizer.
Dropout \citep{JMLR:v15:srivastava14a} is also an effective regularization technique that works by dropping connections between neurons during training with a specified probability $p$.
Using the dropout layer just before the bottleneck layer proved to be most efficient in increasing the robustness of the network to noise as it simulates the noise in latent vector from the RNN network. A dropout probability of 0.35 is used in the network.
Considering $Li$ to be the $i$-th layer of AE, BN to be the batch normalization and DR to be the dropbout, the architecture used in this paper is as follows:
% Two batch-norm layers and one dropout layer is added to AE as shown in table \red{\ref{table:t1, table:t2}} for predicting state from noisy measurements. \red{give refrence to paper explaining droput and batch-norm} Batch normalization \citep{ioffe2015batch} is used to  normalizes the activation distribution i.e. mean zero and variance one. It has multiple advantages like reducing sensitivity to learning rate \citep{arora2018theoretical}, reduces training time and increase stability. Other than its computational advantages it also acts as a regularizer. 
% Dropout\citep{JMLR:v15:srivastava14a} is a quite effective regularization technique that works by dropping connections between neurons during training with a specified probability p. Using dropout layer just before bottleneck layer proved to be most efficient in increasing robustness to noise of network as is simulates the noise in latent vector from RNN network. A dropout probability of 0.35 is used in network.
% Other details like batch size are problem specific and has been furnished while describing the problem.
\[L1\rightarrow BN\rightarrow L2\rightarrow BN\rightarrow DR\rightarrow L3\rightarrow L4\rightarrow L5\rightarrow L5\rightarrow Output\] 

Once the AE trains, we proceed to train the RNN part. The objective here is to learn a mapping between the sequential sensor data and the reduced state variable obtained by using the encoder part of the trained AE.
First, the sensor data passes through RNN. It helps in capturing information from the sequential data. After that, the RNN outputs are mapped to the reduced states by using a feed-forward neural network.
Reduced states of the training outputs are obtained by using the trained AE (encoder part).
The parameters $\bm \theta_M$ (see Eq. (\ref{eq:map})) corresponds to the RNN and the feedforward neural network and are obtained by solving the following optimization problem
\begin{equation}\label{eq:loss_rnn}
    \bm \theta_M^* = \arg \min_{\bm \theta} \sum_{i=1}^{N_{\text{samp}}} \left\| h^t - \mathcal N \left( \bm s^{t-k : t}; \bm \theta_M \right) \right\| + \lambda\left\| \bm \theta_M  \right|_2^2,
\end{equation}
Where $\mathcal N$ represents the combined RNN and feed-forward neural network mapping. The second term in Eq. (\ref{eq:loss_rnn}) represents $L^2$ regularization and is adopted to avoid overfitting.
$\lambda$ is a tuning parameter and needs to be tuned manually.
Similar to AE, the optimization problem is solved by using ADAM optimizer \citep{Kingma}. We have used weight decay of $10^{-5}$ and a learning rate of $0.0008$.
Early stopping is used, which also acts as a regularizer \citep{Goodfellow}.
RNN training is schematically shown in Fig. \ref{fig:schematic}(b).
The steps involved in training the proposed (ARE) are shown in Algorithm \ref{alg:train}.
\begin{algorithm}
\caption{Training ARE}\label{alg:train}
\textbf{Library data generation:} Perform experiment or run CFD simulations to generate a library of data, $\mathcal D = \left[ \bm s, \bm w \right]$ where,
\[ \bm s = [\bm s^1_{1:N_s}, \ldots, \bm s^{N_t}_{1:N_s},\;\; \bm w = \left[ \bm w^1, \ldots, \bm w^{N_t} \right], \]
where $\bm w^i \in \mathbb R^{N_w}$. \\
\textbf{Initialize:} Initialize $\bm \theta_g$ and $\bm \theta_M$. Provide sequence length $k+1$ for RNN, learning rate parameters, network architectures and number of training epochs.\\
Train AE for $\bm w = \left[ \bm w^1, \ldots, \bm w^{N_t} \right]$.\\
$\bm a^t \leftarrow \omega(\bm w^t; \bm \theta_w)$\Comment*[r]{Reducing dimensionality using AE}
Train RNN for sequence length $k+1$\Comment*[r]{Eq. (\ref{eq:loss_rnn})}
\end{algorithm}

\subsection{State estimation using the proposed approach}

\begin{figure}
    \centering
    \includegraphics[width = \textwidth]{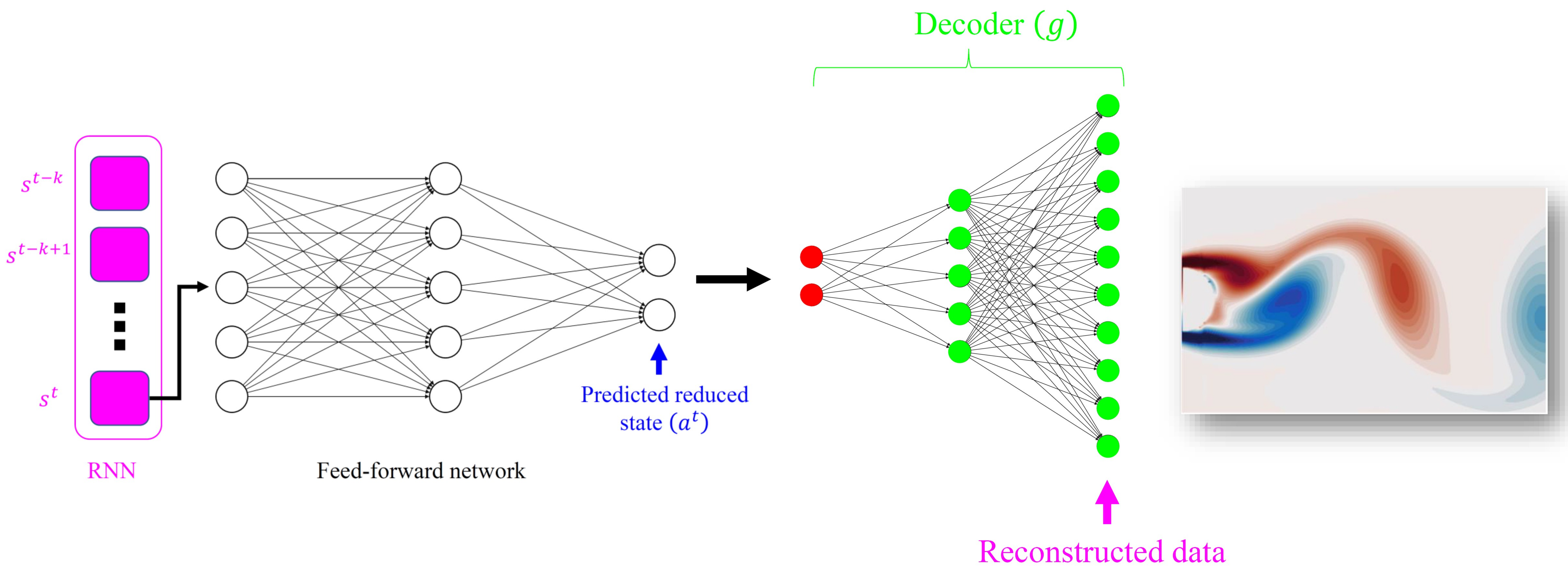}
    \caption{Schematic representation of the proposed ARE for state estimation}
    \label{fig:are_pred}
\end{figure}

Once the proposed ARE is trained by following the procedure detailed in Algorithm \ref{alg:train}, one can use it for estimating the state.
A trained ARE performs state estimation in two simple steps.
In the first step, the trained RNN is used for estimating the reduced state 
based on the sensor measurement.
Once the reduced state has been estimated, the decoder part of the AE is used to project the reduced state onto the original state. For clarity of readers, the steps for predicting state using ARE are shown in Algorithm \ref{alg:are_pred}.
A schematic representation of the same is also shown in Fig. \ref{fig:are_pred}.

\begin{algorithm}
\caption{State estimation using ARE}\label{alg:are_pred}
\textbf{Pre-requisite: }Trained ARE model, $\bm \theta = \left[ \bm \theta_M, \bm \theta_g \right]$ and sensor data $\bm s_*^{t-k: t}$. \\
$\bm a^t_* \leftarrow \mathcal N(\bm s_*^{t-k: t}; \bm \theta_M)$\Comment*[r]{Reduced state prediction using RNN}
$\bm w^t_* \leftarrow = g(\bm a^t_*; \bm \theta_g)$\Comment*[r]{Full-state estimation using decoder part of AE}
\end{algorithm}

\section{Numerical experiments}
\label{sec:ne}
In this section, two examples are presented to illustrate the performance of the proposed approach.
The examples selected are well-known benchmark problems in the fluid mechanic's community.
For both examples, we have considered that the sensor measures vorticity, and the objective is to reconstruct the vorticity field.
We present case studies by varying the number of sensors and the number of sequences available.
To illustrate the excellent performance of our approach, a comparison with another state-of-the-art method has been provided. Comparison among results is carried out based on a qualitative and quantitative metric.
To be specific, visual inspection is used as a qualitative metric, and the relative error is used as a quantitative metric,
\begin{equation}\label{eq:error}
    \epsilon = \frac{\left\| \bm \omega^n (\bm \theta^*) - \bm \omega^n_r (\bm \theta^*) \right\|_2}{\left\| \bm \omega^n (\bm \theta^*) \right\|_2}\times 100.
\end{equation}
where $\epsilon$ represents the error, $\bm \omega^n (\bm \theta^*)$ is the true state and $\bm \omega^n_r (\bm \theta^*)$ is the state vector predicted using the proposed approach.
$\left\| \cdot \right\|_2$ represents the $l_2$ norm. 
The dataset for solving the state estimation problems is generated using OpenFoam \citep{jasak2009openfoam}. The proposed approach has been implemented using PyTorch \citep{paszke2017automatic}. The software associated with the proposed approach, along with the implementation of both the examples, will be made available on acceptance of the paper.

\subsection{Periodic Vortex shedding past a cylinder}
\label{subsec:eg1}
\begin{figure}
    \centering
    \subfigure[]{
    \includegraphics[width=0.5\textwidth]{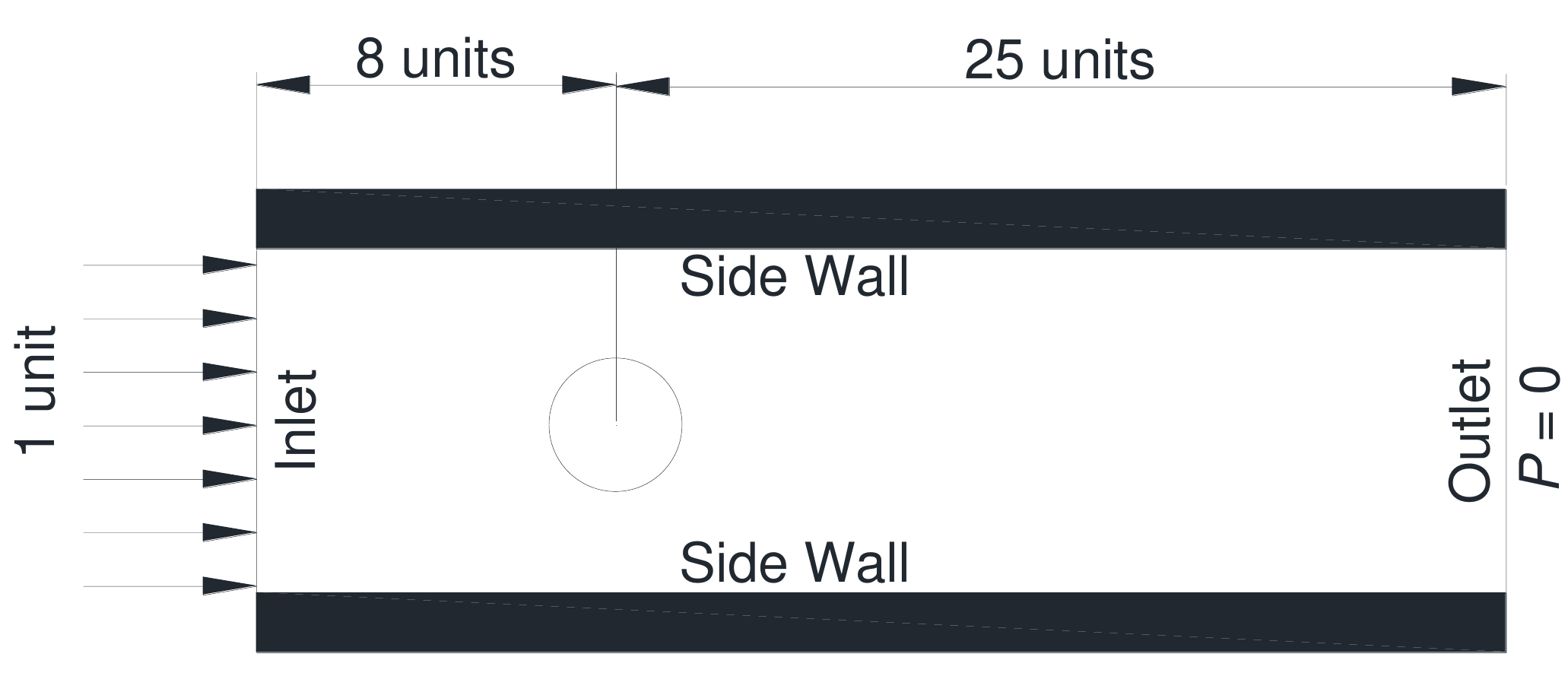}}
    \subfigure[]{
    \includegraphics[width=0.45\textwidth]{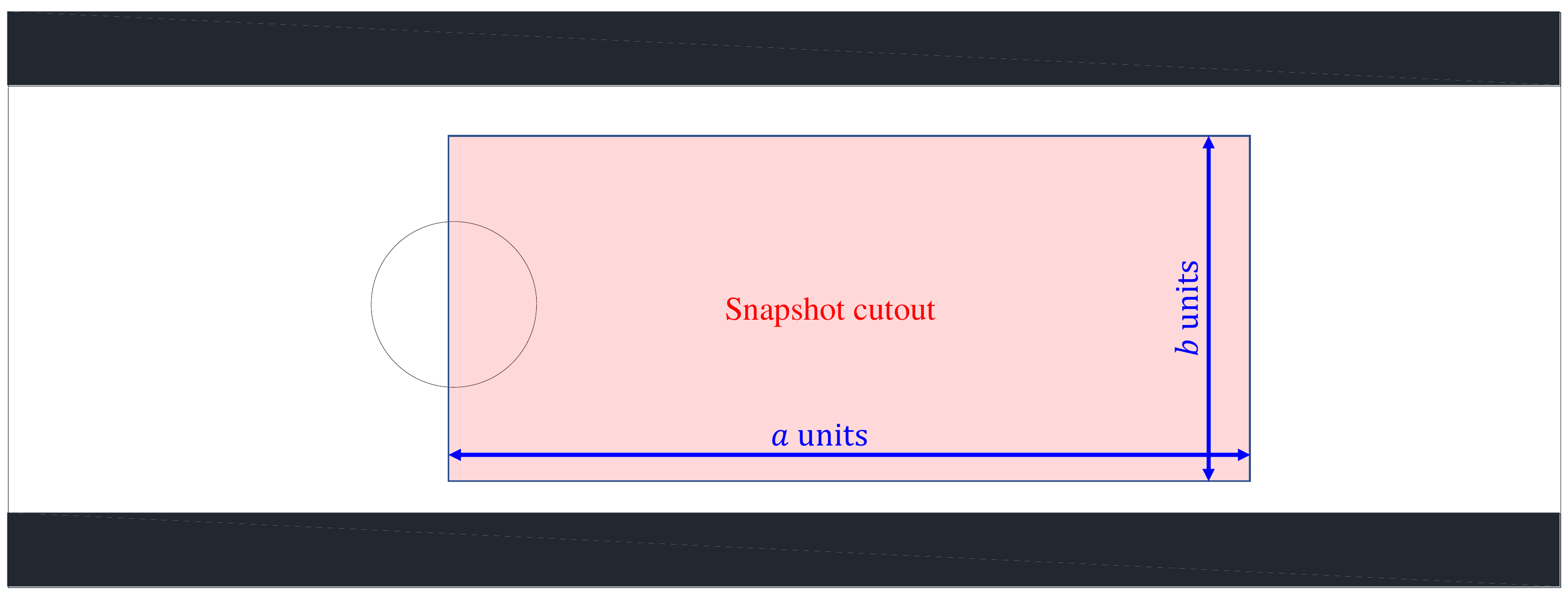}}
    \caption{(a) Schematic representation of the computational domain with boundary conditions at the inlet and the outlet. The cylinder is having a diameter of 1 unit. A no-slip boundary is considered at the cylinder wall. Zero pressure gradient at the inlet and zero velocity gradient at the outlet are considered. (b) Schematic of the problem domain with snapshot cutout of $a \times b$. For periodic vorticity problem, $a = 6$ units and $b = 4$ units. For transient flow problem, $a = 12$ units and $b = 6$ units. \it{The schematics are not to scale}.}
    \label{fig:domain}
\end{figure}

As the first example, we consider two-dimensional flow past a circular cylinder at Reynolds' number $Re = 190$.
 It is a well known canonical problem and is characterized by periodic laminar flow vortex shedding.
A schematic representation of the computational domain is shown in Fig. \ref{fig:domain}.
The circular cylinder is considered to have a diameter of $1$ unit. The center of the cylinder is located at a distance of $8$ units from the inlet. The outlet is located at a distance of $25$ units from the center of the cylinder. The sidewalls are at $4$ units distance from the center of the cylinder.
At the inlet boundary, a uniform velocity of $1$ unit along the $X$-direction is applied.
Pressure boundary condition with $P=0$ is considered at the outlet.
A no-slip boundary at the cylinder surface is considered.

The dataset necessary for training the proposed model is generated by using Unsteady Reynolds-averaged Navier Stokes (URANS) simulation in OpenFoam \citep{jasak2009openfoam}. 
The selection of URANS for data generation is based on the fact that results obtained using URANS are highly accurate in $Re\in[30,300]$ \citep{STRINGER20141}.
Nonetheless, the method proposed is not dependent on the fluid simulator used and can be seamlessly used in conjunction with more accurate simulator like DNS.
The overall problem domain is discretized into 63420 elements with finer mesh near the cylinder.
Time step $\delta t = 0.02$ units is considered. 

For training the model, a library of 180 snapshots is generated by running OpenFoam. 
Additional 120 snapshots, 60 for validation and 60 for testing, have also been generated.
Two consecutive snapshots are separated by 10$\delta t$.
Coordinate of the snapshot cutout stretches from $[0,-2]\times [6, 2]$ which is discretized into $168 \times 251$ points in $x$ and $y$ directions (see Fig. \ref{fig:domain}(b)).
The objective here is to recover the complete vorticity-field in the cutout by using the sensor measurements.
Details on the network architecture are provided in Table \ref{tab:na_eg1}.

\begin{table}
    \centering
    % \caption{Network architecture of proposed ARE for periodic vortex shedding problem. BN indicates batch normalization and DR represents dropout. In case of noise-free data, BN and DR are not used. A DR rate of 0.35 is used.}
    \caption{Network architecture of proposed ARE and SD for periodic vortex shedding problem. BN indicates batch normalization and DR represents dropout. HS is the number of features in the hidden state of lstm.}
    
    \label{tab:na_eg1}
    \begin{tabular}{|p{2cm}|p{15cm}|}
        \hline
        Networks & Architecture  \\ \hline
    ARE(Auto-Encoder) & $251\times168\rightarrow BN^* \rightarrow1024\rightarrow BN^* \rightarrow DR(0.35)^* \rightarrow 256 \rightarrow 25 \rightarrow 256 \rightarrow 1024 \rightarrow 168\times251$ \\
    ARE(RNN) & LSTM(HS=50) $\rightarrow 50 \rightarrow 50 \rightarrow 25$ \\ 
    SD & $N_s\rightarrow 35 \rightarrow BN \rightarrow DR(0.1)^* \rightarrow 40 \rightarrow BN\rightarrow 168\times251$ \\ \hline
    \multicolumn{2}{l}{ \footnotesize{Components with $^*$ are not used in case of noise-free data.}}
    \end{tabular}
\end{table}

To illustrate the superiority of the proposed approach, the results obtained using the proposed approach are compared with those obtained using proper orthogonal decomposition-based deep state estimation (PDS) proposed by \cite{nair_goza_2020} and shallow decoder (SD) proposed by \cite{erichson2019shallow}.
In PDS, the first 25 modes are used, which is the same as the number of neurons in the bottleneck layer of the proposed approach.
The feed-forward neural network used in PDS for mapping the sensor measurements to the latent state is the same as the feed-forward network used in ARE. Note that comparison with other popular approaches such as gappy-POD and linear stochastic estimation is not shown as it is already established in \cite{nair_goza_2020} that PDS outperforms both the approaches.
Brief details on PDS, gappy-POD, and linear stochastic estimation are provided in \ref{appA}.

\begin{figure}[hbt!]
\centering
\subfigure[PDS]{
\includegraphics[width=0.4\linewidth,trim=3.6cm 2.0cm 3.6cm 2.0cm,clip]{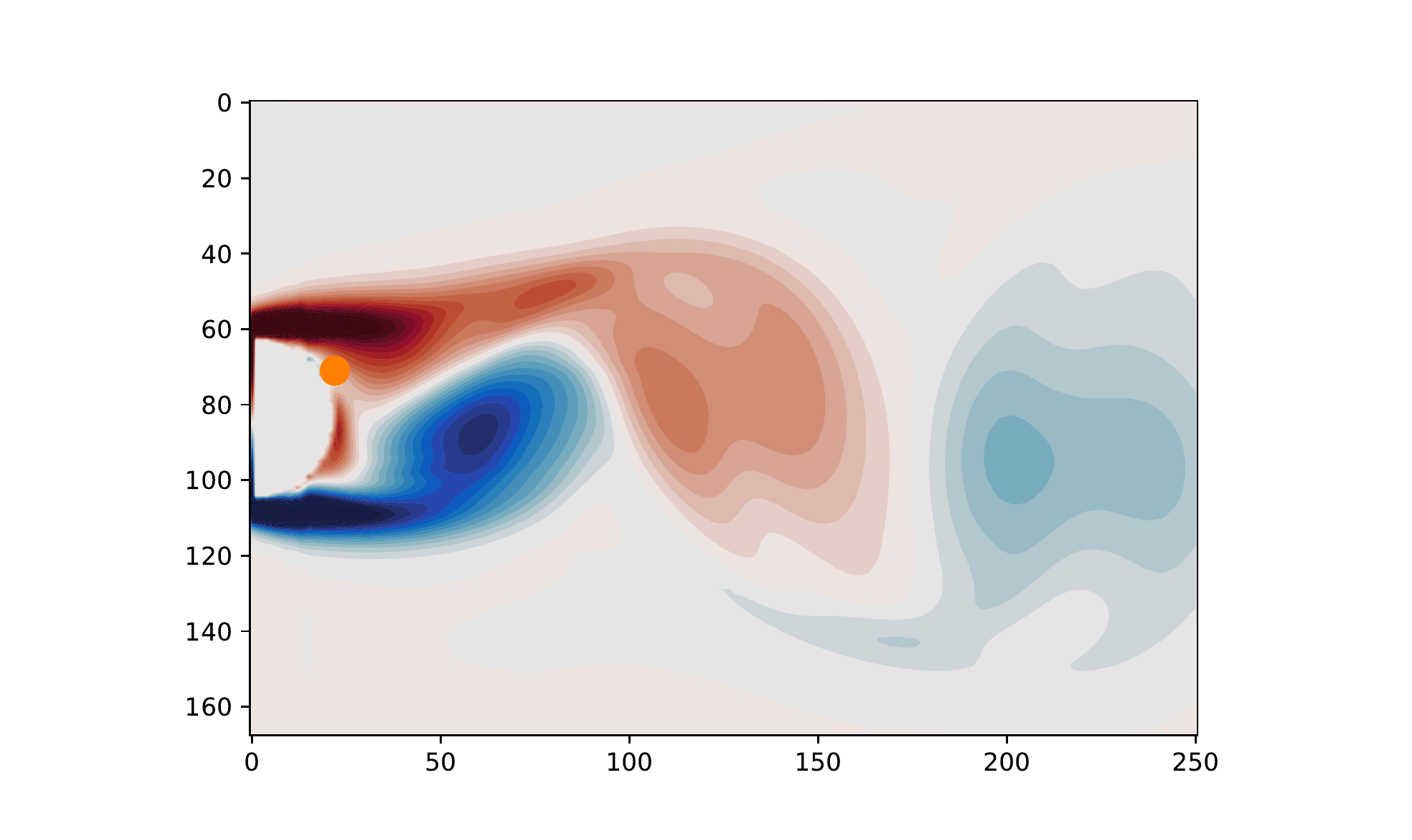}}

\subfigure[ARE]{\includegraphics[width=0.4\linewidth,trim=3.6cm 2.0cm 3.6cm 2.0cm,clip]{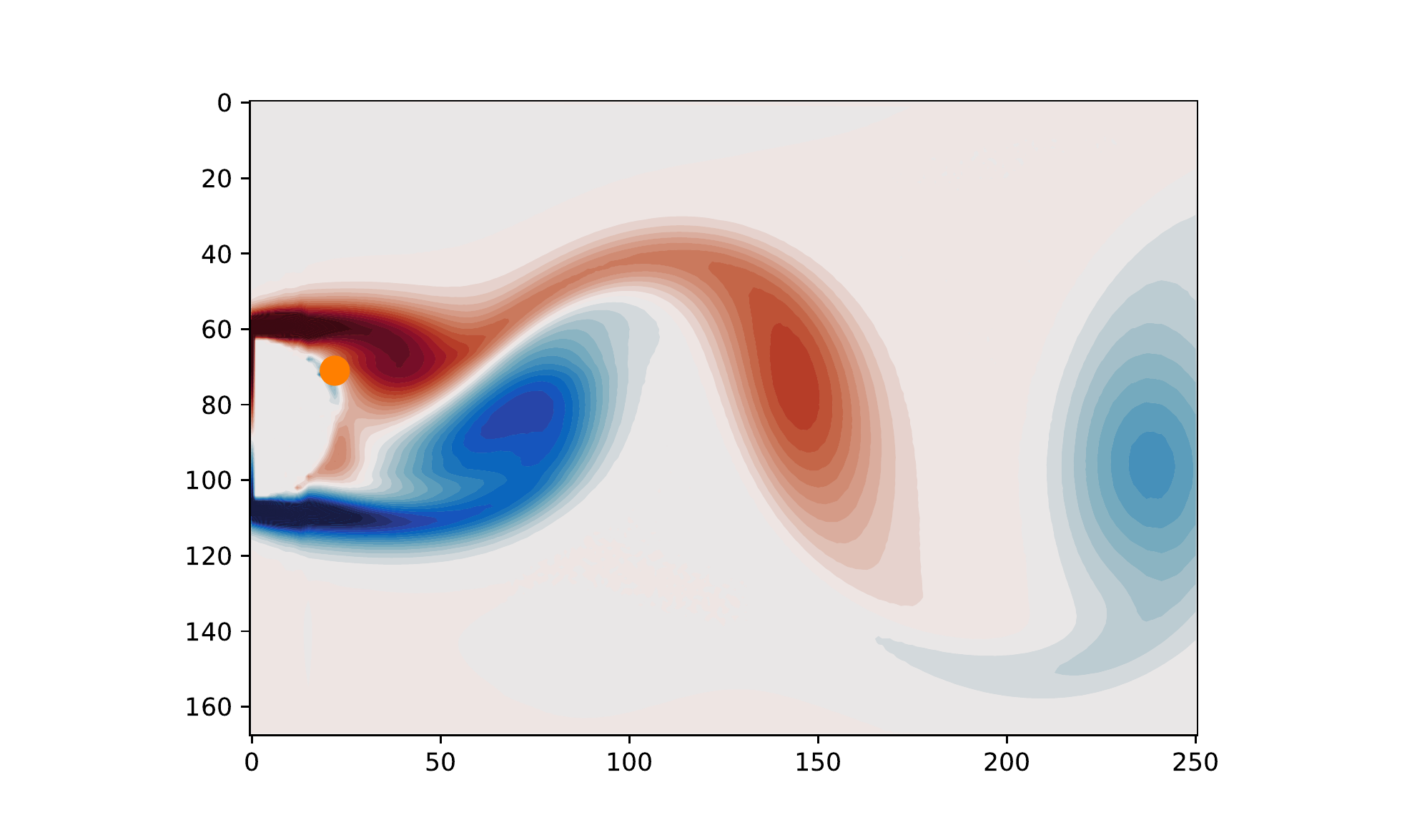}}

\subfigure[SD]{\includegraphics[width=0.4\linewidth,trim=3.6cm 2.0cm 3.6cm 2.0cm,clip]{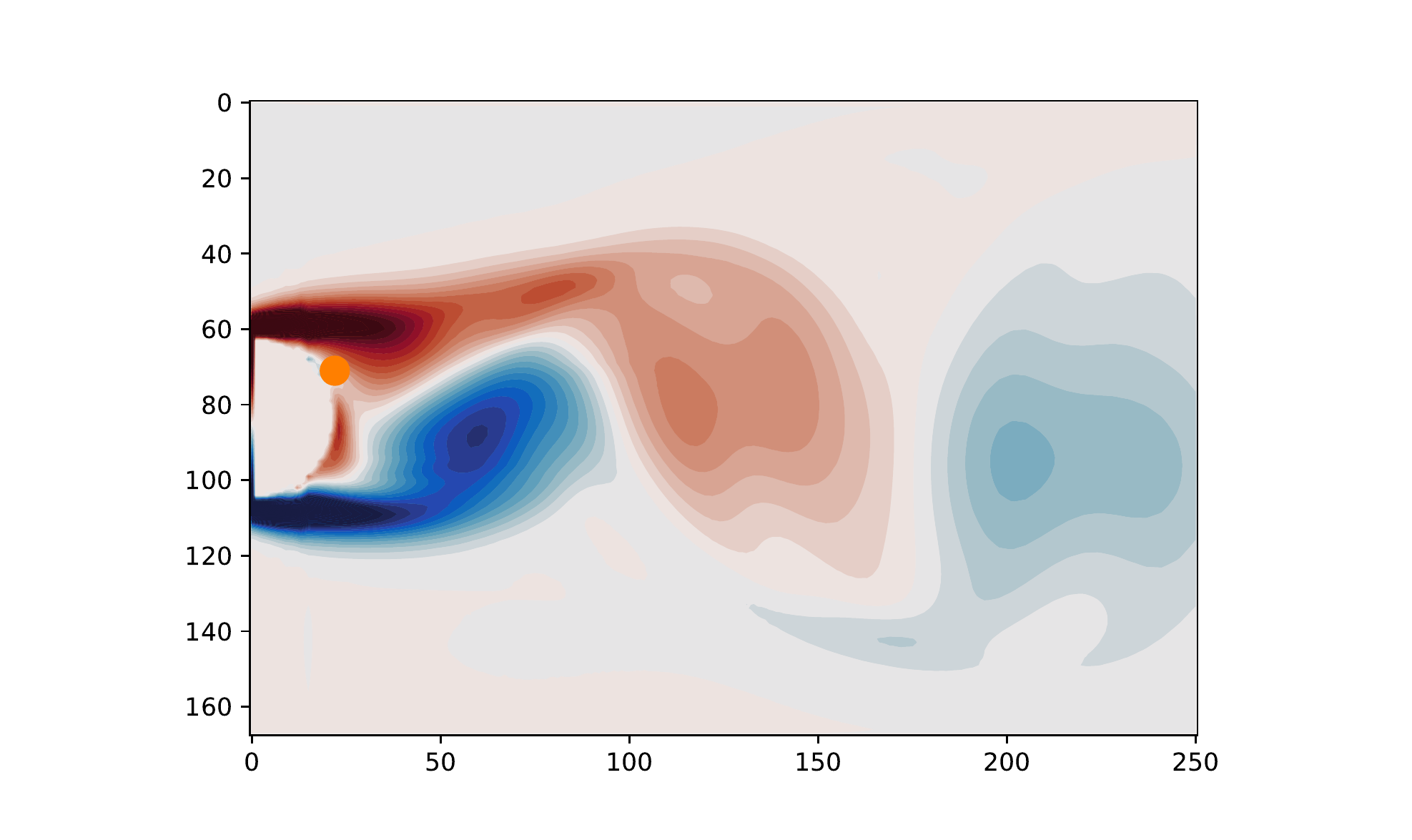}}

\subfigure[Ground Truth]{\includegraphics[width=0.4\linewidth,trim=3.6cm 2.0cm 3.6cm 2.0cm,clip]{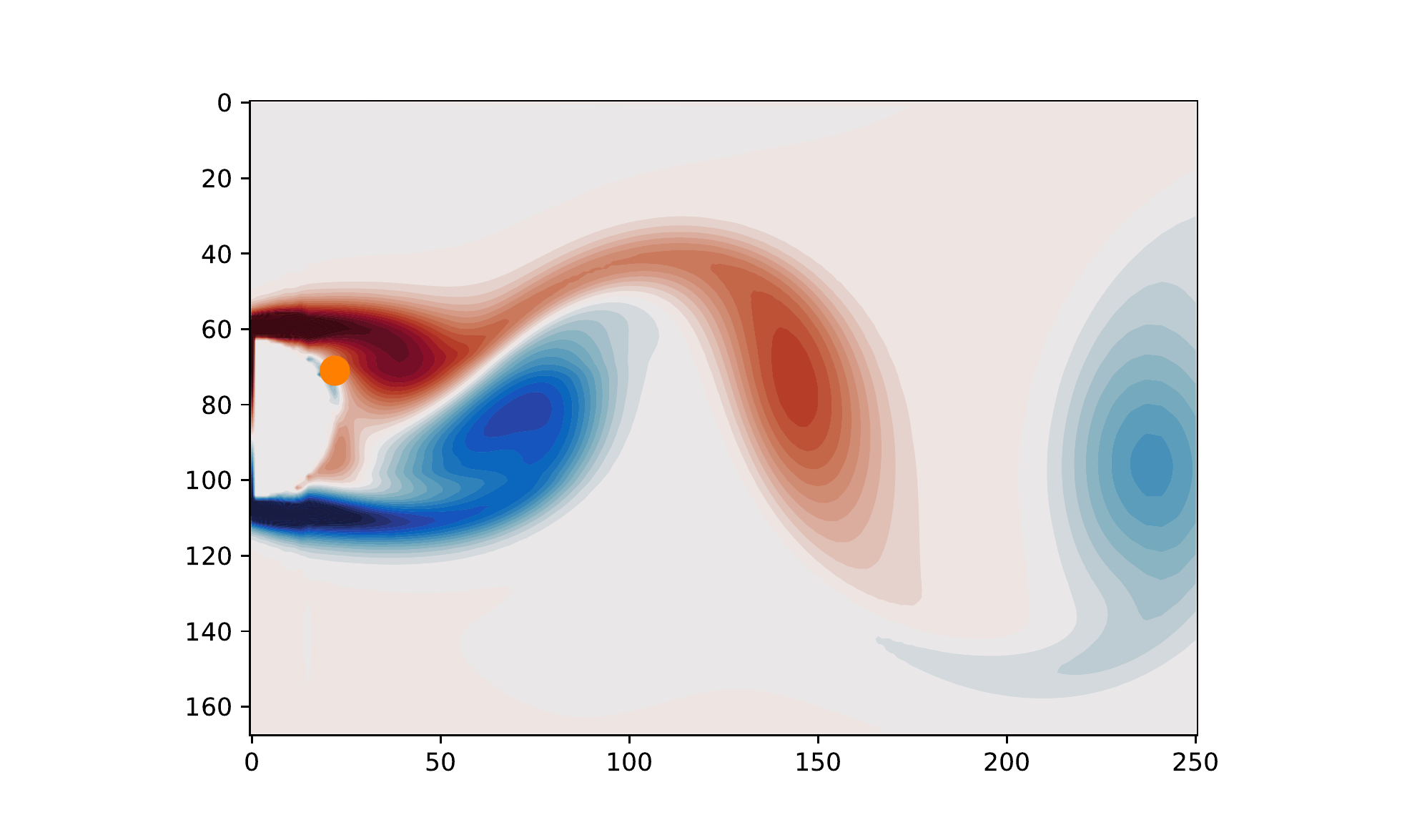}}

\caption{Figure depicts results of periodic vortex shedding for one sensor. Orange dots in the images represents the location  of sensor. SEQ-LEN used for ARE model is 4.}
\label{fig:res1_eg1}
\end{figure}

Fig. \ref{fig:res1_eg1} shows the results obtained using different approaches. It is an idealized case where we have considered the sensor data to be noise free. The sequence length of four is used for training ARE. 
We have considered the extreme case where data from only one sensor is available.
We observe that the proposed ARE, with only one senor, is able to recover the full state accurately. PDS and SD, on the other hand, yields less accurate results.
% \begin{figure}%[h]
% \centering
% \subfigure[PDS]{
% \includegraphics[width=0.49\linewidth,trim=3.6cm 2.0cm 3.6cm 2.0cm,clip]{im_PDSv3_s1_Ds.eps}}
% %\includegraphics[width=0.49\linewidth,trim=3.6cm 2.0cm 3.6cm 2.0cm,clip]{im_PDSv3_s2_Ds.eps}
% \subfigure[ARE]{\includegraphics[width=0.49\linewidth,trim=3.6cm 2.0cm 3.6cm 2.0cm,clip]{im_lstm_s1_sq7_i39_SNR1000_Pr.eps}}
% %\includegraphics[width=0.49\linewidth,trim=3.6cm 2.0cm 3.6cm 2.0cm,clip]{im_lstm_s2_sq7_i39_SNR1000_Pr.eps}
% \subfigure[Ground Truth]{\includegraphics[width=0.49\linewidth,trim=3.6cm 2.0cm 3.6cm 2.0cm,clip]{im_true_39_Ds.eps}}
% %\includegraphics[width=0.49\linewidth,trim=3.6cm 2.0cm 3.6cm 2.0cm,clip]{im_true_39_Ds.eps}
% \caption{Figure depicts results of periodic vortex shedding for one sensor. Orange dots in the images represents the location  of sensor. (First row -PDS, Second row -ARE, Third row -Ground truth) SEQ-LEN used for ARE model is 4.}
% \label{fig:res1_eg1}
% \end{figure}
% \begin{figure}%[h]
% \centering
% \subfigure[PDS]{
% \includegraphics[width=0.49\linewidth,trim=3.6cm 2.0cm 3.6cm 2.0cm,clip]{im_PDSv3_s1_Ds_SNR20.eps}}
% %\includegraphics[width=0.49\linewidth,trim=3.6cm 2.0cm 3.6cm 2.0cm,clip]{im_PDSv3_s2_Ds.eps}
% \subfigure[ARE]{\includegraphics[width=0.49\linewidth,trim=3.6cm 2.0cm 3.6cm 2.0cm,clip]{im_lstm_s1_sq7_i39_SNR20_Pr.eps}}
% %\includegraphics[width=0.49\linewidth,trim=3.6cm 2.0cm 3.6cm 2.0cm,clip]{im_lstm_s2_sq7_i39_SNR1000_Pr.eps}
% \subfigure[Ground Truth]{\includegraphics[width=0.49\linewidth,trim=3.6cm 2.0cm 3.6cm 2.0cm,clip]{im_true_39_Ds.eps}}
% %\includegraphics[width=0.49\linewidth,trim=3.6cm 2.0cm 3.6cm 2.0cm,clip]{im_true_39_Ds.eps}
% \caption{Figure depicts results of periodic vortex shedding for one sensor. The data is corrupted with white Gaussian noise having SNR = 20. Orange dots in the images represents the location  of sensor. SEQ-LEN used for ARE model is 4.}
% \label{fig:res2_eg1}
% \end{figure}
\begin{figure}[hbt!]
\centering
    \subfigure[PDS]{\includegraphics[width=0.4\linewidth,trim=3.6cm 2.0cm 3.6cm 2.0cm,clip]{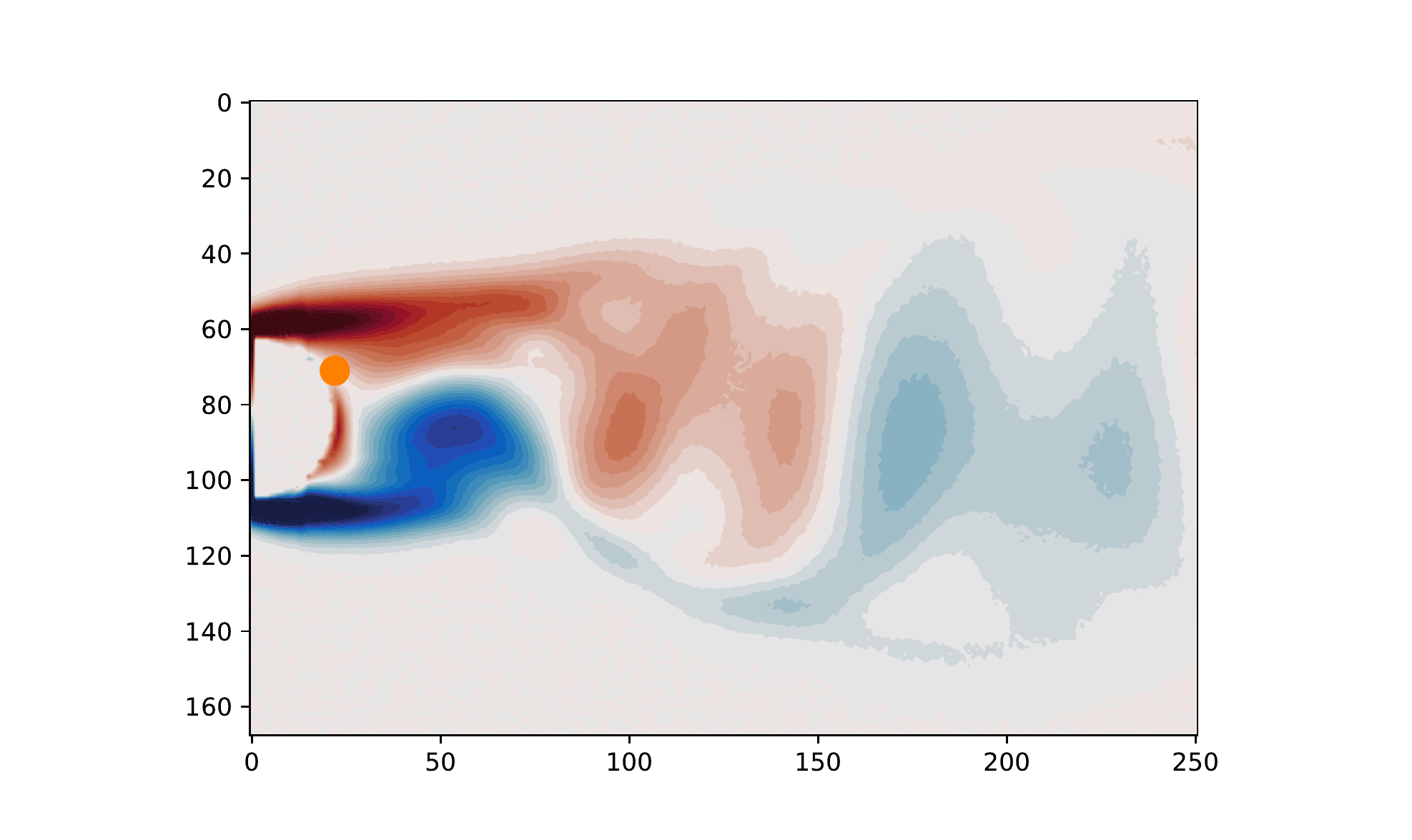}}
    
    \subfigure[ARE]{\includegraphics[width=0.4\linewidth,trim=3.6cm 2.0cm 3.6cm 2.0cm,clip]{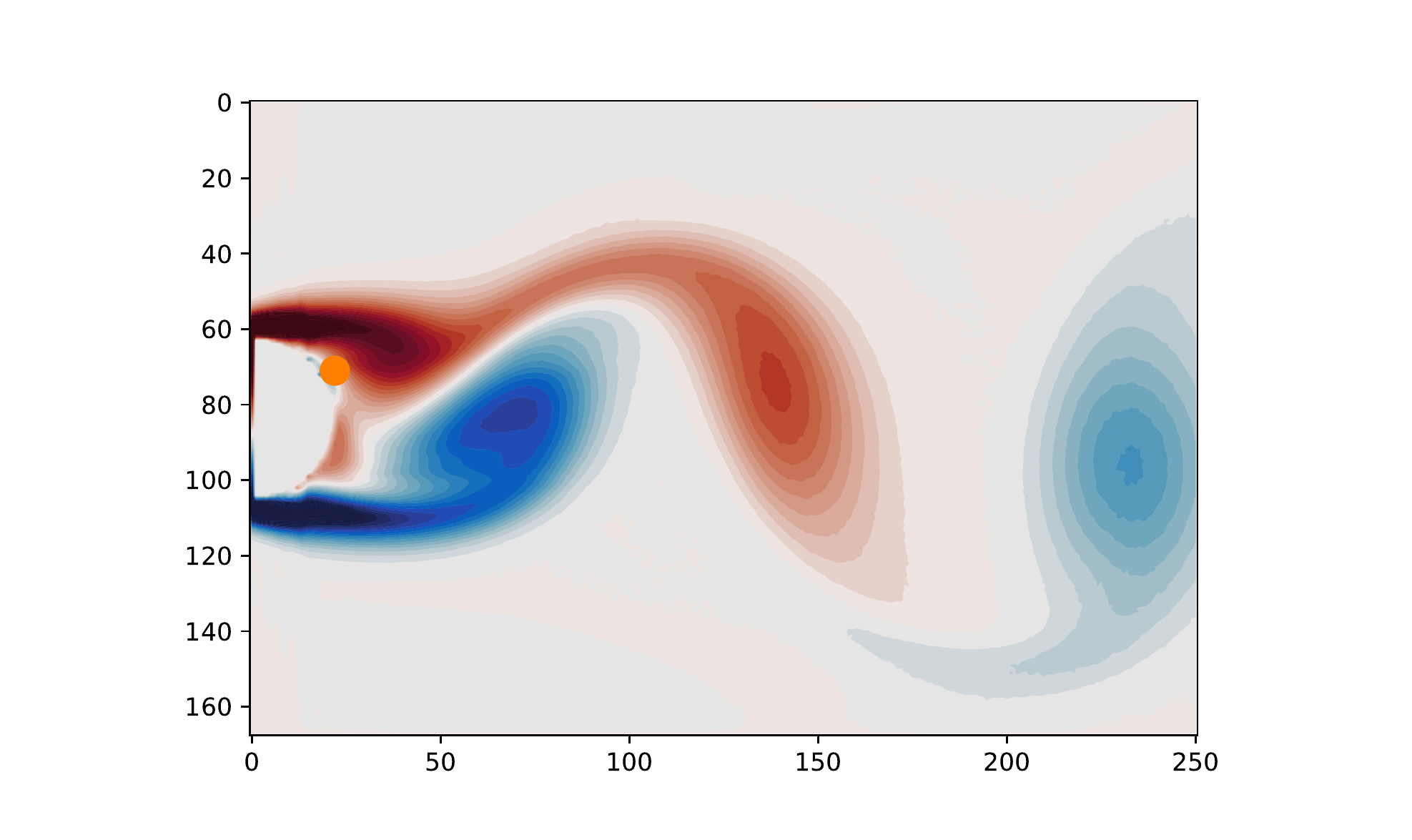}}
    
    \subfigure[SD]{\includegraphics[width=0.4\linewidth,trim=3.6cm 2.0cm 3.6cm 2.0cm,clip]{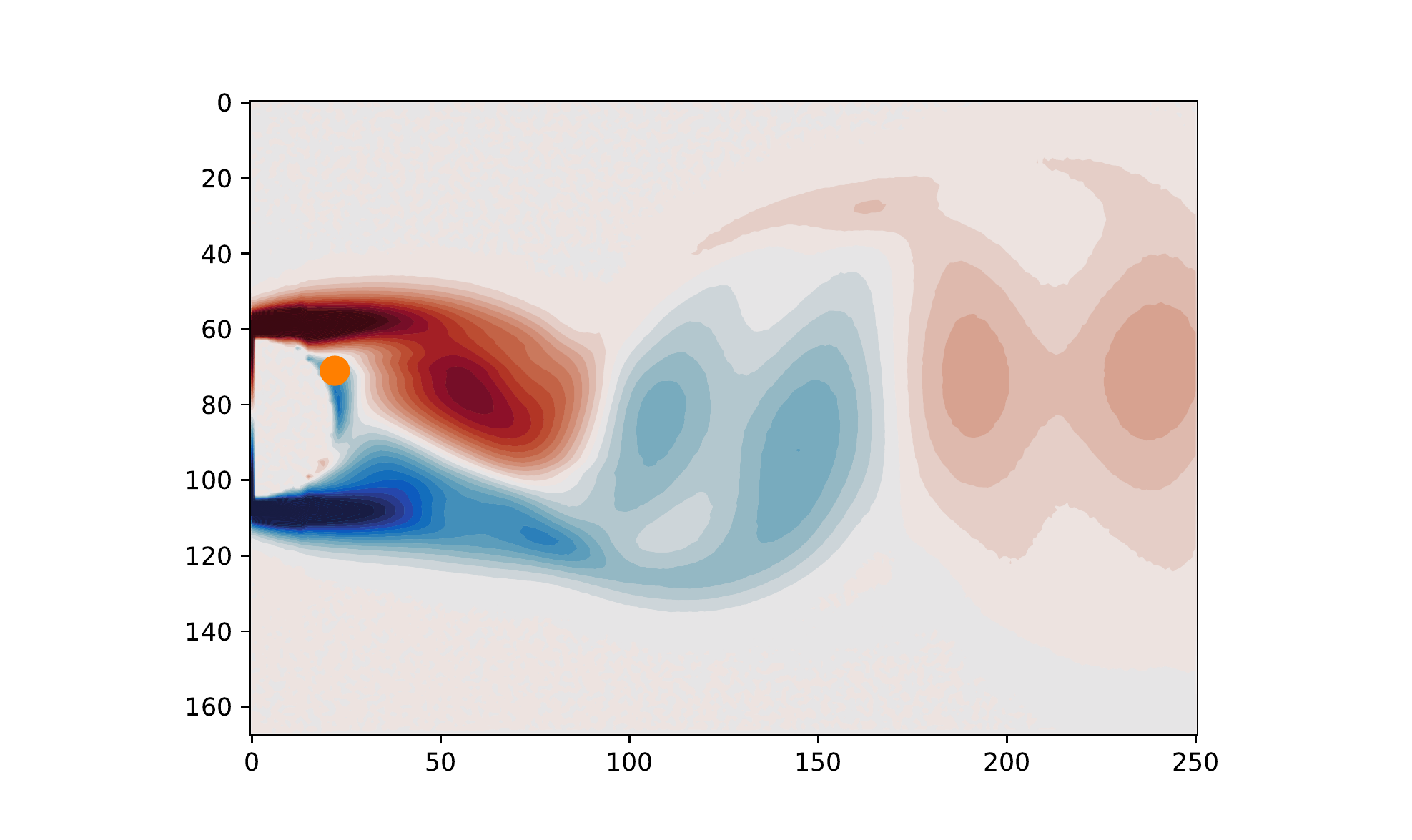}}
    
    \subfigure[Ground Truth]{\includegraphics[width=0.4\linewidth,trim=3.6cm 2.0cm 3.6cm 2.0cm,clip]{im_true_periodic_mthd__snsr_are_s1_sq7_i39.pdf}}

\caption{Figure depicts results of periodic vortex shedding for one sensor. The data is corrupted with white Gaussian noise having SNR = 20. Orange dots in the images represents the location  of sensor. SEQ-LEN used for ARE model is 4.}
\label{fig:res2_eg1}
\end{figure}

\begin{figure}[hbt!]
    \centering
    \subfigure[]{\includegraphics[width = 0.4\textwidth]{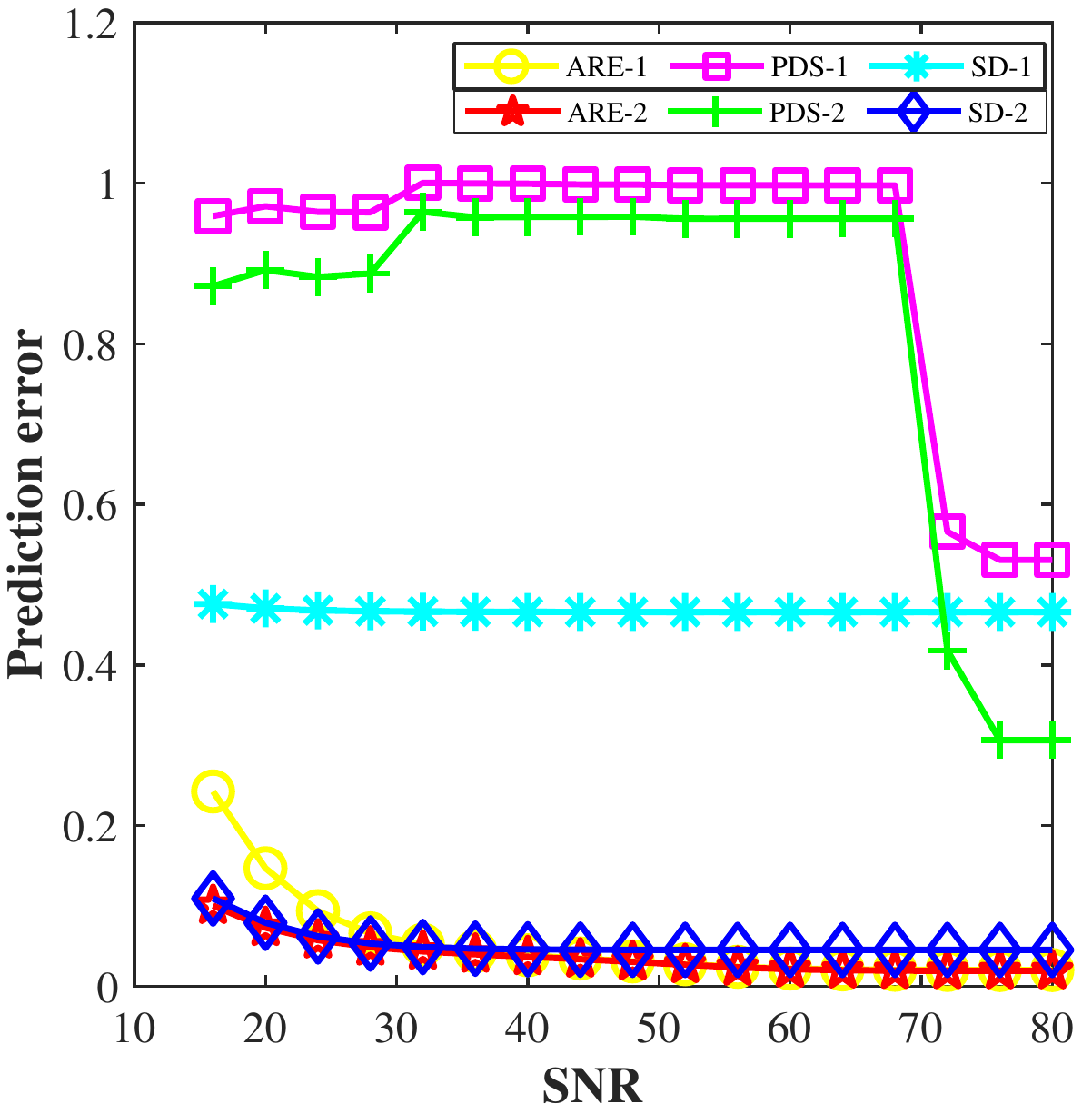}
    \label{fig:res3_eg1a}}
    \subfigure[]{\includegraphics[width = 0.4\textwidth]{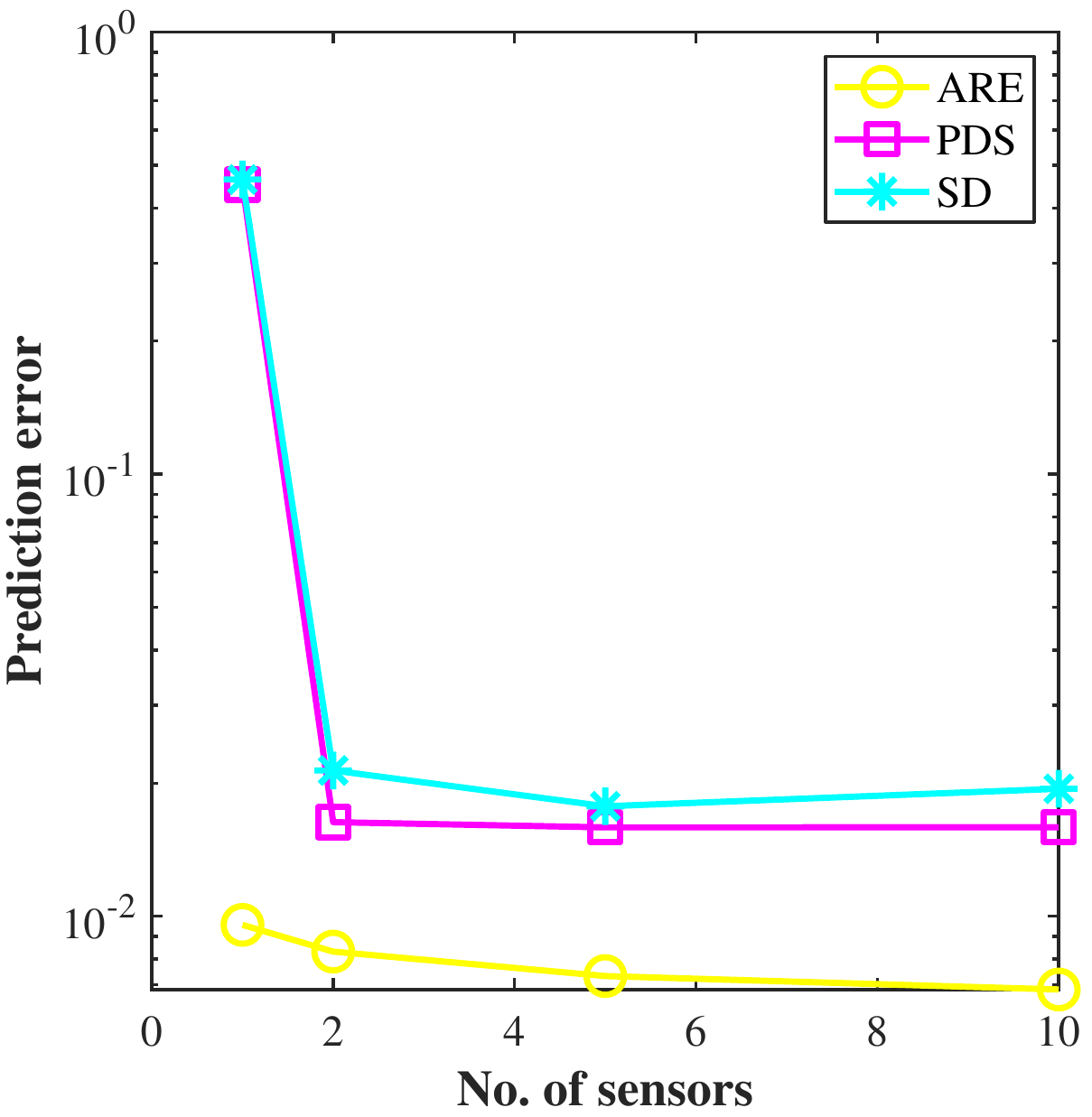}
    \label{fig:res3_eg1b}}
    \caption{(a) Performance of the proposed approach ARE, SD and PDS for periodic vortex shedding problem at different noise level. For both ARE and PDS, results corresponding to one and two sensors are presented. (b) Performance of proposed ARE, SD and PDS with increase in number of sensors. Data used is noise free.}
    \label{fig:res3_eg1}
\end{figure}

Fig. \ref{fig:res2_eg1} shows results corresponding to the
case where the sensor is corrupted by white Gaussian noise.
This is a more realistic case. Again, a sequence length of 4 is considered and it is assumed that data from only one sensor is available.
In this case also, we observe that ARE yields highly accurate results and outperforms PDS and SD.
The effect of noise on the proposed ARE is shown in Fig. \ref{fig:res3_eg1a}. 
We observe that both ARE (with one and two sensors) and SD (with two sensors) yield the best results. The fact that the proposed approach is able to correctly predict the state from only one sensor data (noisy) is really impressive. SD with one sensor and PDS predicted results are significantly less accurate as compared to the proposed approach. 
% Periodic testing with and without stats
% \begin{figure}
%     \centering
%     % \subfigure[noisy mean and std.]{\includegraphics[width = 0.3\textwidth]{error_periodic_mthd__snr_noise_s.pdf}}
%     % \subfigure[no mean and std.]{\includegraphics[width = 0.3\textwidth]{error_periodic_mthd__snr_no_noise.pdf}}
%     % \subfigure[pure mean and std.]{\includegraphics[width = 0.3\textwidth]{error_periodic_mthd__snr_pure_s.pdf}}
%     \caption{\red{new} Performance of the proposed approach (ARE) and PDS for periodic vortex shedding problem at different noise level. For both ARE and PDS, results corresponding to one and two sensors are presented.}
%     \label{fig:res3_eg1}
% \end{figure}
Next, we investigate the effect of varying the number of sensors. Fig. \ref{fig:res3_eg1b} shows the performance of different methods with the increase in the number of sensors.
The data considered is idealistic with no noise.
Cases corresponding to one, two, five, and ten sensors are presented.
For all the cases, the proposed ARE is found to yield the best result. 
Results obtained using SD and PDS follows a similar trend.
% \begin{figure}[h]
%     \centering
%     % \includegraphics[width = 0.6\textwidth]{Prmodelvssensor.pdf}
%     \includegraphics[width = 0.4\textwidth]{error_periodic_mthd__snsr.pdf}
%     \caption{Performance of proposed ARE and PDS with increase in number of sensors.}
%     \label{fig:res4_eg1}
% \end{figure}
We also carried out an additional case study where we considered that sensor data from both past and future is available.
A bi-directional RNN (B-RNN) based ARE developed for the same. However, due to the paucity of space, the same is not presented here. Those interested can refer to \ref{appB} for details on the same.

\subsection{Transient Flow past a cylinder}
\label{subsec:eg2}
As the second example, we consider the problem involving transient flow past a cylinder.
Because of the transient nature of the flow, this is much more challenging than the periodic vortex shedding problem in Section \ref{subsec:eg1}.
The problem domain, meshing, and solution strategy for this problem are considered the same as the periodic vortex shedding problem.
However, unlike the previous problem, 
we have considered variation in the Reynolds' number. This exponentially increases the complexity of the problem.

The training library for this problem was created by running OpenFoam. Total 1200 snapshots consisting of $400$ snapshots at $Re = [180, 190, 200]$ were generated. Given the fact that $Re$ considered for this problem, resides in $[30,300]$, we have used URANS for generating the data \citep{STRINGER20141}.
Validation and test set consists of $400$ sequential snapshots at $Re = [185, 195]$.
Similar to the previous problem, the snapshots were separated by a time interval of $10 \delta t$.
Co-ordinate of snapshot cutout stretches from $\left[ 0, -3 \right] \times \left[12.0, 3.0 \right]$ (see Fig. \ref{fig:domain}(b)).
The cutout is discretized into $252$ and $502$ points in the $x$ and $y$ direction, respectively. Similar to the previous example, the objective here is to recover the vorticity field based on sensor measurements.
Similar to the previous example, results obtained have been compared with PDS and SD.
Details on the network architecture used for this problem are provided in Table \ref{tab:na_eg2}.

\begin{table}[hbt!]
    \centering
    \caption{Network architecture of proposed ARE for transient flow problem. HS is the number of features in the hidden state of lstm.}
    \label{tab:na_eg2}
    \begin{tabular}{|p{2cm}|p{15cm}|}
        \hline
        Networks & Architecture  \\ \hline
    ARE(Auto-Encoder) & $252\times502\rightarrow BN^* \rightarrow2000\rightarrow BN^* \rightarrow DR(0.35)^* \rightarrow300\rightarrow25\rightarrow300\rightarrow2000\rightarrow252\times 502$ \\
    ARE(RNN) & LSTM(HS=50) $\rightarrow 200 \rightarrow 200 \rightarrow 25$ \\
    SD & $N_s\rightarrow350\rightarrow BN\rightarrow DR(0.1)^* \rightarrow 400 \rightarrow BN \rightarrow 252\times502$ \\ \hline
    \multicolumn{2}{l}{\footnotesize{Components with $^*$ are not used in case of noise-free data.}}
    \end{tabular}
\end{table}
% ================= old Transient method images ===============

% \begin{figure}%[h]
% \centering
% \subfigure[PDS]{\includegraphics[width=0.49\linewidth,trim=2.5cm 2.1cm 1.8cm 2.0cm,clip]{im_PDS_s1_D3.eps}}
% % \includegraphics[width=0.49\linewidth,trim=2.5cm 2.1cm 1.8cm 2.0cm,clip]{im_PDS_s2_D3.eps}
% \subfigure[ARE]{\includegraphics[width=0.49\linewidth,trim=2.5cm 2.1cm 1.8cm 2.0cm,clip]{im_lstm_s1_sq7_i335_SNR1000_Tr.eps}}
% % \includegraphics[width=0.49\linewidth,trim=2.5cm 2.1cm 1.8cm 2.0cm,clip]{im_LSTMv3_s2_sq11_D3.eps}
% \subfigure[Ground truth]{\includegraphics[width=0.49\linewidth,trim=2.5cm 2.1cm 1.8cm 2.0cm,clip]{im_true_335_D3.eps}}
% % \includegraphics[width=0.49\linewidth,trim=2.5cm 2.1cm 1.8cm 2.0cm,clip]{im_true_350_D3.eps}
% \caption{Figure depicts results of Transient Flow  for one sensor. Orange dots in the images represents the location  of sensor. SEQ-LEN used for ARE model is 4}
% \label{fig:res1_eg2}
% \end{figure}
% ===========================================================

\begin{figure}[hbt!]
\centering
\subfigure[PDS]{\includegraphics[width=0.42\linewidth,trim=2.5cm 2.1cm 1.8cm 2.0cm,clip]{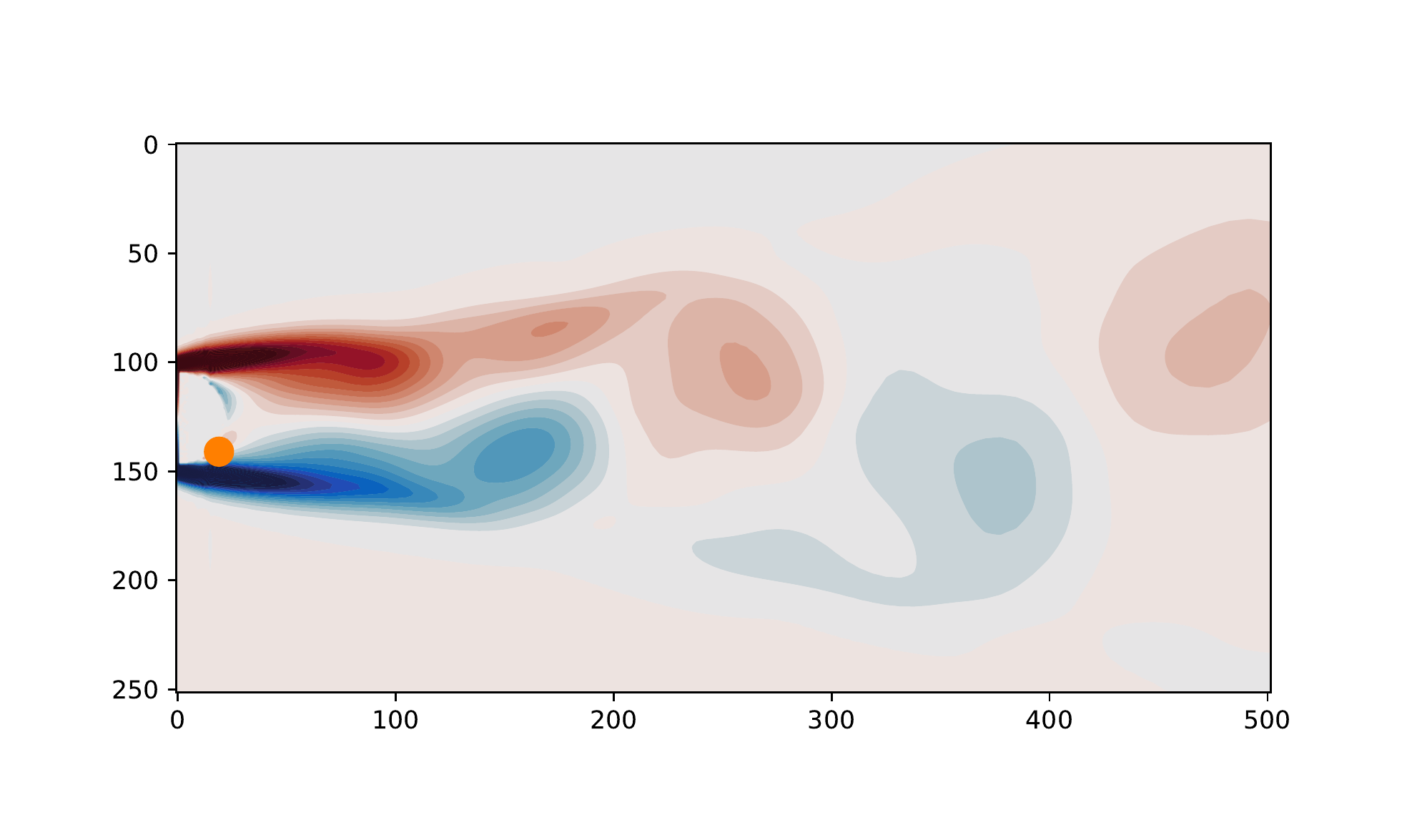}}
\subfigure[ARE]{\includegraphics[width=0.42\linewidth,trim=2.5cm 2.1cm 1.8cm 2.0cm,clip]{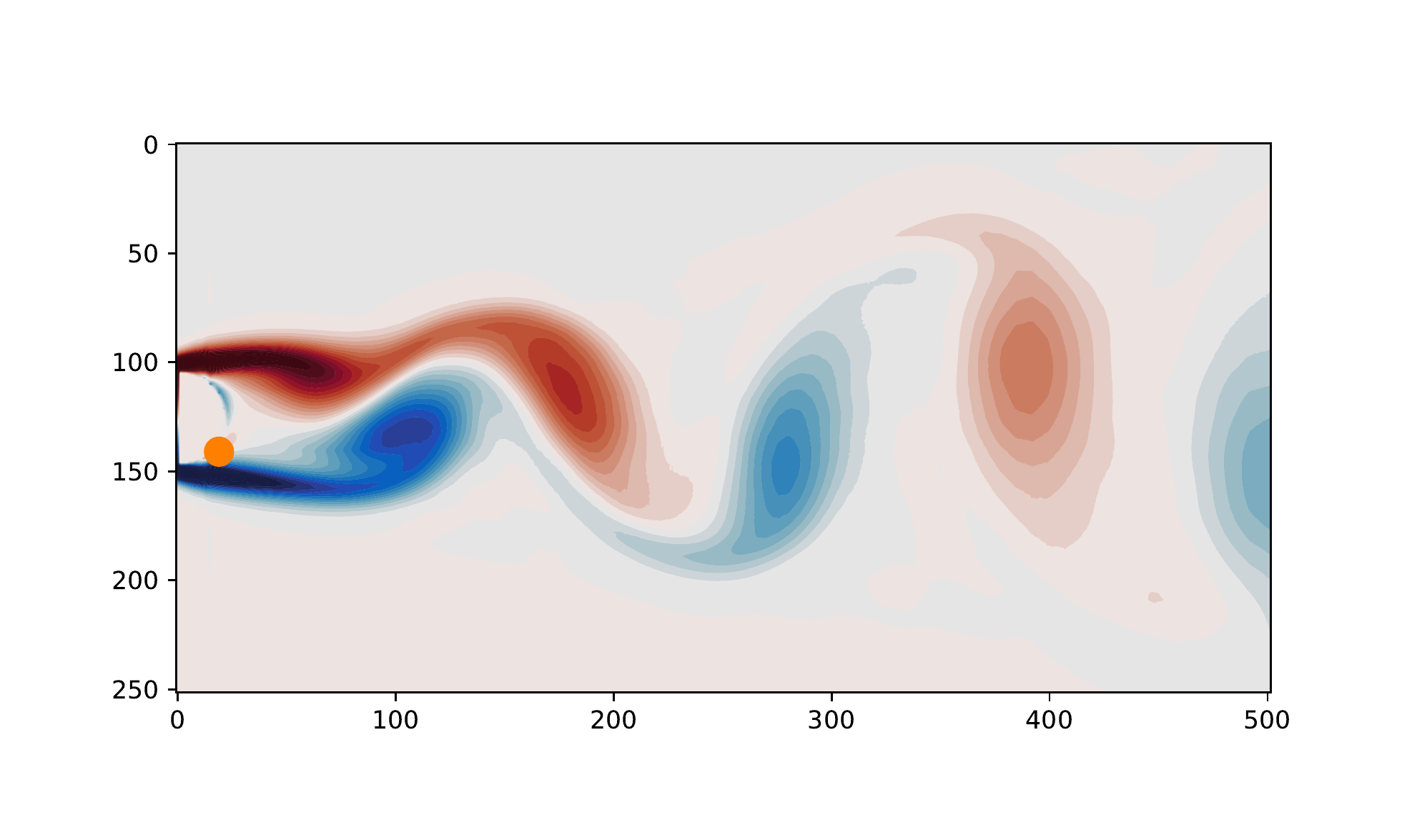}}
\subfigure[SD]{\includegraphics[width=0.42\linewidth,trim=2.5cm 2.1cm 1.8cm 2.0cm,clip]{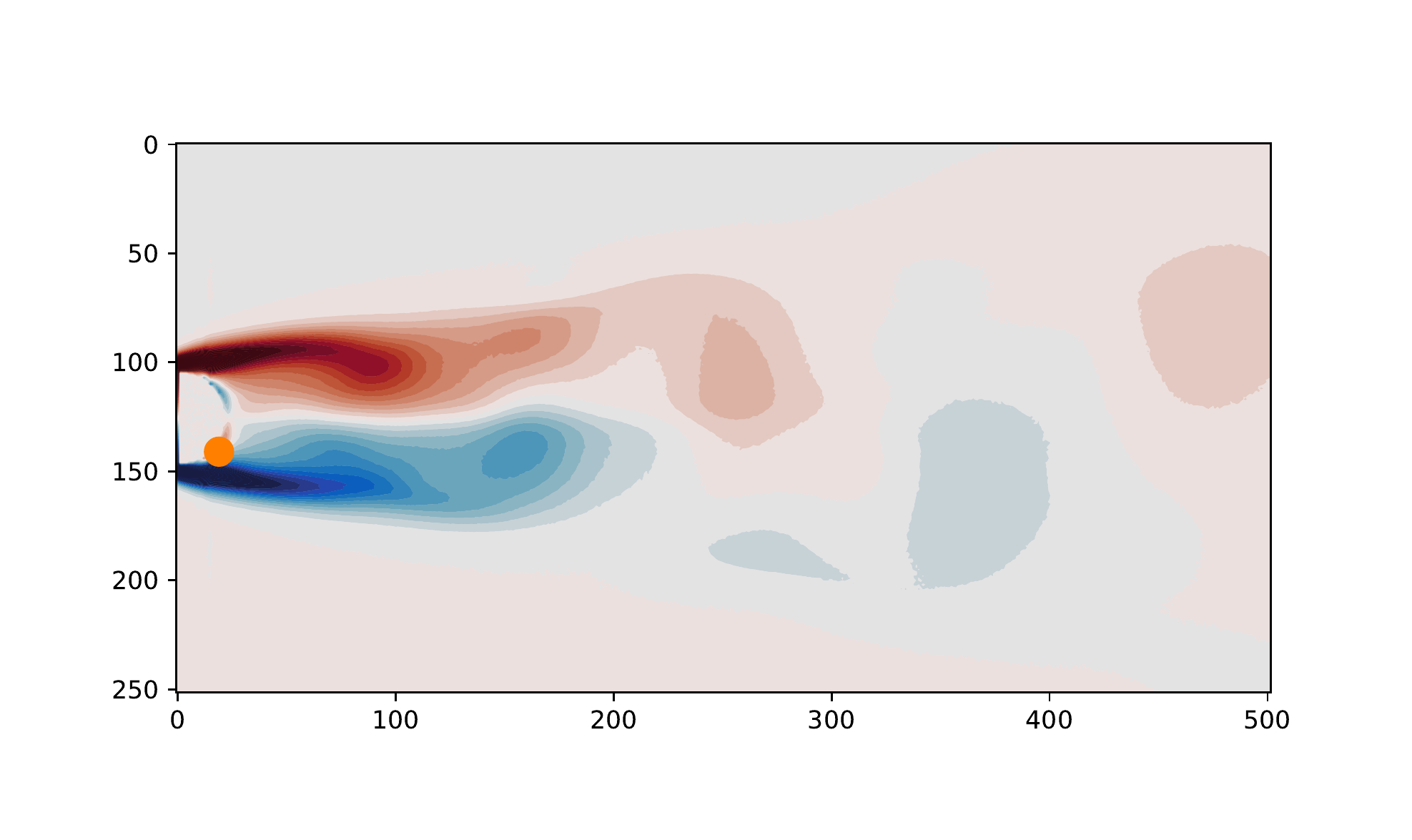}}
\subfigure[Ground truth]{\includegraphics[width=0.42\linewidth,trim=2.5cm 2.1cm 1.8cm 2.0cm,clip]{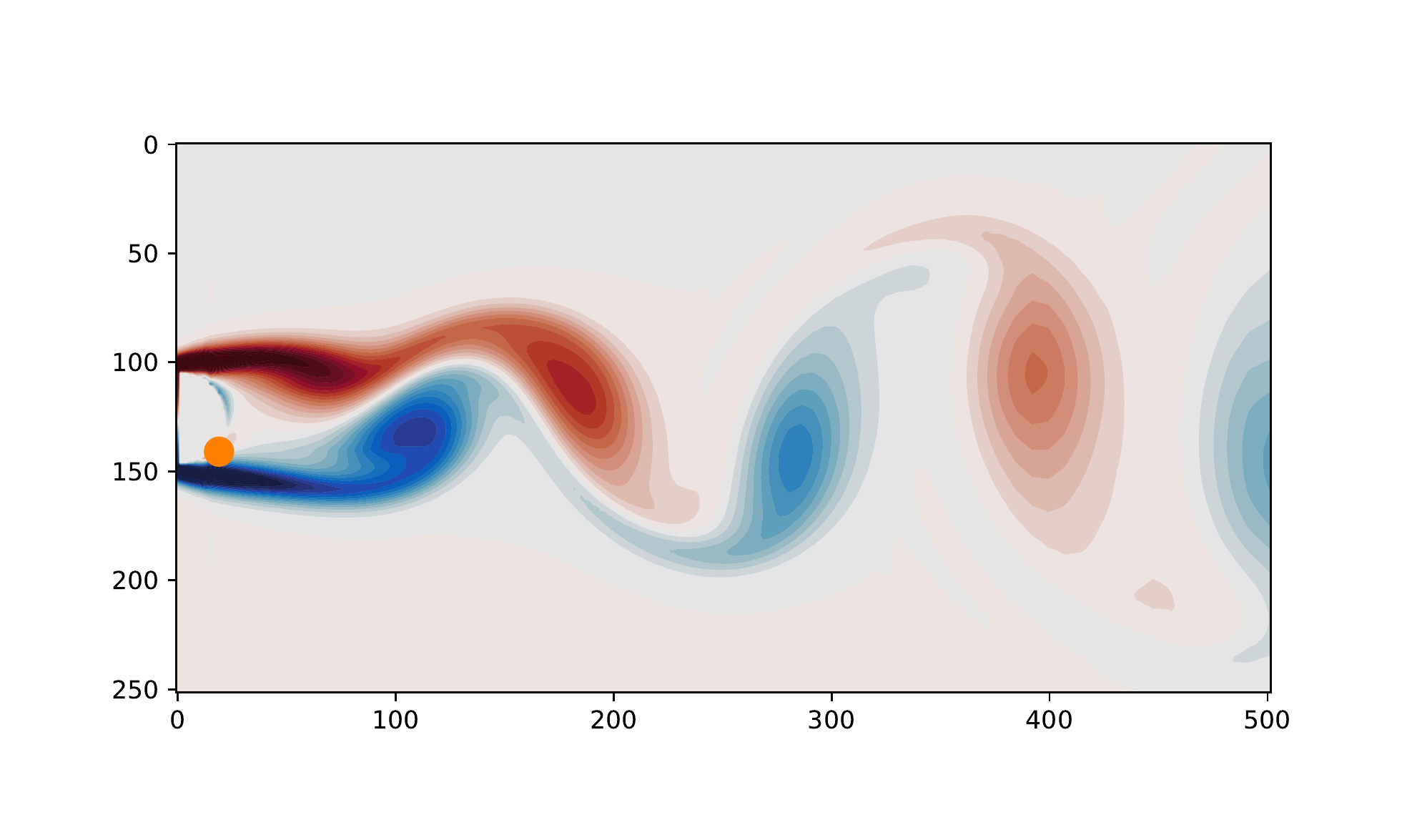}}
\caption{Figure depicts results of Transient Flow  for one sensor. Orange dots in the images represents the location  of sensor. SEQ-LEN used for ARE model is 4}
\label{fig:res1_eg2}
\end{figure}

\begin{figure}[hbt!]
\centering
\subfigure[PDS]{\includegraphics[width=0.42\linewidth,trim=2.5cm 2.1cm 1.8cm 2.0cm,clip]{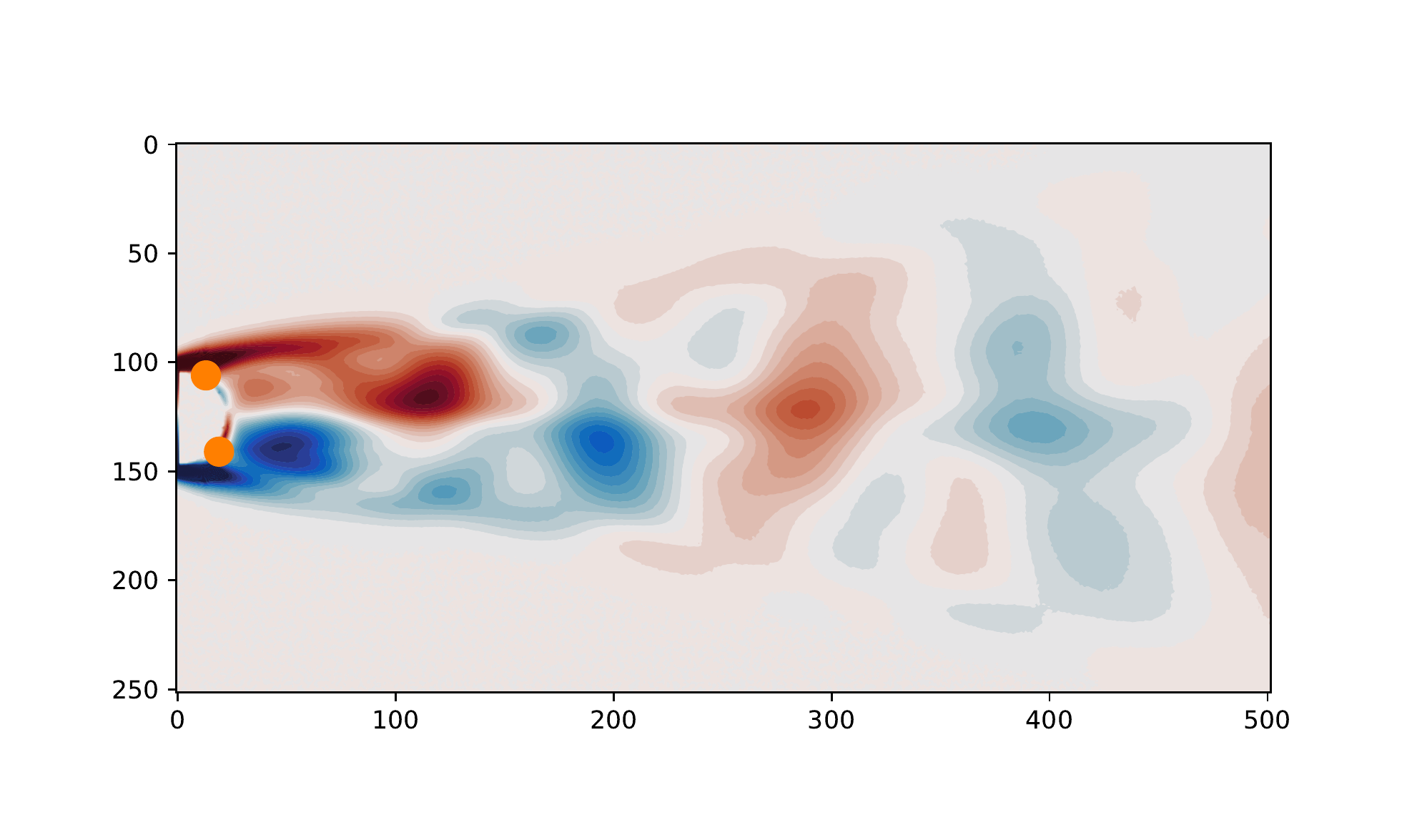}}
\subfigure[ARE]{\includegraphics[width=0.42\linewidth,trim=2.5cm 2.1cm 1.8cm 2.0cm,clip]{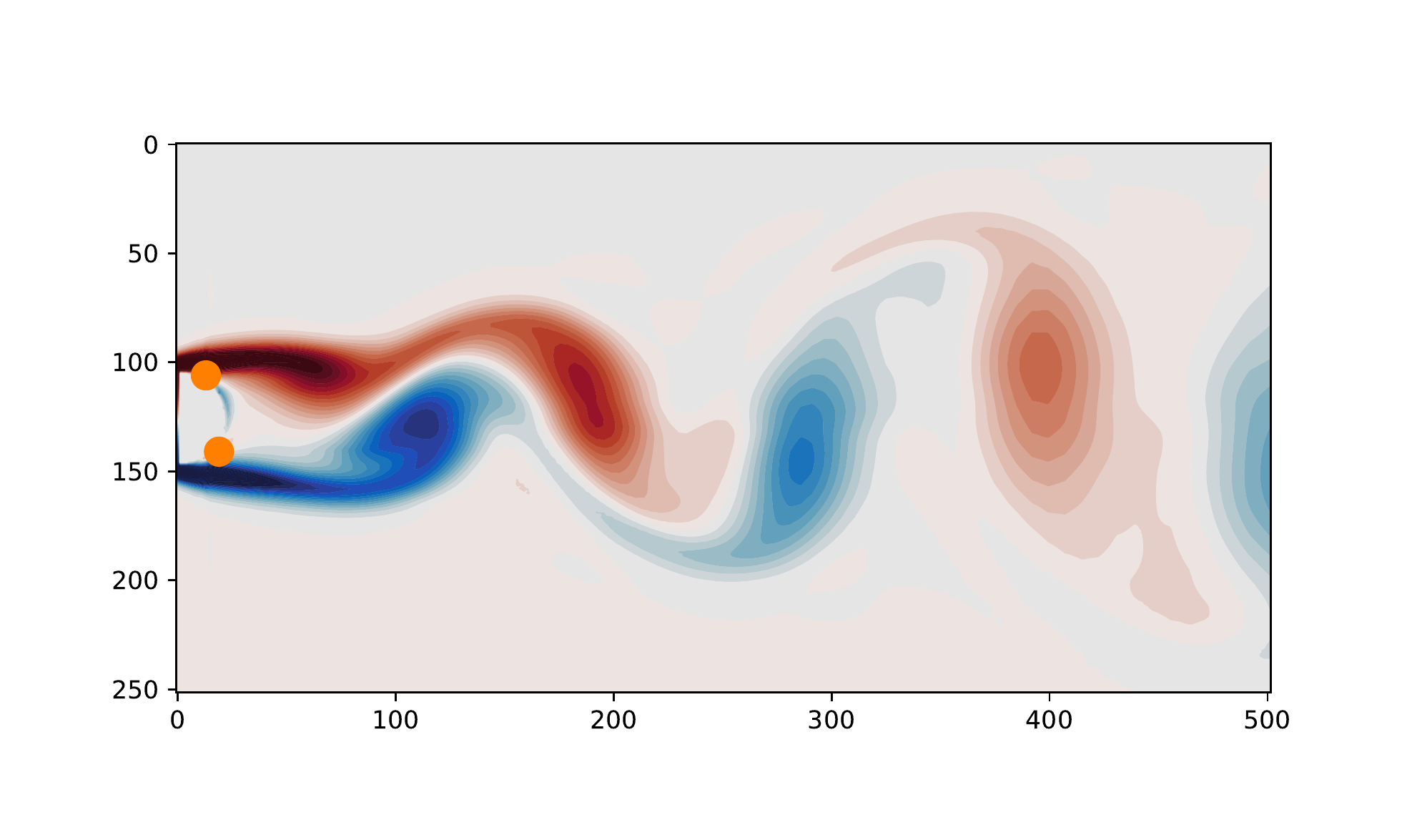}}
\subfigure[SD]{\includegraphics[width=0.42\linewidth,trim=2.5cm 2.1cm 1.8cm 2.0cm,clip]{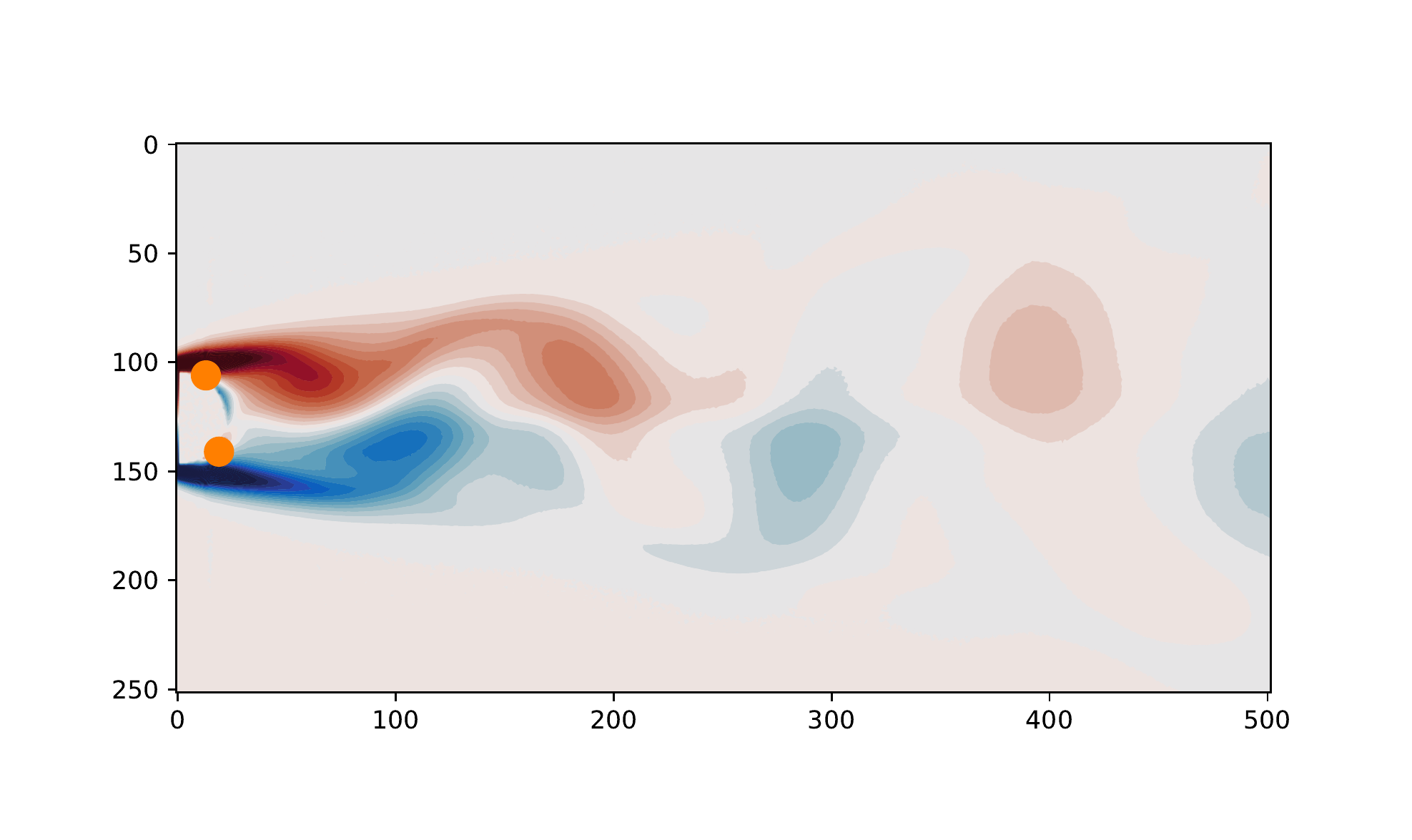}}
\subfigure[Ground truth]{\includegraphics[width=0.42\linewidth,trim=2.5cm 2.1cm 1.8cm 2.0cm,clip]{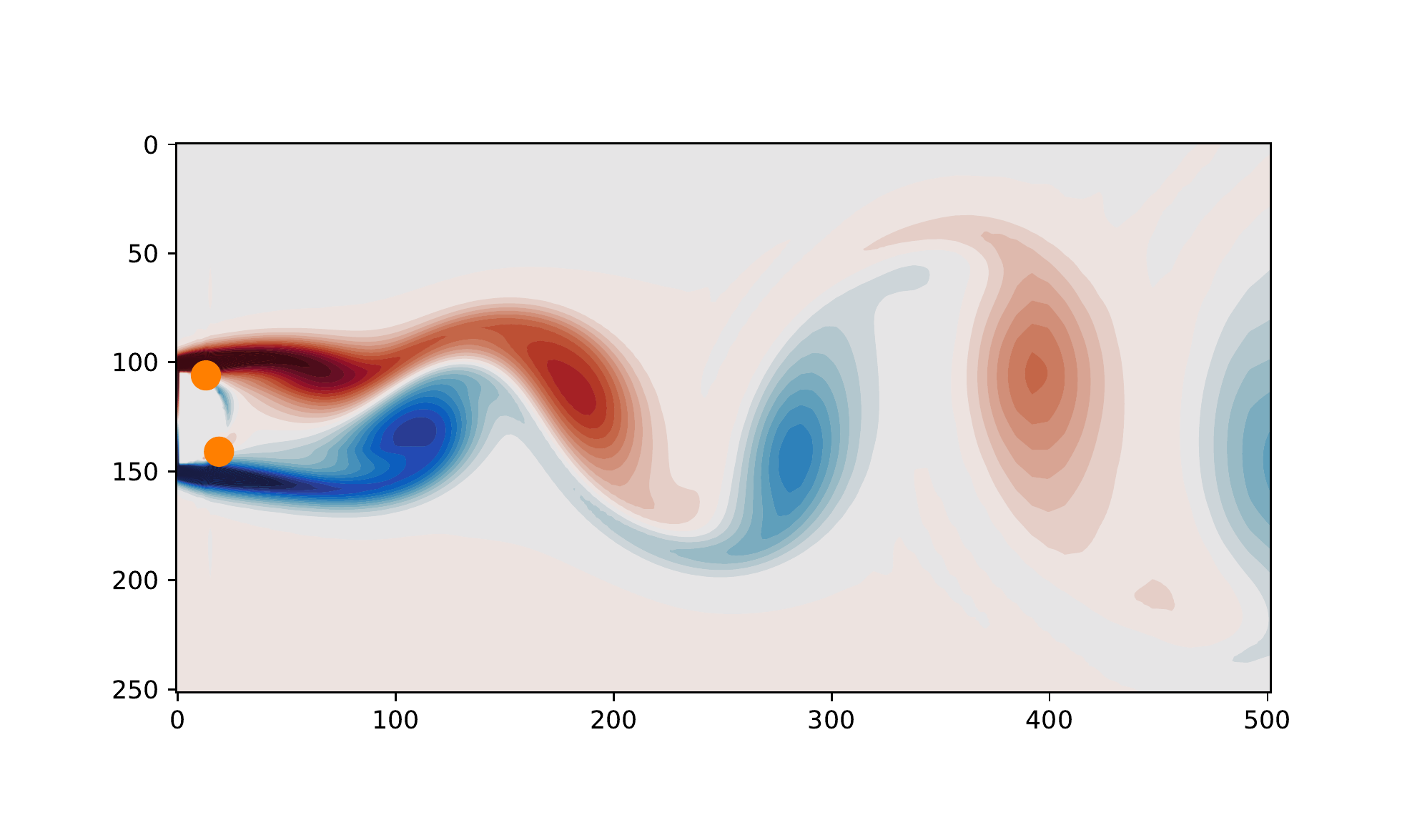}}
\caption{Figure depicts results of Transient Flow  for two sensors. The data is corrupted with white Gaussian noise with SNR = 28. Orange dots in the images represents the location  of sensor. SEQ-LEN used for ARE model is 4}
\label{fig:res2_eg2}
\end{figure}

Fig. \ref{fig:res1_eg2} shows the reconstructed vorticity field using ARE, SD and PDS. Ground truth has also been reported. It is an idealized case where we have considered the sensor data to be noise-free. A sequence length of four is used for ARE. We have considered the extreme case where data from only one sensor is available. For this problem also, the proposed is able to recover the full vorticity field accurately. PDS and SD, on the other hand, fails to accurately recover the vorticity field.
Next, we consider a more realistic scenario where the sensor data is corrupted by noise. Fig. \ref{fig:res2_eg2} shows the results corresponding to noisy sensor measurement.
In this case, we have considered that measurements from two sensors are available.
For this case also, the results obtained using the proposed ARE is found to be superior as compared to those in the literature.

% ========= old Transient method images with SNR ===============
% \begin{figure}%[h]
% \centering
% \subfigure[PDS]{\includegraphics[width=0.49\linewidth,trim=2.5cm 2.1cm 1.8cm 2.0cm,clip]{im_PDSv3_s2_Tr_SNR28.eps}}
% % \includegraphics[width=0.49\linewidth,trim=2.5cm 2.1cm 1.8cm 2.0cm,clip]{im_PDS_s2_D3.eps}
% \subfigure[ARE]{\includegraphics[width=0.49\linewidth,trim=2.5cm 2.1cm 1.8cm 2.0cm,clip]{im_lstm_s2_sq7_i335_SNR28_Tr.eps}}
% % \includegraphics[width=0.49\linewidth,trim=2.5cm 2.1cm 1.8cm 2.0cm,clip]{im_LSTMv3_s2_sq11_D3.eps}
% \subfigure[Ground truth]{\includegraphics[width=0.49\linewidth,trim=2.5cm 2.1cm 1.8cm 2.0cm,clip]{im_true_335_D3.eps}}
% % \includegraphics[width=0.49\linewidth,trim=2.5cm 2.1cm 1.8cm 2.0cm,clip]{im_true_350_D3.eps}
% \caption{Figure depicts results of Transient Flow  for two sensors. The data is corrupted with white Gaussian noise with SNR = 28. Orange dots in the images represents the location  of sensor. SEQ-LEN used for ARE model is 4}
% \label{fig:res2_eg2}
% \end{figure}
% ===========================================================

\begin{figure}[hbt!]
    \centering
    \includegraphics[width = 0.4\textwidth]{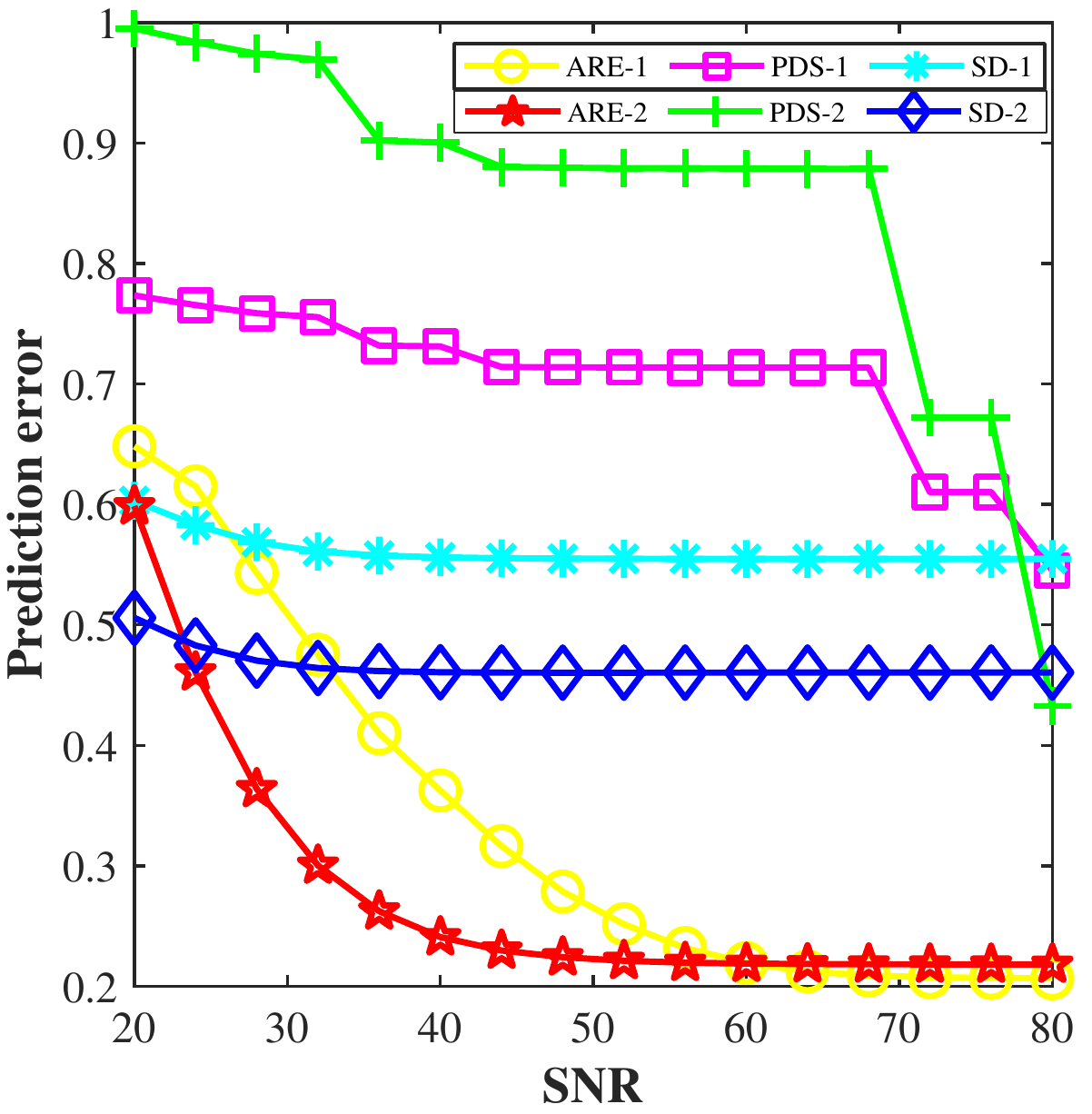}
    \caption{Performance of the proposed approach ARE, SD and PDS for the transient flow problem at different noise level. ARE1 corresponds to results with one sensor and ARE2 corresponds to results with 2 sensors.}
    \label{fig:res3_eg2}
\end{figure}

The effect of noise on the performance is shown Fig. \ref{fig:res3_eg2}.
Overall, results obtained using the proposed ARE are found to be more accurate as compared to those obtained using PDS and SD. The improved accuracy of ARE can be attribute to the fact that it learns from a sequence of data. 
Surprisingly, we observe that PDS with one sensor yields superior result as compared to PDS with two sensors. This is because the POD modes for two sensor cases is susceptible to the noise in the data. As noise in the data reduces, the POD modes are identified correctly and performance of PDS with two sensors is found to be superior as expected.
The proposed ARE and SD are immune to such overfitting phenomenon.

\begin{figure}[hbt!]
    \centering
    \subfigure[]{\includegraphics[width = 0.4\textwidth]{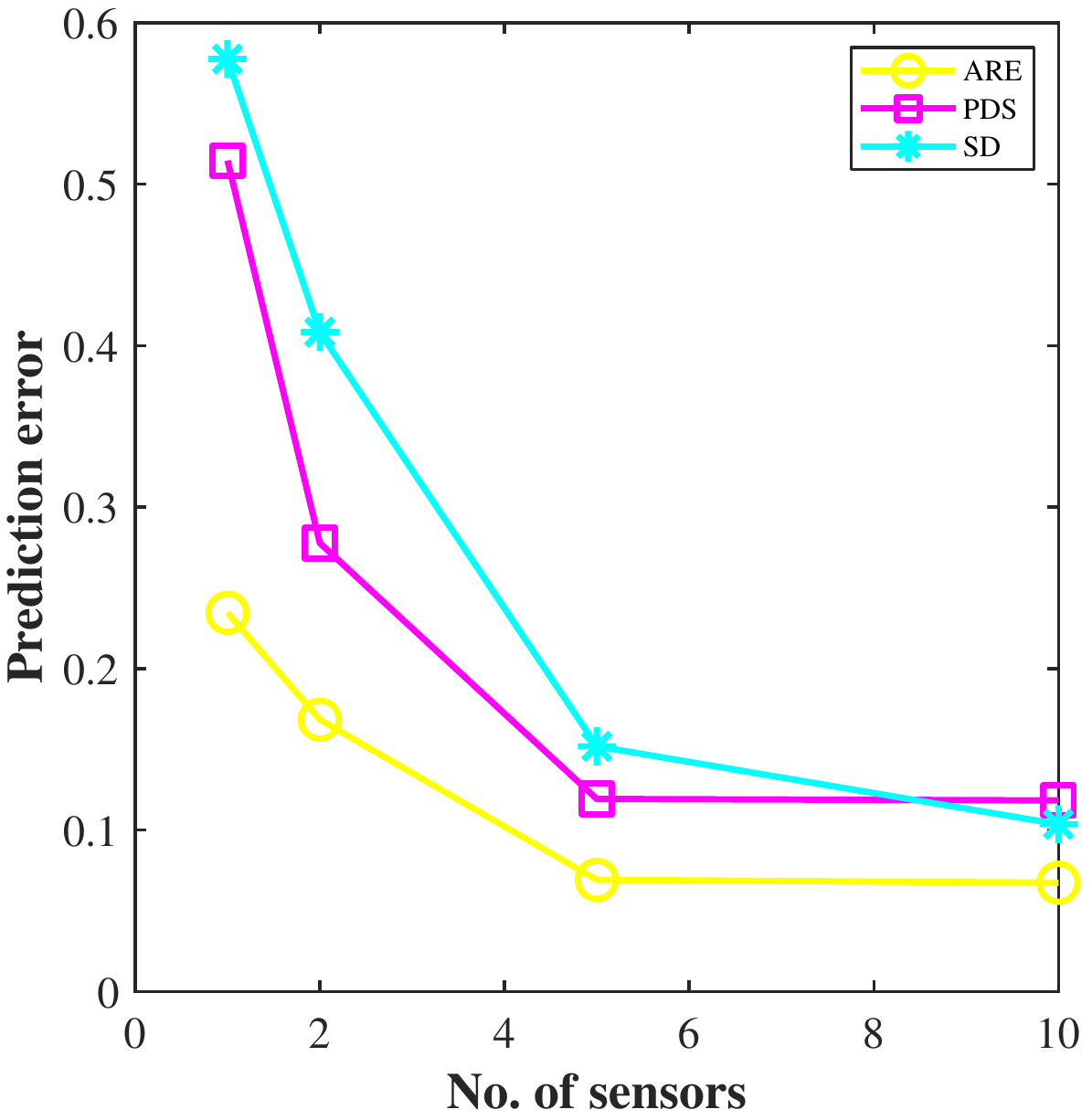}
    \label{fig:res4_eg2a}}
    \subfigure[]{\includegraphics[width = 0.46\textwidth]{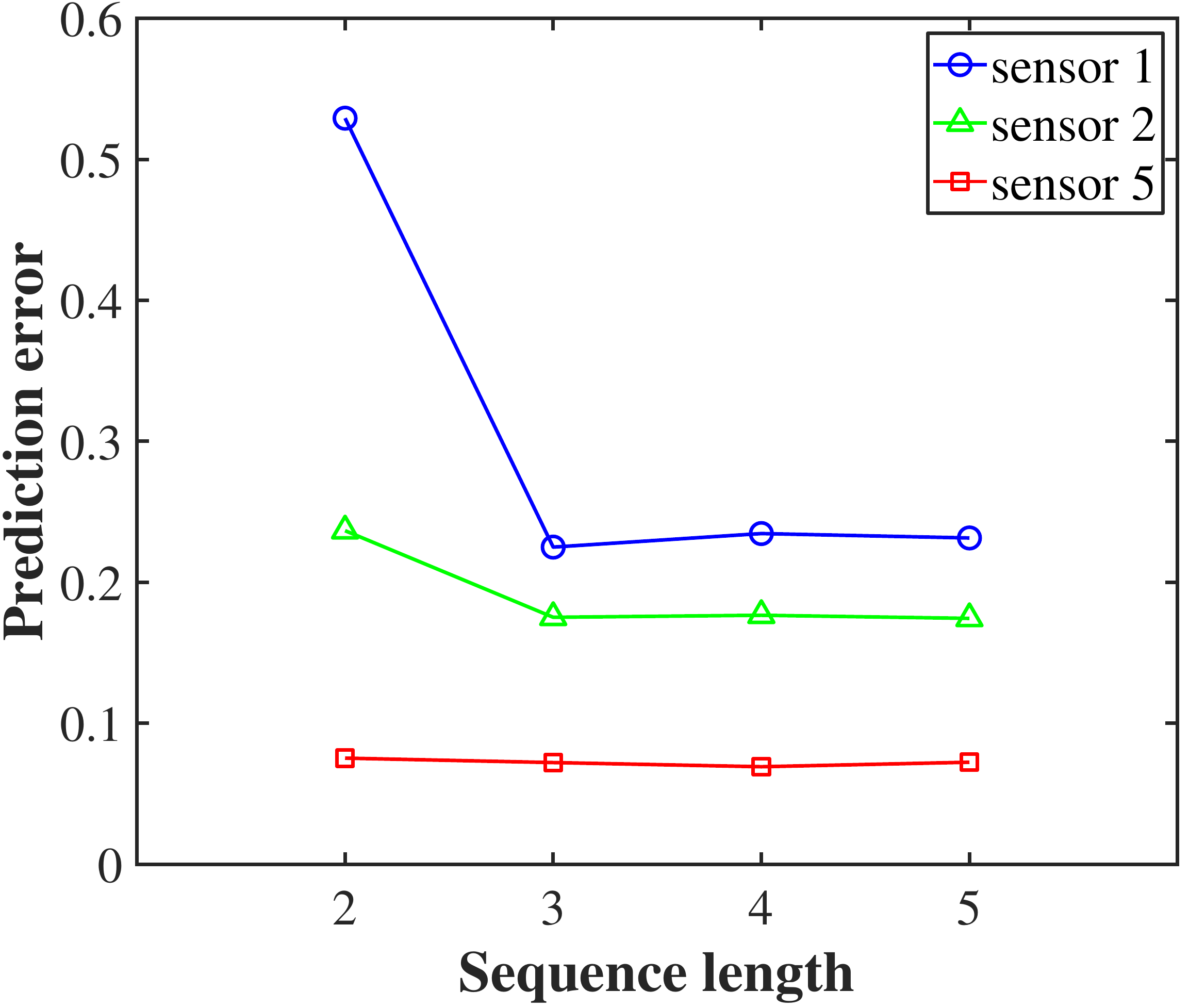}
    \label{fig:res4_eg2b}}
    \caption{(a) Performance of PDS, SD, and proposed ARE with increase in number of sensors. (b) Performance of proposed ARE with increase in sequence length. Data used in both is without noise.}
    \label{fig:res4_eg2}
\end{figure}

We also investigate the effect of a number of sensors and sequence length considered in the proposed approach. Fig. \ref{fig:res4_eg2a} shows the performance of different methods with an increase in the number of sensors.  Similar to previous example, cases corresponding to one,  two,  five, and ten sensors are presented. As expected, increase in number of sensors results in a improved accuracy.
As for accuracy, for all the four cases, the proposed approach yields the best result outperforming both PDS and SD. Fig. \ref{fig:res4_eg2b} illustrates the performance of the proposed approach corresponding to the different sequence length. 
As expected, initially, the results are found to improve with an increase in the sequence length.
ARE reaches a saturation point at sequence length three, and no significant improvement in results is observed on further increasing the sequence length.

% \begin{figure}
%     \centering
%     \includegraphics[width = 0.6\textwidth]{Trmodelvssensor.pdf}
%     \caption{Performance of PDS and proposed ARE with increase in number of sensors.}
%     \label{fig:res4_eg2}
% \end{figure}

% \begin{figure}
%     \centering
%     \includegraphics[width = 0.6\textwidth]{TrOnlineSeq1.pdf}
%     \caption{Performance of proposed ARE with increase in sequence length.}
%     \label{fig:res5_eg2}
% \end{figure}

Lastly, similar to previous example, we have also considered the case where both past and future data are available. B-RNN based ARE is used for solving the same. For details on the same, interested readers may refer \ref{appB}.

\subsection{NOAA Optimum Interpolation (OI) SST V2 Dataset}
As the last example, we test our approach on the well-known Sear Surface Temperature (SST) dataset. SST Data is publicly available and uploaded by the Physical Sciences Division at National Oceanic and Atmospheric Administration (NOAA). The data is made available by the respective organization after the analysis of in-situ and satellite observation of the SST. The resolution of the data available corresponding to the variable ``time'' is Weekly, Monthly and monthly long-term mean. The resolution corresponding to the spatial domain is of 1 deg latitude and 1 deg longitude, which translates to 180 x 360 grid points. In this study we make use of weekly mean data from the year 1990 on-wards. The data is said to be centred around Wednesday for this period. Data for initial 400 weeks was chosen as the training data-set. Following 100 weeks data was chosen as validation data-set and following 200 weeks data for the testing. The problem description here is to reconstruct the SST data on the spatial domain with the help of limited sensor data measurements. Details on the network architecture used for this problem are provided in Table \ref{tab:na_eg3}.

\begin{table}[h]
    \centering
    \caption{Network architecture of proposed ARE and SD for SST dataset. BN indicates batch normalization and DR represents dropout. HS is the number of features in the hidden state of lstm.}
    \label{tab:na_eg3}
    \begin{tabular}{|p{2cm}|p{15cm}|}
        \hline
        Networks & Architecture  \\ \hline
    ARE(Auto-Encoder) & $44219\rightarrow BN^* \rightarrow 512\rightarrow BN^* \rightarrow DR(0.35)^*\rightarrow 256\rightarrow 25\rightarrow 256\rightarrow 512\rightarrow 44219$ \\
    ARE(RNN) & LSTM(HS=50) $\rightarrow 100 \rightarrow 100 \rightarrow 25$ \\ 
    SD & $N_s\rightarrow350\rightarrow BN\rightarrow DR(0.1)^* \rightarrow400\rightarrow BN\rightarrow 44219$ \\ \hline
    \multicolumn{2}{l}{\footnotesize{Components with $^*$ are not used in case of noise-free data.}}
    \end{tabular}
\end{table}

\begin{figure}[h]
    \centering
    \subfigure[sensors 2]{\includegraphics[width = 0.3\textwidth]{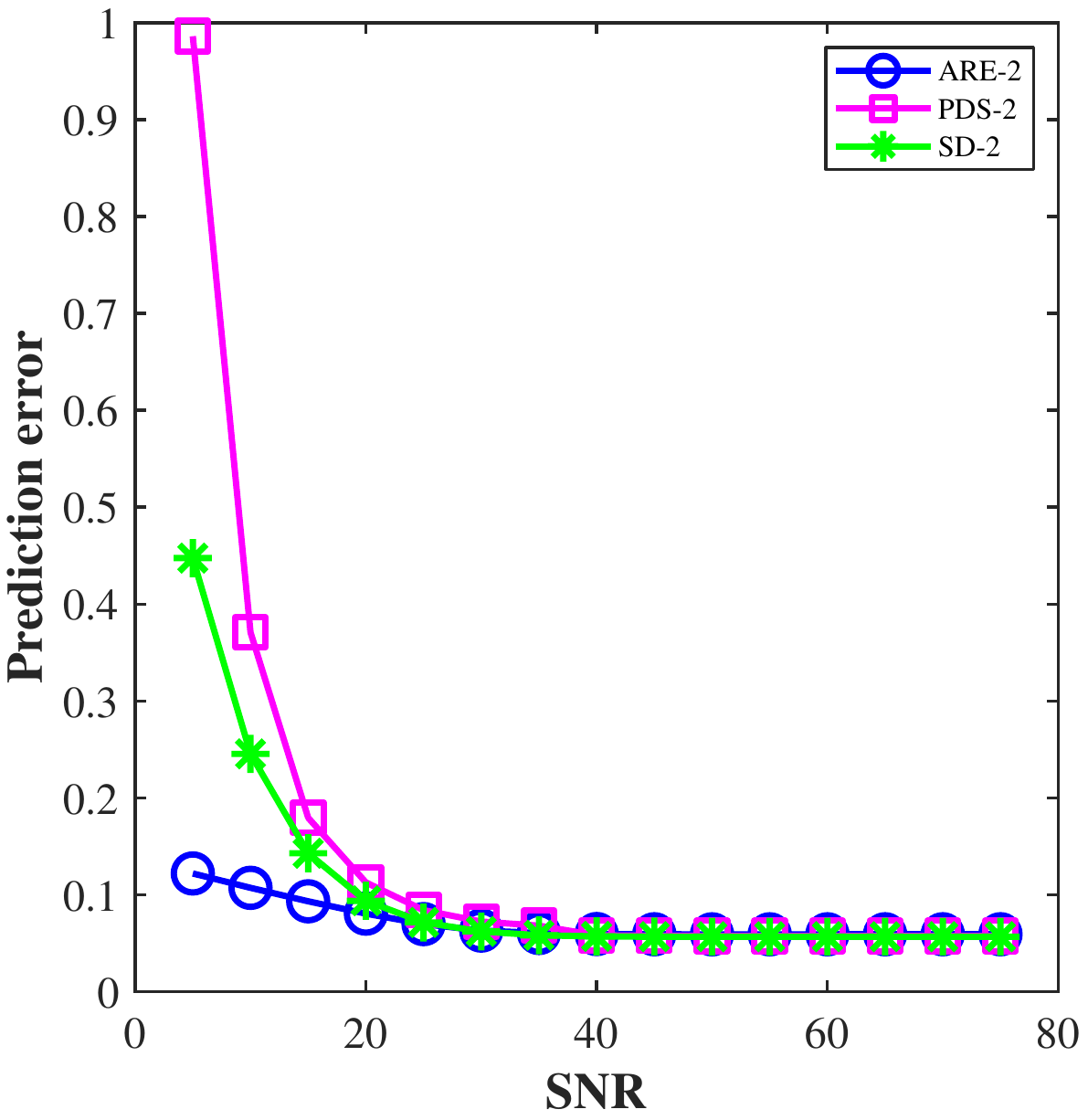}}
    \subfigure[sensors 4]{\includegraphics[width = 0.3\textwidth]{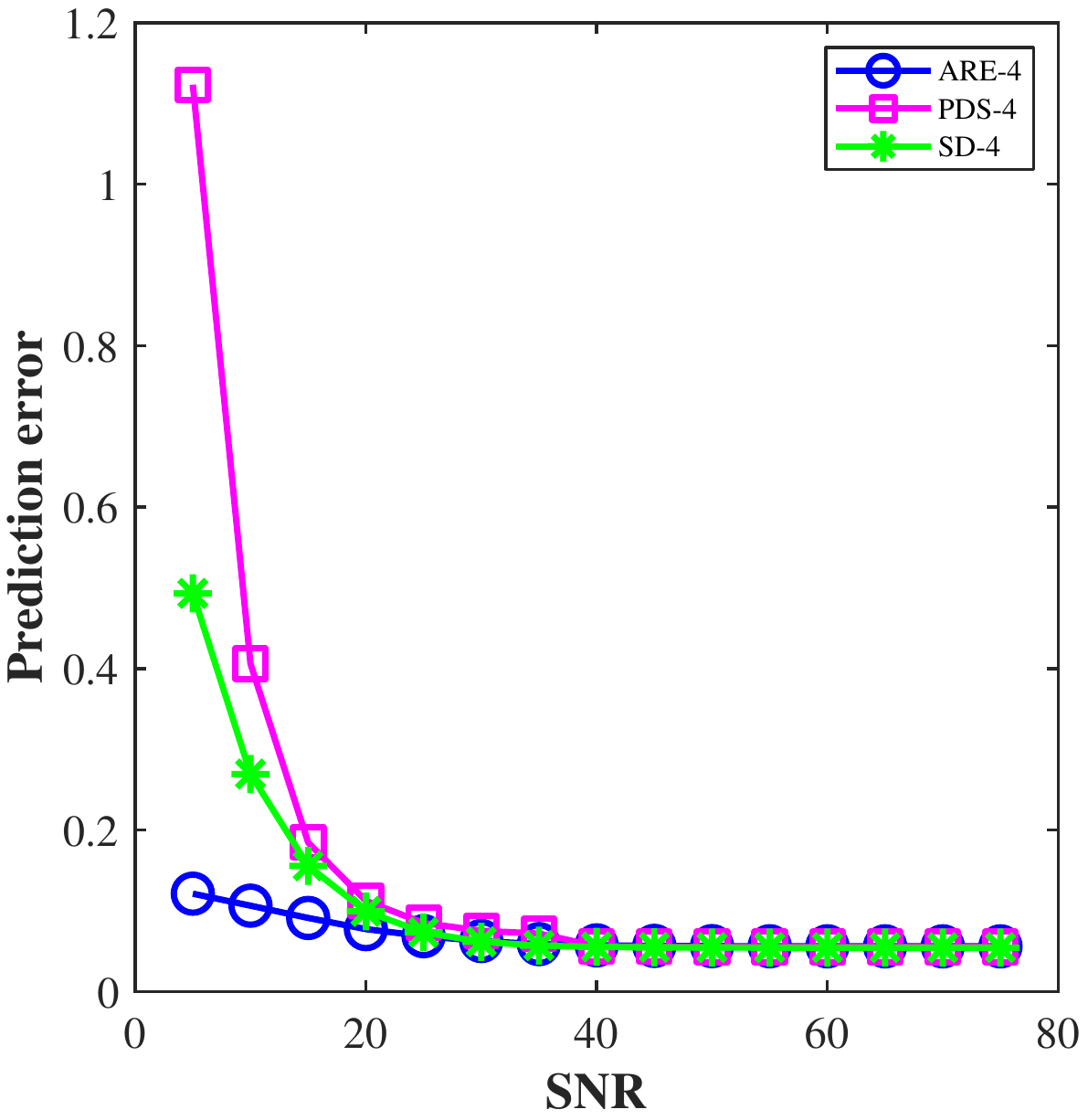}}
    \subfigure[sensors 8]{\includegraphics[width = 0.3\textwidth]{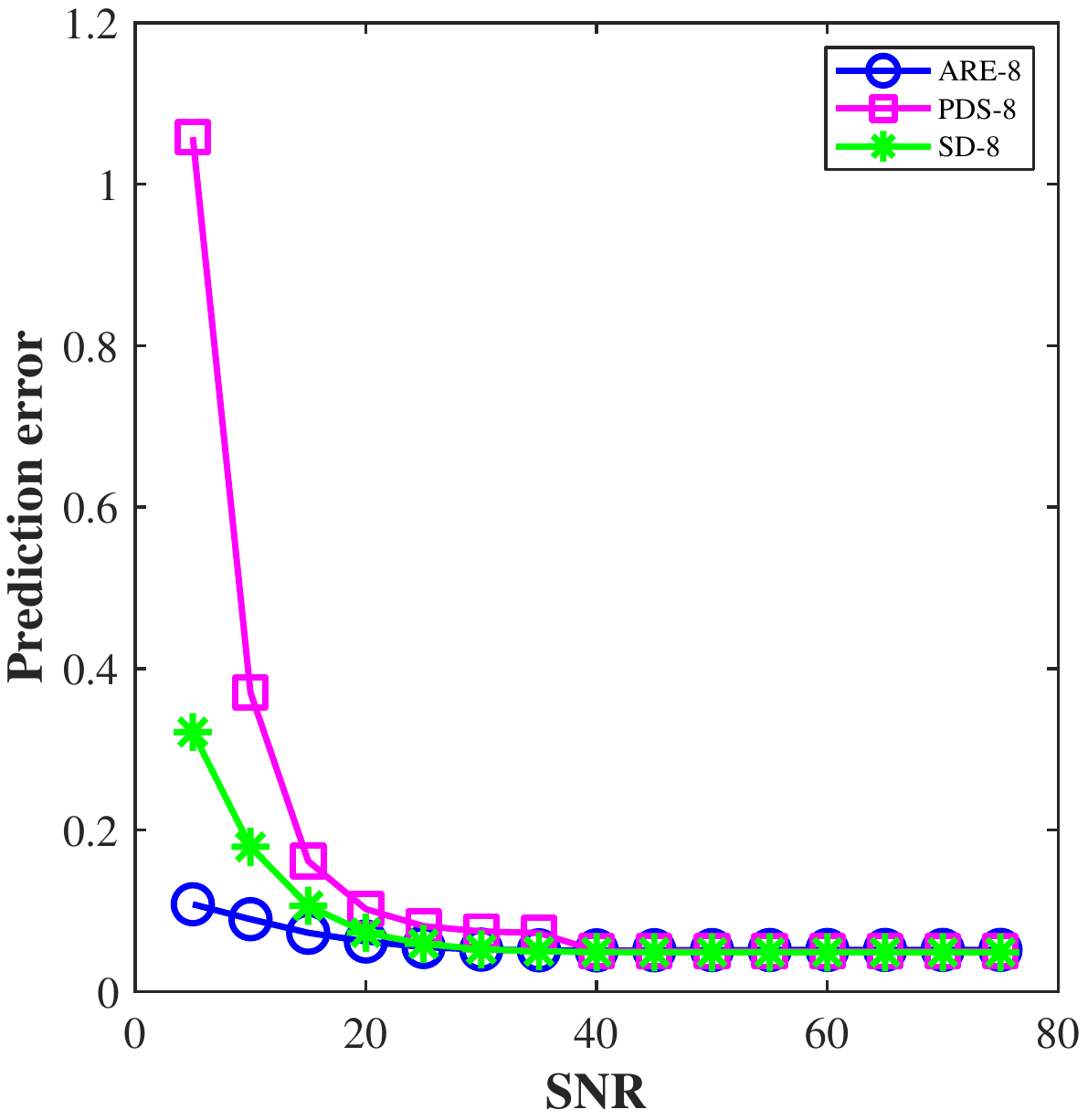}}
    
    \caption{Change in performance of different approaches is presented with increasing noise using three cases}
    \label{fig:sst_mthd__snr}
\end{figure}

\begin{figure}[t]
\centering
\subfigure[PDS]{\includegraphics[width=0.42\linewidth,trim=2.5cm 2.1cm 1.8cm 2.0cm,clip]{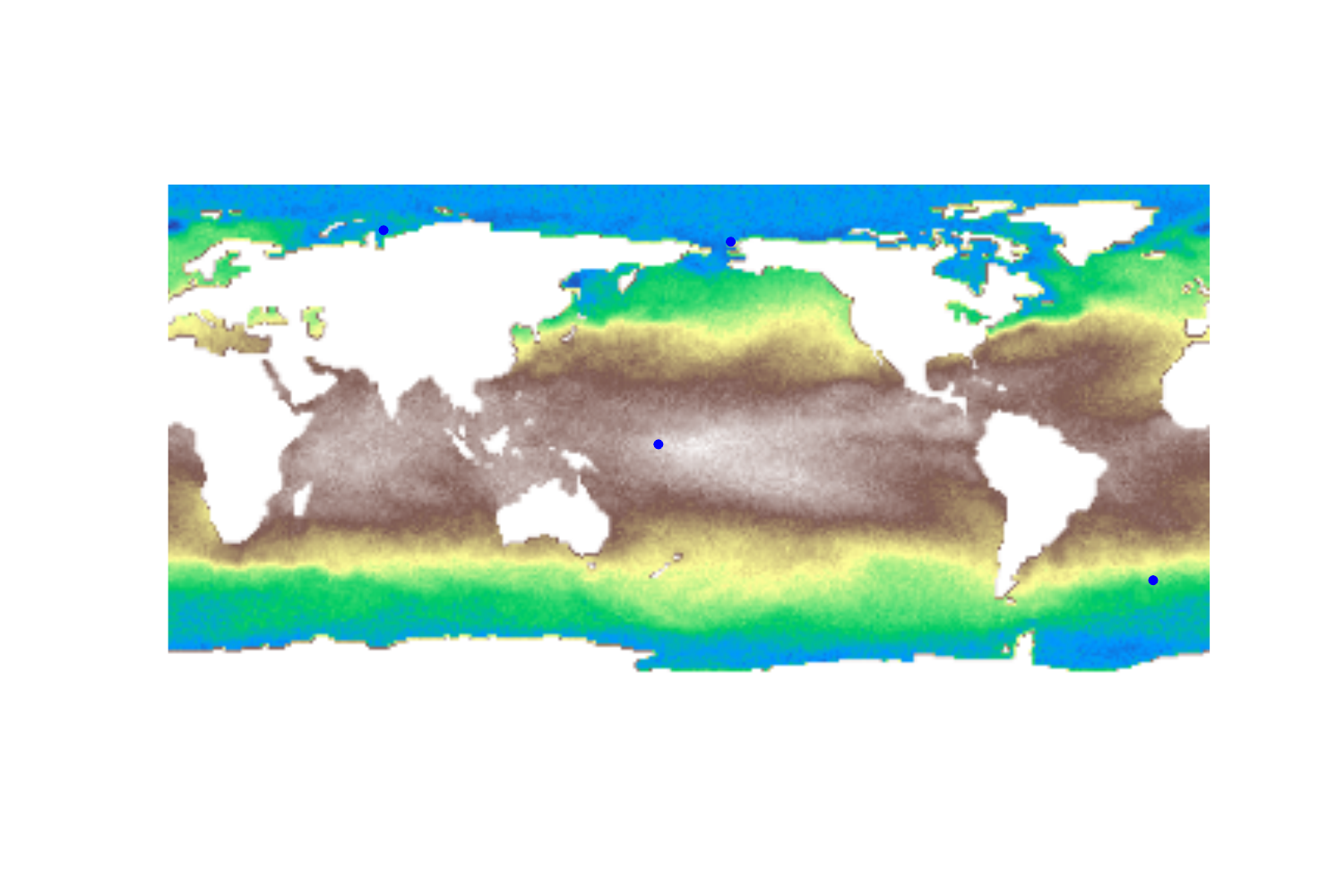}}
\subfigure[ARE]{\includegraphics[width=0.42\linewidth,trim=2.5cm 2.1cm 1.8cm 2.0cm,clip]{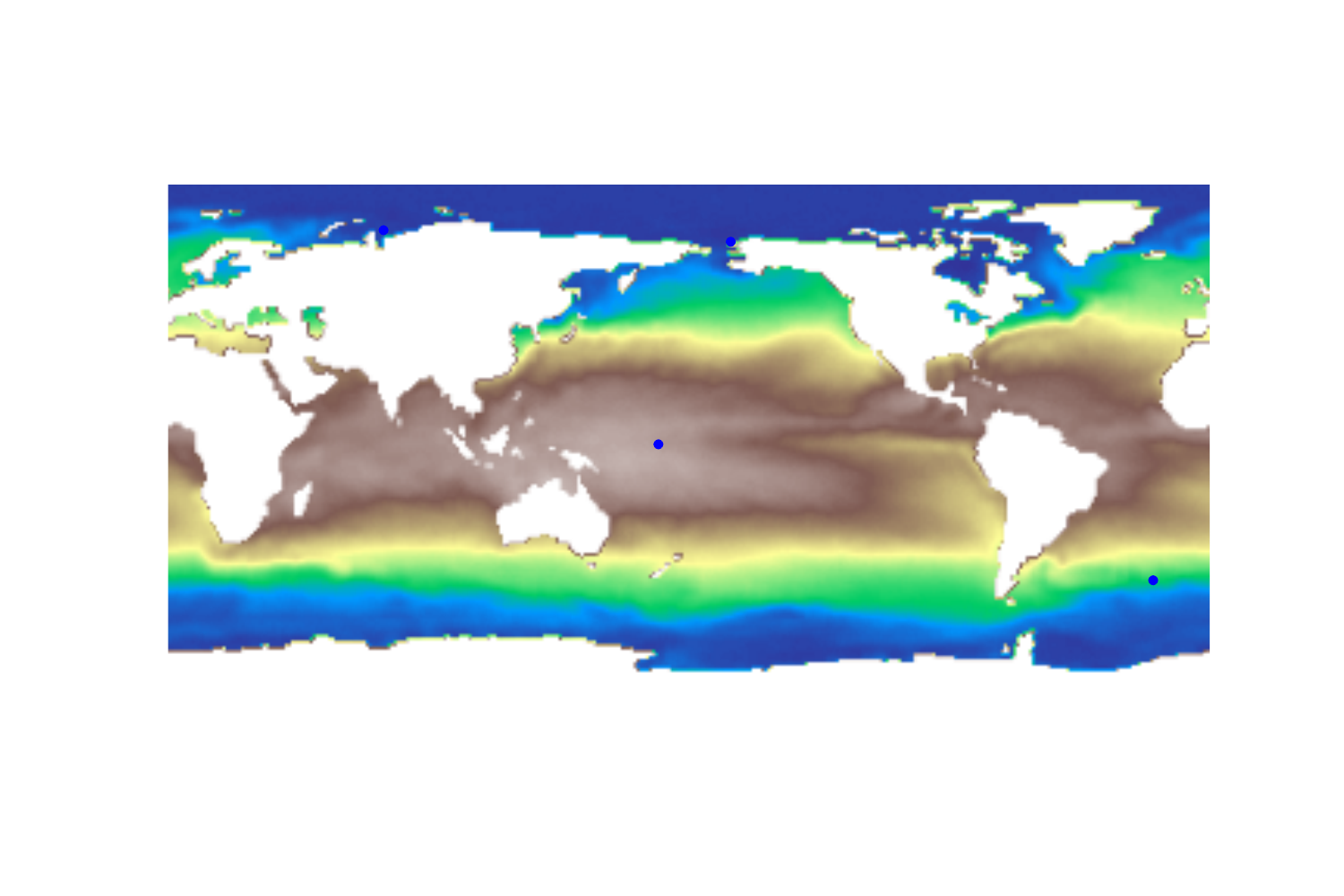}}
\subfigure[SD]{\includegraphics[width=0.42\linewidth,trim=2.5cm 2.1cm 1.8cm 2.0cm,clip]{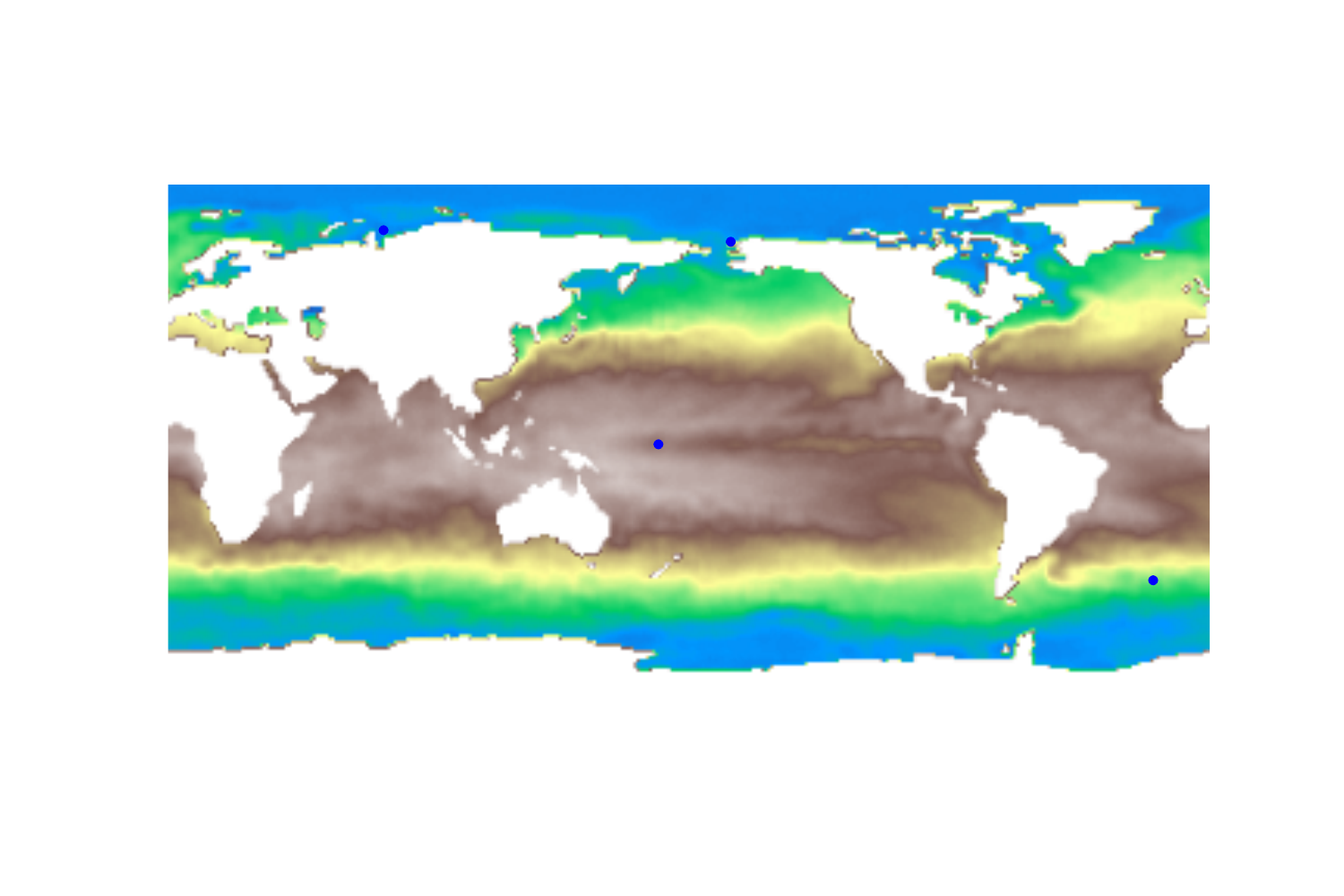}}
\subfigure[Ground truth]{\includegraphics[width=0.42\linewidth,trim=2.5cm 2.1cm 1.8cm 2.0cm,clip]{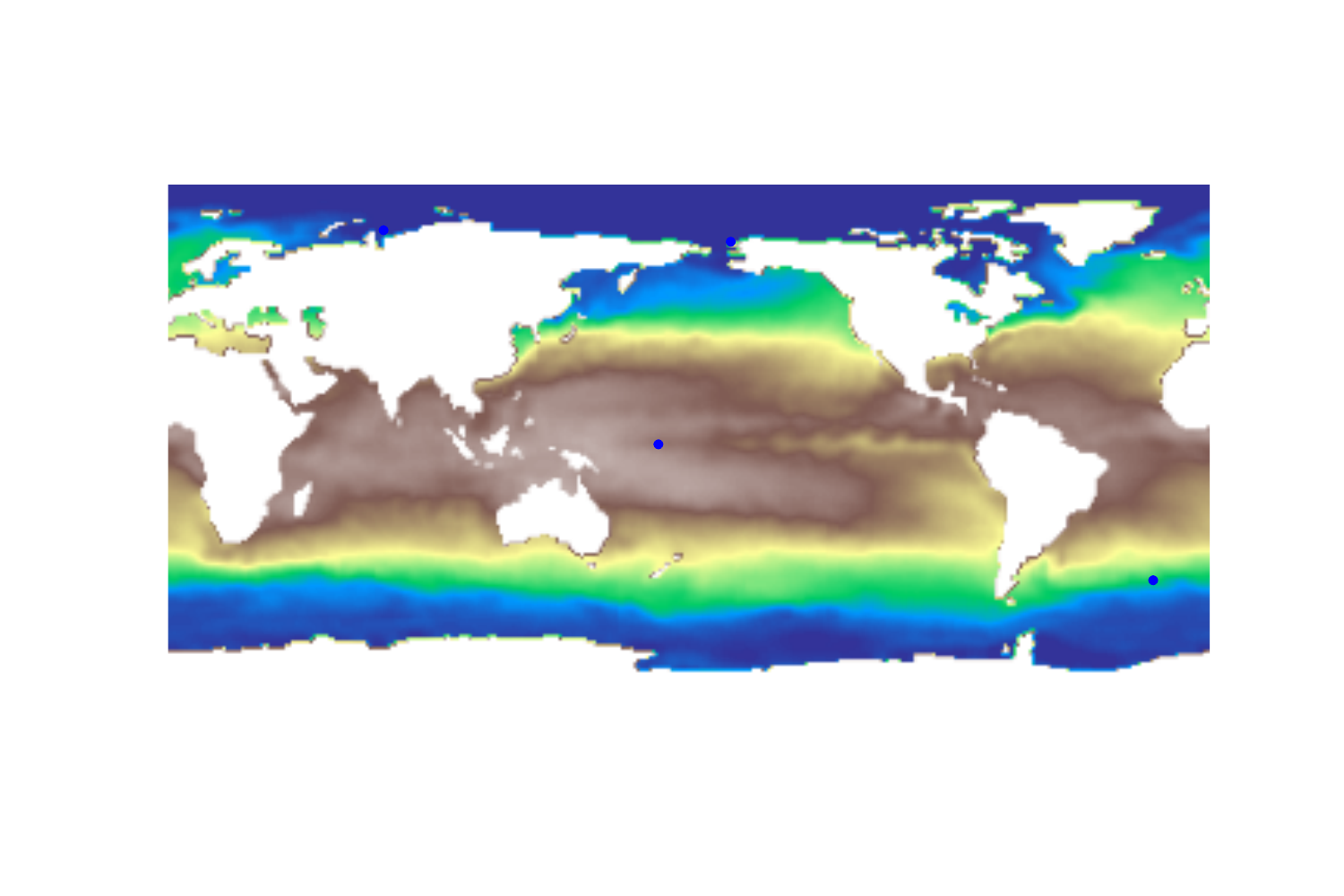}}
\caption{Figure depicts results of SST with 4 sensors and at noise of SNR 20. Blue dots in the images represents the location  of sensor. SEQ-LEN used for ARE model is 4}
\label{fig:res3_eg3}
\end{figure}

Fig. \ref{fig:sst_mthd__snr} shows the comparison of results between three different architectures we chose for our study i.e., PDS, SD, and ARE. To show the applicability of the techniques in real world scenarios, we make use of corrupted or noisy data at different signal to noise ratios and at three different sensor measurements i.e. 2, 4 and 8. The proposed problem reconstructs the SST data over the globe accurately for different sensor data input. We observe, with decreasing noise in the data, either of the approaches perform well; however, when noise in the data is significant, the proposed ARE is found to be significantly more accurate as compared to PDS and SD. Even with higher number of sensors, the proposed approach is found to be superior (at higher noise level) as compared to PDS and SD.
For better visualization, the contour plot of reconstructed sea surface temperature is shown in Fig. \ref{fig:res3_eg3}. We have considered SNR to be 20 and assumed data to be available from four sensors. Sequence length of four is used for the proposed ARE.
It is clearly visible that the reconstructed field using the proposed ARE is more accurate compared to the PDS and SD reconstructed fields.

\begin{figure}[ht!]
    \centering
    \subfigure[SNR 10]{\includegraphics[width = 0.3\textwidth]{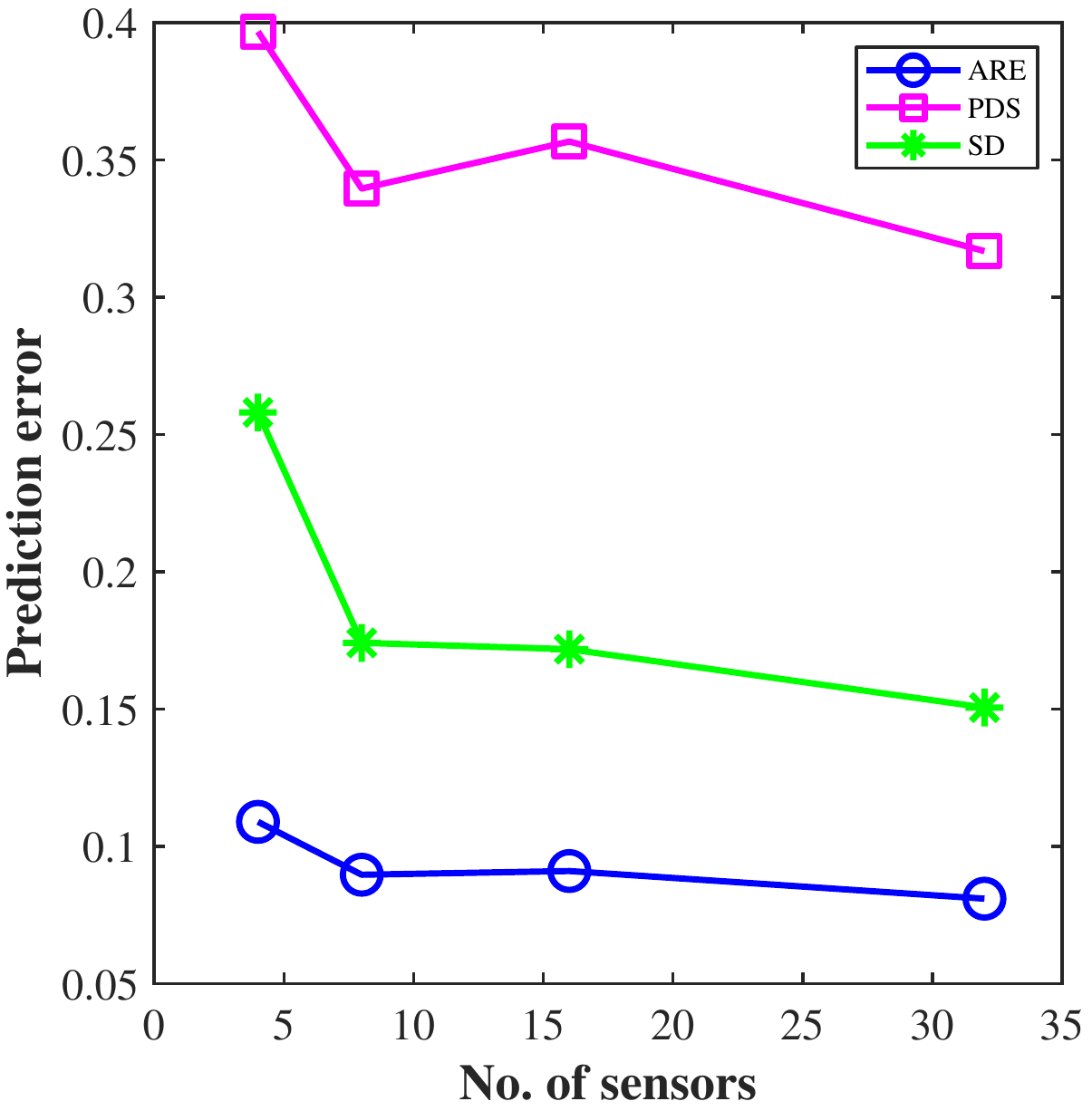}}
    \subfigure[SNR 20]{\includegraphics[width = 0.3\textwidth]{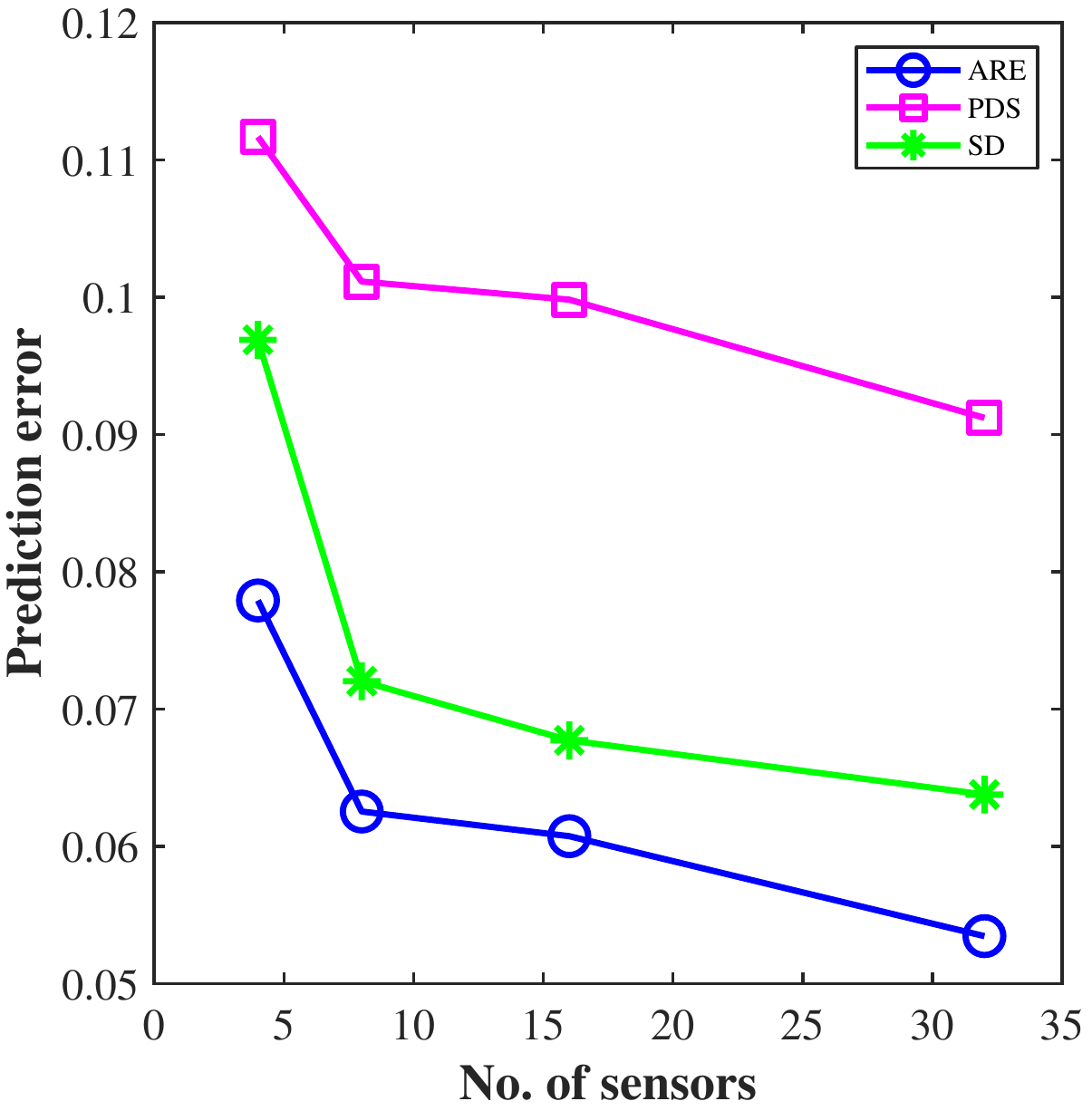}}
    \subfigure[SNR 30]{\includegraphics[width = 0.3\textwidth]{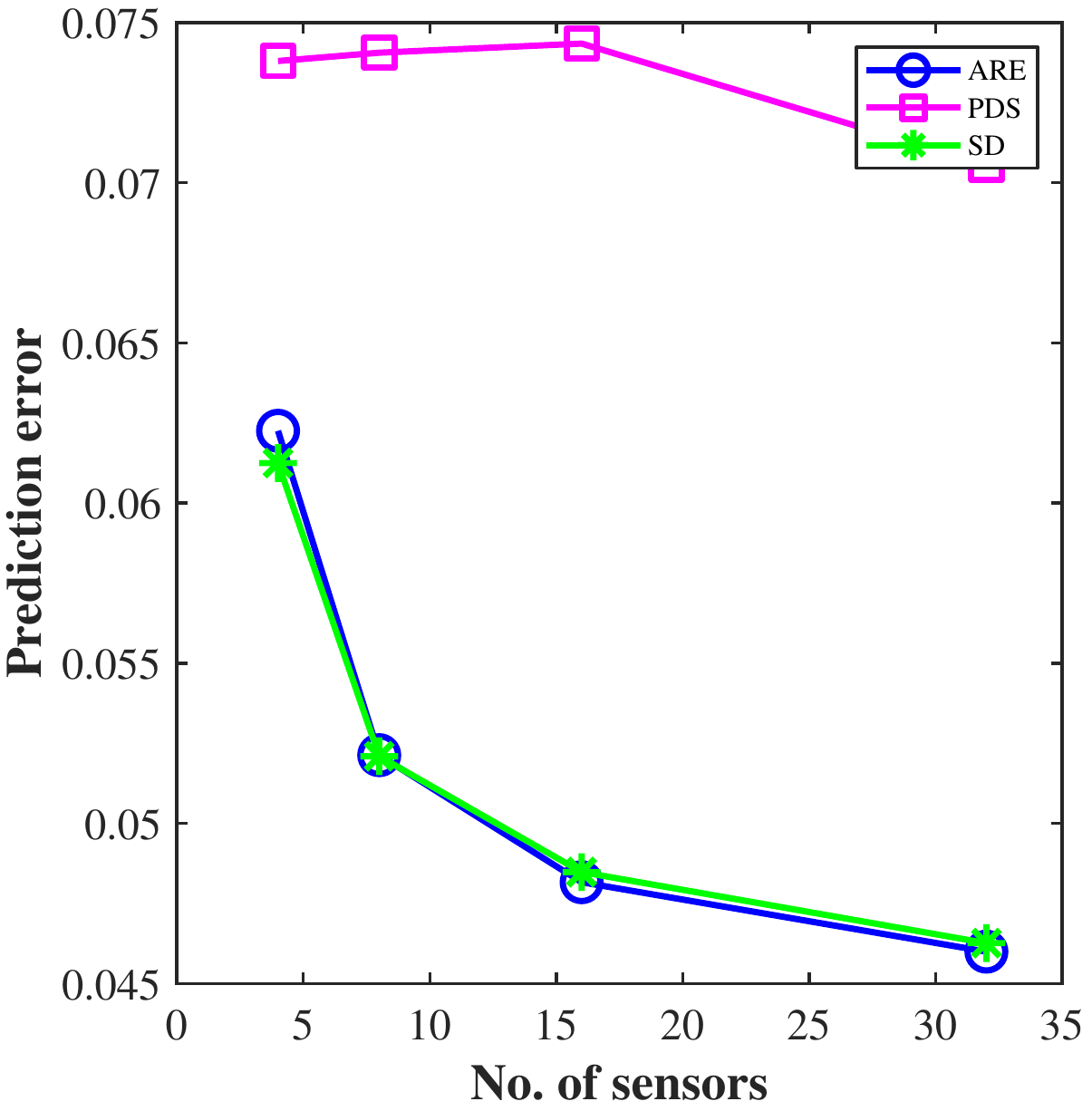}}
    
    \caption{Performance of the three model is compared with increasing no. of sensors at SNR {10, 20, 30}}
    \label{fig:sst_mthd__snsr_with_noise}
\end{figure}

\begin{figure}[t]
    \centering
    \subfigure[sensor 2]{\includegraphics[width = 0.3\textwidth]{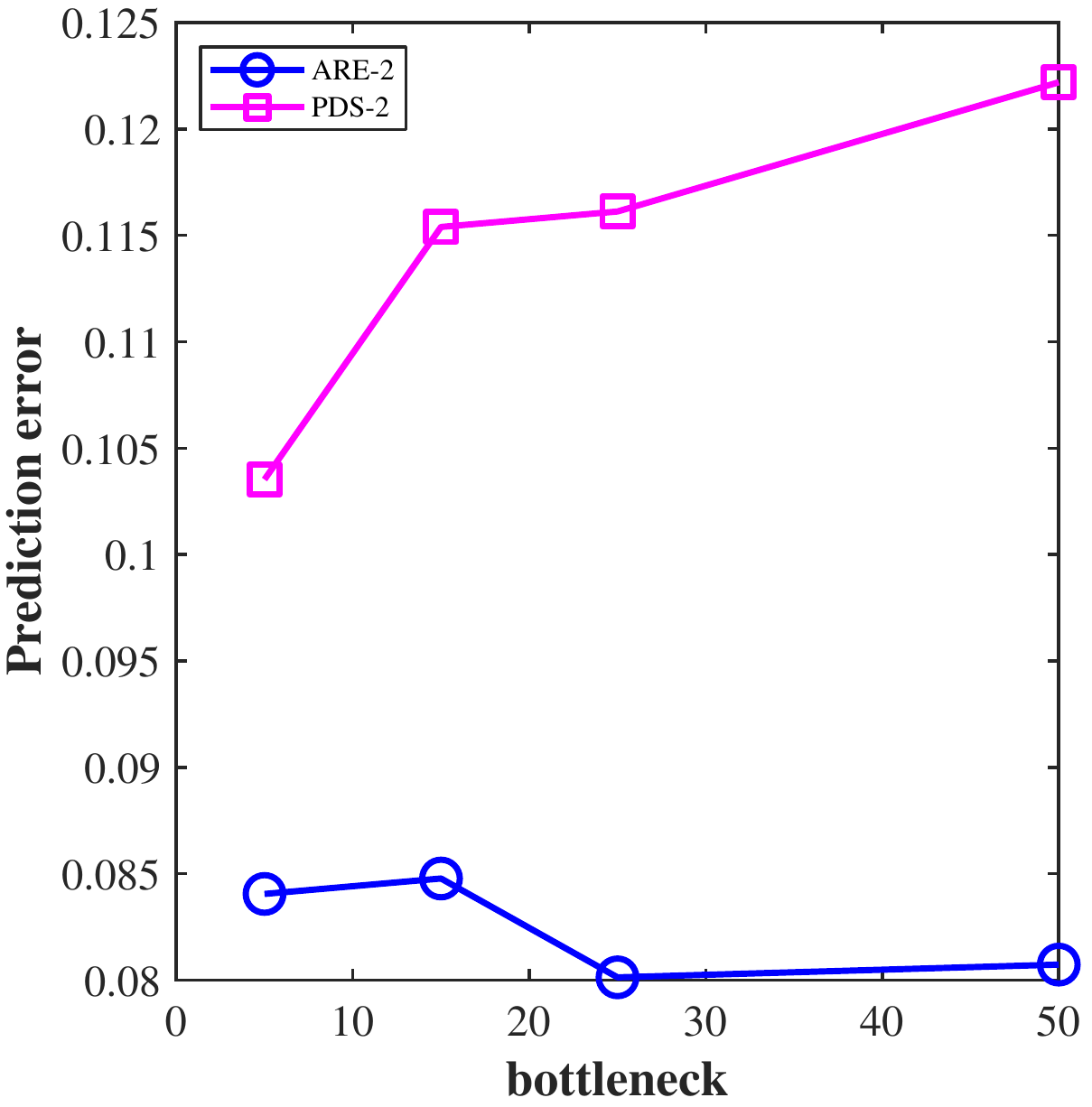}}
    \subfigure[sensor 4]{\includegraphics[width = 0.3\textwidth]{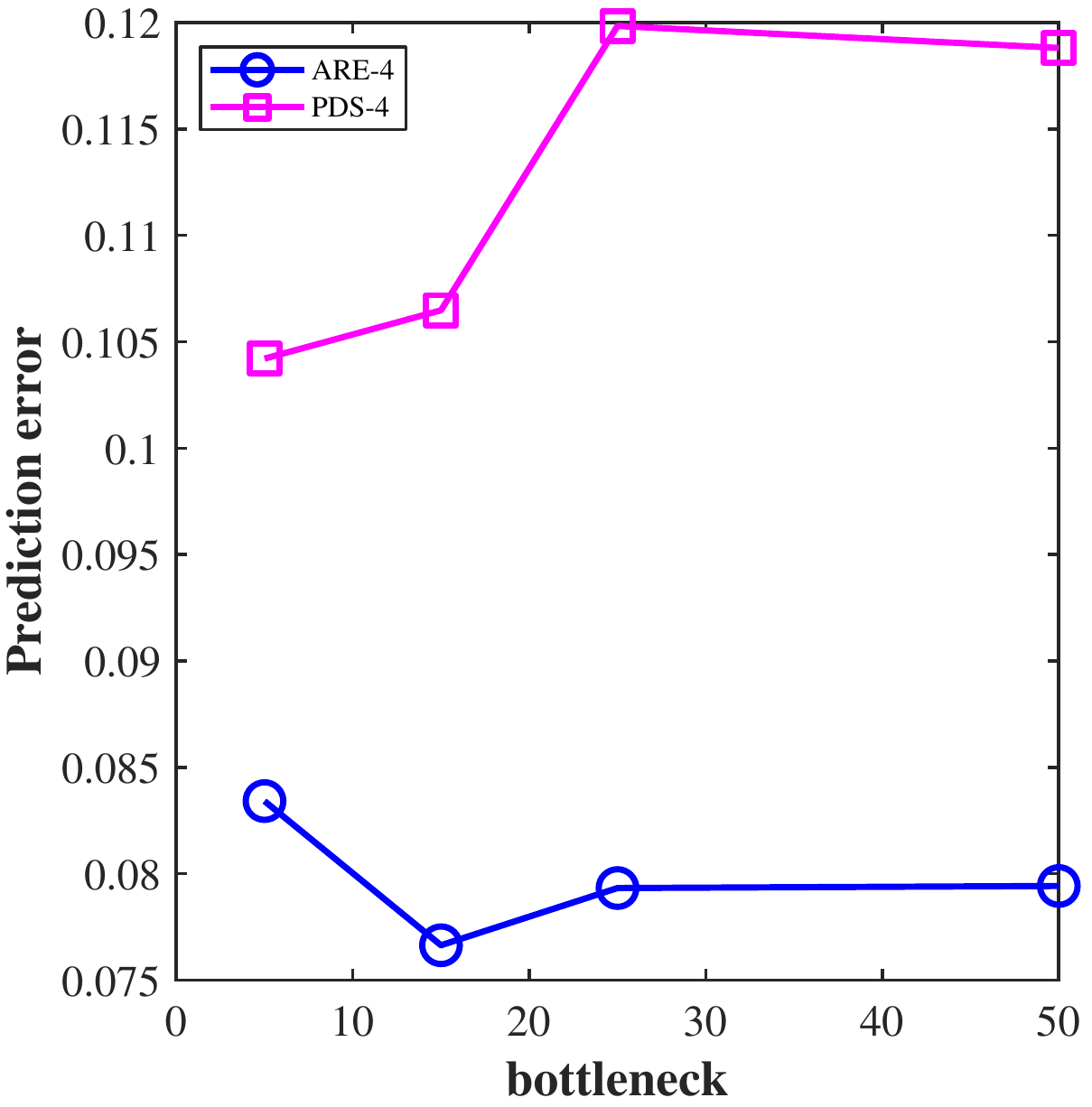}}
    \subfigure[sensor 8]{\includegraphics[width = 0.3\textwidth]{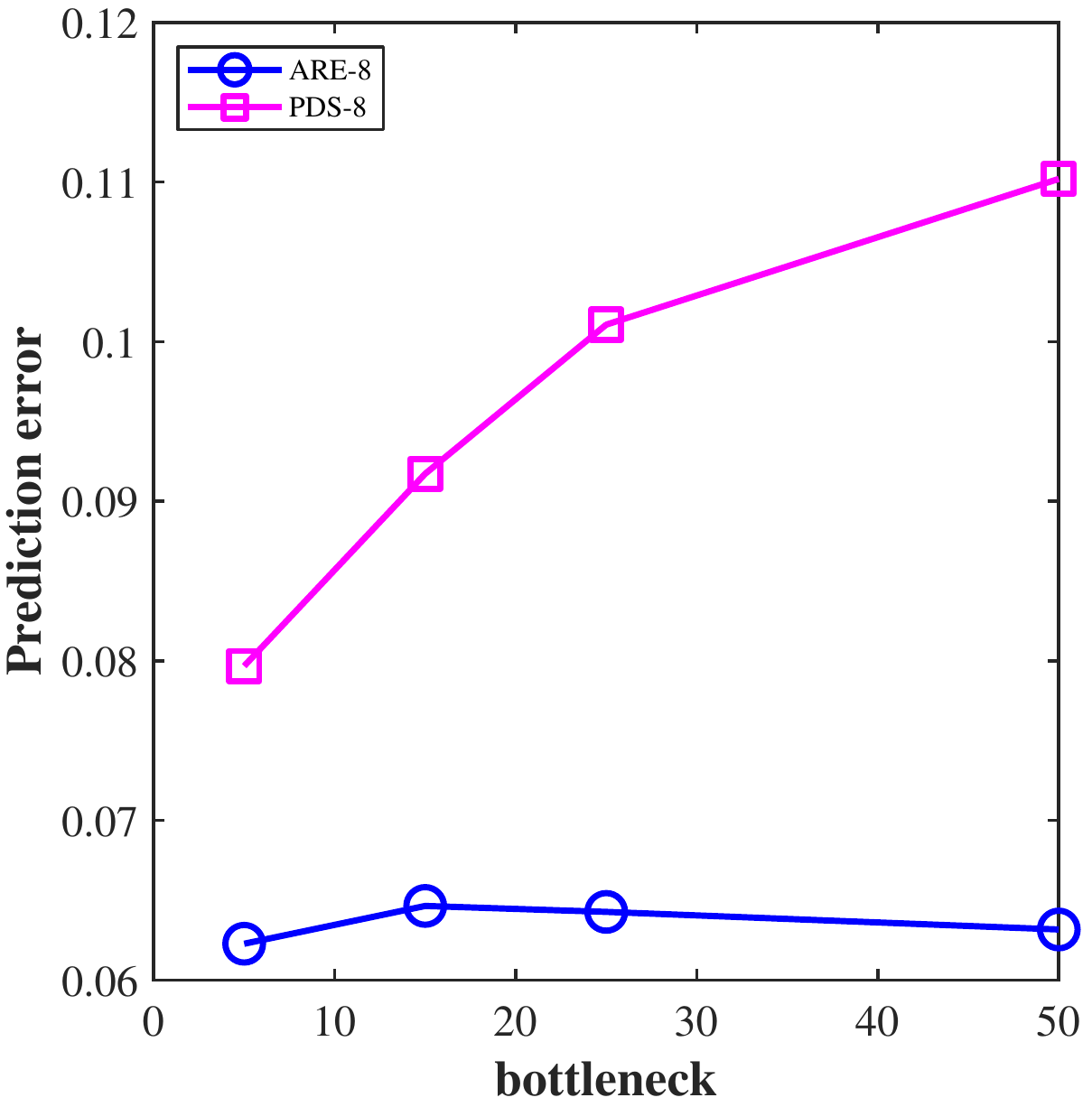}}
    
    \caption{Performance of ARE and PDS is compared at different number of neurons in bottleneck layer with noise of SNR 20}
    \label{fig:sst_mthd__bn}
\end{figure}

In Fig. \ref{fig:sst_mthd__snsr_with_noise},  we compare the results of the approaches corresponding to the increasing number of sensors at three different SNR ratios. The investigation further strengthens our claim for the robustness of our proposed approach, ARE. As the number of sensors increases, it is evident that all the approaches tend to present better results. For the sake of comparison, we considered corrupted data at three different SNR,  10, 20 and 30, since the approaches mentioned perform more or less the same without noise. At SNR 10, the proposed approach tends to outperform its counterparts by a larger margin compared to SNR 30. As the corruption in data increases, ARE tends to perform much better than its counterparts. SD performs almost similar to ARE at lower noise level and better than PDS in all cases. 

Lastly, we investigate the sensitivity of the proposed approach to the number of neurons in the bottleneck layer.
For PDS, this corresponds to the number of POD modes.
The concept of bottleneck layer doesn't exist for SD and hence, the same is not reported.
Results obtained are shown in Fig. \ref{fig:sst_mthd__bn}. The results are presented corresponding to cases with 2, 4 and 8 sensors. The SNR is kept fixed at 20. The number of neurons in the bottleneck layer is varied in $[5,50]$.
We observe that the proposed approach is less sensitive to the number of neurons in the bottleneck layer, although slight improvement can be observed as the layer size is increased. This observation presents an insightful understanding of our approach which could be translated to a computationally less expensive approach, with decreased layer size of our bottleneck layer. As for PDS, we observe that the accuracy reduces with increase in the number of POD modes.
This seems counter-intuitive as increase in POD modes is expected to reduce the error. However, on a closer examination, we observed that with more number of POD modes, the number of parameters in neural network increases, resulting in overfitting. This can be addressed by increasing the number of sensors.
Overall, the results presented in this section illustrate the accuracy and robustness of the proposed approach in state estimation from limited sensor.

\section{Conclusions}
\label{sec:conc}
In this work, we introduced a novel deep learning based approach for state estimation.
The proposed approach uses an auto-encoder as a reduced-order model and recurrent neural network to map the sensor measurements to the reduced state.
The proposed framework is superior to existing state estimation frameworks in two aspects.
First, auto-encoder, being a nonlinear manifold learning framework, is superior to the usually used proper orthogonal decomposition.
Secondly, unlike existing state estimation frameworks, the proposed approach utilizes present and past sensor measurements using the recurrent neural network.
It results in improved accuracy, specifically for cases where limited sensors are deployed.
Experiments performed on simulation of flow past a cylinder and sea-surface data illustrated the capability of the proposed approach in learning from sequential data.
Comparison carried out with respect to proper orthogonal decomposition-based deep state estimation and shallow-decoder showed the superior accuracy of the proposed approach. 
Moreover, the proposed approach was also found to be robust to the noise in the sensor measurements.

Utilizing sequential information can prove beneficial in many other state estimation tasks. Future work can be aimed at exploiting effects of other models used on sequential information such as transformers, comparing between other auto-encoder network variations, the effect of varying time step between measurements, developing models capable of transfer learning, i.e., trained on smaller time step but can use larger time step. Work on utilizing, recently developed a physics-informed neural network for solving such problems can also be pursued in the future.

\textbf{Acknowledgements: } 
The authors would like to thank Dr. Arghya Samanta and Nirmal J Nair for the useful discussions during this paper's preparation. SC acknowledges the financial support of the I-Hub Foundation for Cobotics (IHFC) and seed grant provided through Faculty initiation grant, IIT Delhi.

% \bibliographystyle{unsrt}
% % Note the spaces between the initials
% \bibliography{sample}

\appendix
\section{Proper orthogonal decomposition based deep state estimation (PDS)}\label{appA}

In this section, we briefly provide the theoretical background of PDS. We note that development of PDS is motivated from gappy-POD and linear stochastic estimation and hence, the discussion in this section also starts from gappy-POD and then proceeds to PDS via linear stochastic estimation.

Let training set consist of $N_{t}$ flattened data images, $\mathbf W=(\bm w_1, \bm w_2, \ldots, \bm w_{N_{t}})$, where $\bm w_t \in \mathbb{R}^{N_w}$ and corresponding sensor measurements $\bm s = (\bm s_1, \bm s_2,\dots,\bm s_{N_{t}})$. As already state in Section \ref{sec:ps}, the aim in state estimation is to recover full state $\bm w$ via sensor measurements $\bm s$.
\subsection{Gappy-POD}
Consider, $\mathbf H \in \mathbb{R}^{N_s\times N_w}$ is measurement operator that maps full state to measurements and have ones as sensor location and zeros otherwise. Mathematically,  
\begin{equation} \label{eu_eqn111} 
\bm s = \mathbf{H} \bm w, 
\end{equation}
where $\bm w$ can be approximated by linear combination of $k$ modes. Modes used here are $k$ most dominant modes, $\bm \Phi$, and are obtained by proper orthogonal decomposition of training data $\mathbf W$ and selecting first $k$ singular vectors,
\begin{equation} 
    \bm w \approx \sum_{i=1}^{k}\phi_i a_i = \bm \Phi \bm A, 
\end{equation}
where
\begin{equation} \label{eq:POD}
    \mathbf W \approx \bm \Phi D B^* , truncated\hspace{2mm}k-rank,
\end{equation}
and 
\begin{equation}  
\bm s \approx \mathbf H \bm \Phi a.
\end{equation}
In Eq. (\ref{eq:POD}), $\bm \Phi \in \mathbb{R}^{N_w\times k}$. During testing $\bm a$ is obtained by solving the following minimization problem.
\begin{equation} \label{eu_eqn11} 
\bm a \in  \underset{\tilde{\bm a}}{\arg\min} \left\|\bm s-\mathbf H \bm \Phi \tilde{\bm a} \right\||_2^2.
\end{equation}
Solution to this problem is obtained by taking Moore-Penrose pseudo inverse 
\begin{equation} 
\bm a =(\mathbf H \bm \Phi)^+ \bm s
\end{equation}
This approach requires previous knowledge of operator $\mathbf H$. This operator is only available for simple systems but is often unknown for systems of practical interest.

\subsection{Linear stochastic estimation}
Linear Stochastic Estimation overcomes this issue by defining another operator which will map latent state to sensor. Operator $P : \mathbb{R}^{k}\mapsto \mathbb{R}^{N_s}$ is learned from training data via the following minimization problem. 
% \begin{equation} \label{eu_eqn}
% P = \bm s( \bm \Phi^+ \mathbf W)^+
% \end{equation}
\begin{equation} 
P \in \underset{\tilde{P}}{\arg\min}||\bm s-\tilde{P}\bm A||^2_2,
\end{equation}
where $\mathbf S \in \mathbb{R}^{N_s\times N_{t}}$ and $A\in \mathbb{R}^{k\times N_{t}}$.
This approach completely skips over the operator $\mathbf H$; instead, $P$ is
considered to be the empirical estimate of the linear operator $\mathbf H\Phi$.
Subsequently, $\bm a $ is obtained as
\begin{equation} \label{eq:jk} 
\bm a \in  \underset{\tilde{\bm a}}{\arg\min} ||\bm s - P \tilde{\bm a}||^2_2\end{equation}

\subsection{Deep state estimation}\label{PDS}
Deep state estimation \citep{nair_goza_2020} approache replaces the linear mapping between latent state and sensors with nonlinear mapping to further generalize the approach. It uses a neural network $G: \mathbb{R}^{N_s}\mapsto \mathbb{R}^{bn}$ parametrized by $\theta$ for nonlinear mapping of sensor measurements to approximate embeddings.
\begin{equation} \label{eu_eqn1} \bm a = G(\bm s,\bm \theta)\end{equation}
The neural network is trained as 
\begin{equation}
\bm \theta \in \underset{\tilde{\bm \theta}}{\arg\min}\sum_{i=1}^{N_{t}}||\bm a^{(i)}-G(\bm s^{(i)},\tilde{\bm \theta})||_2^2,
\end{equation}
where
\begin{equation} \label{eq:op}\bm a \in  \underset{\tilde{\bm a}}{\arg\min} ||\bm w-\Phi \tilde{\bm a}||^2_2. \end{equation}
$\bm \Phi$ as before are the POD-modes. During testing trained neural network $G$ approximate embeddings for sensor measurements which is used to recover full state.
We refer to this method as POD based deep state estimation (PDS).

\section{Bi-directional RNN}\label{appB}

The basic idea of bidirectional recurrent neural nets (B-RNN) \citep{650093,Baldi} is to present each training sequence to two separate recurrent nets namely forward and backward, both of which are connected to the same output layer. (In some cases a third network is used in place of the output layer, but here we have used the simpler model). This means that for every point in a given sequence, the BRNN has complete, sequential information about all points before and after it. In this work, we have used B-RNN for solving a special state estimation problem where sensor measurements from both past and future states are available.

\begin{figure}
    \centering
    \includegraphics[width = \textwidth]{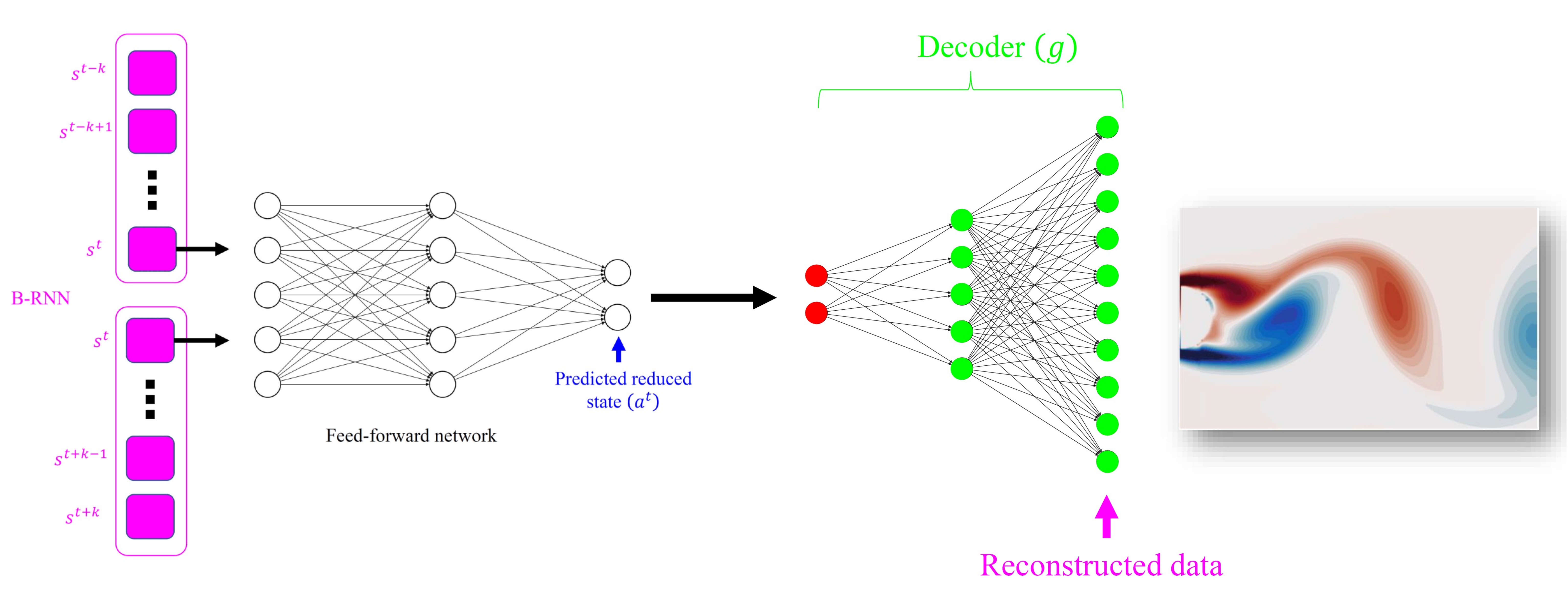}
    \caption{Schematic representation of bi-directional ARE. This is useful when both past and future sensor measurements are available.}
    \label{fig:B-RNN}
\end{figure}

A Schematic representation of B-RNN used is shown in the Fig. \ref{fig:B-RNN}. It is modified to stop propagating information after the middle time step in forward and backward network. This lowers computational time and reduce number of training weights. Thus architecture proposed is capable of using information from both ahead and behind in time.   
\begin{figure}[t]
    \centering
    \subfigure[Periodic vortex shedding]{
    \includegraphics[width = 0.4\textwidth]{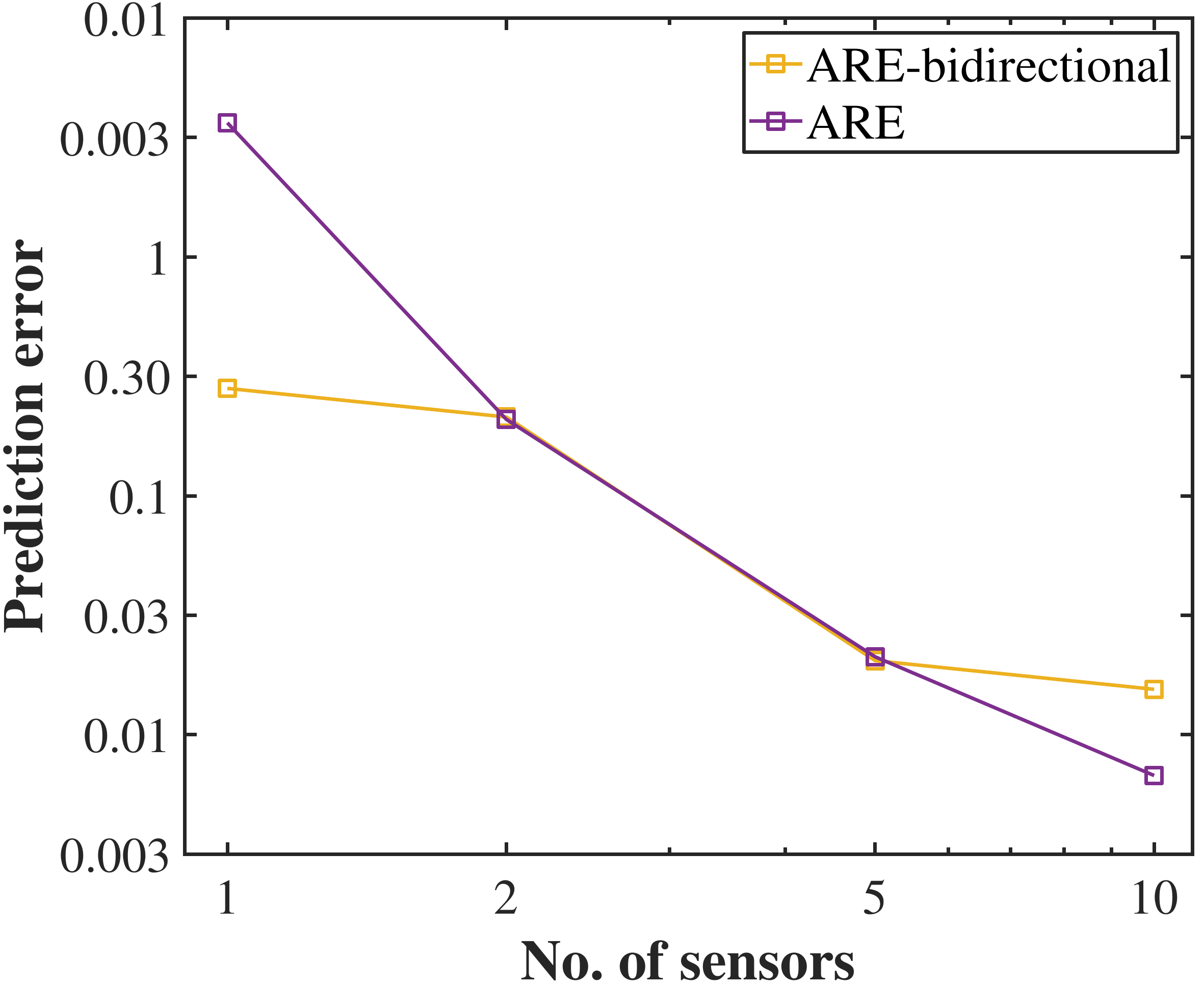}}
    \subfigure[Transient flow]{
    \includegraphics[width = 0.4\textwidth]{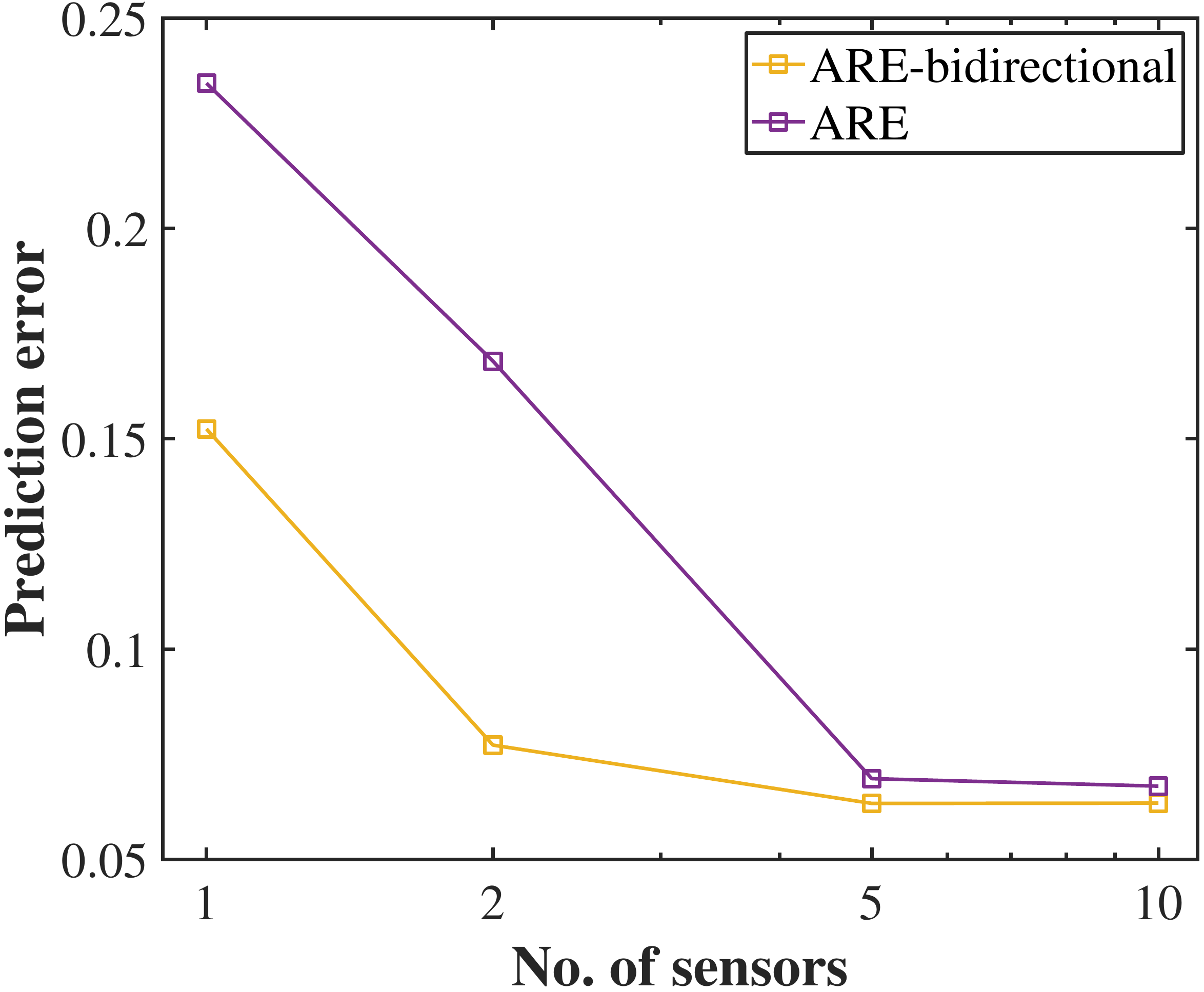}}
    \caption{Comparison between B-RNN based ARE and RNN based ARE}
    \label{fig:brnn_vsrnn}
\end{figure}
To feed sensor measurements $s^{t-k,\ldots,t+k}$ into network, data is splitted into $s^{t-k,\ldots,t}$ and $s^{t,\ldots,t+k}$ as shown in the Fig. \ref{fig:B-RNN}. 
Note that although more accurate, this network can use only be used when delay in time of reconstruction is acceptable also it uses odd sequence length of sensor measurements. Output of forward and backward network is concatenated and passed to feedforward network which yields the final estimated latent representation of middle sensor measurement $s^t$. Thus training of $\mathcal N$ can be formulated as follows
\begin{equation}\label{eq:loss_rnn2}
    \bm \theta_M^* = \arg \min_{\bm \theta} \sum_{i=1}^{N_{\text{samp}}} \left\| h^t - \mathcal N \left( \bm s^{t-k:t+k}; \bm \theta_M \right) \right\| + \lambda\left\| \bm \theta_M  \right|_2^2,
\end{equation}
where $\theta_M$ are parameters of combined RNN and feed-forward network $\mathcal N:  \mathbb{R}^{N_s*(2k+1)}\mapsto \mathbb{R}^{h}$.

\subsection{Results}

Fig. \ref{fig:brnn_vsrnn} shows a comparative assessment between B-RNN based ARE, and RNN based ARE.
Results corresponding to one, two, five, and ten sensors are presented.
For one sensor, results obtained using B-RNN based ARE significantly outperforms those obtained using RNN based ARE.
However, as the number of sensors increases, the results obtained using the two approaches becomes identical.
One counter-intuitive result is obtained for the periodic vortex shedding problem where the result obtained using B-RNN based ARE found to be worse than that obtained using RNN based ARE.
It is probably because the neural network parameters for B-RNN has converged to a local minimum.

\end{document}